# The neutron and its role in cosmology and particle physics

Dirk Dubbers*

*Physikalisches Institut, Universität Heidelberg, Philosophenweg 12,
D-69120 Heidelberg, Germany*

Michael G. Schmidt †

*Institut für Theoretische Physik, Universität Heidelberg, Philosophenweg 16,
D-69120 Heidelberg, Germany*

Experiments with cold and ultracold neutrons have reached a level of precision such that problems far beyond the scale of the present Standard Model of particle physics become accessible to experimental investigation. Due to the close links between particle physics and cosmology, these studies also permit a deep look into the very first instances of our universe. First addressed in this article, both in theory and experiment, is the problem of baryogenesis, the mechanism behind the evident dominance of matter over antimatter in the universe. The question how baryogenesis could have happened is open to experimental tests, and it turns out that this problem can be curbed by the very stringent limits on an electric dipole moment of the neutron, a quantity that also has deep implications for particle physics. Then we discuss the recent spectacular observation of neutron quantization in the earth's gravitational field and of resonance transitions between such gravitational energy states. These measurements, together with new evaluations of neutron scattering data, set new constraints on deviations from Newton's gravitational law at the picometer scale. Such deviations are predicted in modern theories with extra-dimensions that propose unification of the Planck scale with the scale of the Standard Model. These experiments start closing the remaining "axion window" on new spin-dependent forces in the submillimeter range. Another main topic is the weak-interaction parameters in various fields of physics and astrophysics that must all be derived from measured neutron decay data. Up to now, about 10 different neutron decay observables have been measured, much more than needed in the electroweak Standard Model. This allows various precise tests for new physics beyond the Standard Model, competing with or surpassing similar tests at high-energy. The review ends with a discussion of neutron and nuclear data required in the synthesis of the elements during the "first three minutes" and later on in stellar nucleosynthesis.

*dubbers@physi.uni-heidelberg.de
† M.G.Schmidt@thphys.uni-heidelberg.de

**CONTENTS**

Sections marked with an asterisk* contain complementary material not included in the RMP article.

# I. INTRODUCTION

In the morning, a weight watcher sees the mass of typically 40 kilograms of protons, 32 kilograms of neutrons, and 22 grams of electrons. Neutrons are as ubiquitous on earth as are protons and electrons. In the laboratory, we can produce free electrons simply by heating a wire and free protons by igniting a hydrogen gas discharge, but things are not so easy for neutrons. Neutrons are tightly bound in nuclei and free neutrons are unstable with a lifetime of about 15 minutes, so there are very few free neutrons in our environment. To set neutrons free we need a nuclear reaction, a MeV rather than an eV process, and to obtain neutron fluxes competitive for modern scientific experimentation we need big nuclear installations.

As its name says, the outstanding property of the neutron is its electric neutrality. In fact, the neutron is the only long-lived massive particle with zero charge (if we neglect the small neutrino mass). This makes neutrons difficult to handle, as one cannot easily accelerate, guide, or focus them with electric fields. Furthermore, one can detect slow neutrons only in a destructive way, via nuclear reactions that produce secondary ionizing reaction products. Once detected, neutrons are gone, in contrast to charged particles, which leave an ionization track that we can detect in a destructive-free manner. Another disadvantage of neutrons is that fluxes available are much lower than photon, electron, or proton fluxes.

So why take the pain and work with neutrons when they are so difficult to produce and to handle? The answer is, although the neutron's neutrality is a burden, it is also its greatest asset (we find a similar situation for the neutrino: its neutrality to all charges except the weak charge is both its greatest asset and, for the experimenter, a heavy burden). When the usually dominant long-range Coulomb interaction is not present, other more subtle effects come to the forefront. Being hadrons (i.e., particles composed of quarks), neutrons respond to all known interactions: the gravitational, the weak, the electromagnetic, and the strong interaction. They, presumably, also respond to any other "fifth" interaction that might emanate, for example, from the dark matter that pervades the universe. These non-Coulombic interactions are all either weak, short-ranged, or both. In addition, with no Coulomb barriers to overcome, neutrons can easily penetrate condensed matter (including nuclear matter), with often negligible multiple scattering, and for our purposes are essentially point-like probes with negligible polarizability.

In condensed matter research, which is a main application of slow neutrons, neutrons have other decisive advantages. By definition, thermal and cold neutrons have kinetic energies that match the typical energies of thermal excitations in the condensed state. By pure coincidence, the size of the neutron's mass is such that, at thermal velocities, their de-Broglie wavelengths are in the Ångstrom range and thus match typical atomic distances as well. Hence, neutrons can give both structural information (by elastic diffraction) and dynamical information (by inelastic spectroscopy). Neutrons are the only penetrating particles with this feature, which enables them to tell us both "where the atoms are", and, at the same time, "what the atoms do". Dynamical information is provided separately for any length scale as chosen by the experimenter via momentum transfer (selected, for example, by the scattering angle). In the specialist's language: Neutrons provide us with the full atomic space-time correlation functions of the material under study, both for self-correlations of the atoms (from incoherent neutron scattering) and for their pair-correlations (from coherent neutron scattering).

By another unique coincidence (also linked to the neutron mass via the nuclear magneton), the size of the neutron's magnetic moment is such that amplitudes of magnetic electron scattering and nuclear scattering are of similar magnitude. This makes neutrons useful also for structural and dynamical studies of the magnetic properties of matter. On the instrumentation side, the magnetic interaction enables powerful techniques for spin-polarization, magnetic guidance, or confinement of neutrons. Furthermore, slow neutrons generally distinguish easily between elements of similar atomic number, and often between isotopes of the same element, the latter feature making powerful contrast variation techniques applicable. All these are properties that neutrons have and x-rays do not. Neutron-particle physicists should be well acquainted with these particularities of neutron-material science if they want to exploit fully the potential of the field.

## A. Background

Neutron-particle physics is part of the growing field of low-energy particle physics. However, is not high-energy physics, which strives for the largest energies in colliding beams of electrons, protons, or ions up to naked uranium, the synonym for particle physics? Certainly, the preferred method for the study of a new phenomenon is direct particle production at beam energies above the threshold where the process really happens. If an unknown process is mediated by a heavy particle of mass $M$, then this particle can be produced if the energy supplied in a reaction exceeds $Mc^2$. A well-known example is the discovery of electroweak unification in the 1980s when proton collision energies of hundreds of GeV became available for the first time to produce the photon's heavy brethrens $Z^0$ and $W^{\pm}$ that mediate the weak interaction.

If the available beam energy is not sufficient to make the process happen that we are after, then we must use a different method. We can see particles that herald a new process as "virtual" particles already below their threshold for direct production. The appearance of virtual particles of mass $M$ then is signalled, in accordance with the Heisenberg uncertainty principle, by a propagator of the type

$$\text{Propagator} \propto \frac{1}{(p/c)^2 - M^2}, \quad (1.1)$$

with four-momentum transfer $p$, for the case of spinless ("scalar") exchange particles. We then see the particle as a virtual precursor of the real process, via an energy-dependent rise of the relevant cross section. For example, the top quark, a particle that was still missing in an otherwise complete multiplet, was seen virtually before it was actually produced more than a decade ago. Such virtual effects, however, are large only if one is not too far away from threshold, so high-energy experiments see new processes through their virtual effects only when they are almost there. However, we want to explore processes also at scales $M^2 \gg (p/c)^2$ far beyond the Tera-eV scale accessible today, which is probably necessary to understand the deepest problems of nature and the universe. These virtual effects then are very small and in our example become

$$\text{Propagator} \propto \frac{1}{M^2}, \quad (1.2)$$

which is independent of reaction energy.

This is where low-energy particle physics comes into play. For these processes far beyond the present Standard Model of particle physics, the signal size $\propto 1/M^2$ is the same, no matter whether we work at the high-energy frontier of order Tera-eV or at the low-energy frontier of order pico-eV. Of course, we should search where discovery potential is highest.

The precision reached in high-energy experiments is impressive, but the sensitivity for small signals reached in low-energy experiments is almost incredible. In slow-neutron physics, as we shall see, one can detect:

● changes in energy of $10^{-22}$ eV and better, or of about one Bohr cycle per year, as realized in the searches for a neutron electric dipole moment and for neutron-antineutron oscillations;
● changes in relative momentum of $10^{-10}$ and better, via deflections of neutrons by several Å over a 10 m length of beam, as realized in the search for a neutron charge, as well as in neutron interferometry;
● changes in neutron spin polarization of $10^{-6}$ and better, as realized in the experiments on neutron-optical parity violation.

In most cases, the small signals searched for in low-energy particle physics violate some basic symmetry like parity or time reversal symmetry. In other cases, one searches for the breaking of some empirical "law" like baryon number conservation, which, as we shall see, would instead signal the restoration of a higher symmetry. In any case, such "exotic" footprints would help us distinguish the true signals from much larger mundane backgrounds, and, furthermore, could help identify the high-energy process responsible for it. The task of the experimenter then is to suppress all accidental asymmetries of the measurement apparatus that may mimic the effect under investigation. For a discussion of the relative merits of high-energy and low-energy physics, see Ramsey-Musolf and Su (2008).

Historically, neutron-particle physics on a large scale started in the middle of the past century at dedicated research reactors in the USA and in Russia. From the mid-seventies onwards, many experiments in neutron-particle physics were done at ILL, France, by various international collaborations. In recent years, several new neutron facilities with strong neutron-particle physics programs came into operation, which brought many newcomers to the field.

B. Scope of this review

Excellent recent reviews on experiments in neutron-particle physics were written by Abele (2008), and by Nico and Snow (2005). Holstein (2009a) edited a series of five "Focus" articles on this topic. Earlier general reviews on neutron-particle physics were written by Pendlebury (1993), and Dubbers (1991a, 1999). Conference proceedings on neutron-particle physics were edited by Soldner *et al.* (2009), and earlier by Arif *et al.* (2005), Zimmer *et al.* (2000a), Dubbers *et al.* (1989), Desplanques *et al.* (1984), and von Egidy (1978). Books on the subject include Byrne's *opus magnum* "Neutrons, Nuclei and Matter" (1994), Alexandrov's "Fundamental Properties of the Neutron" (1992), and Krupchitsky's "Fundamental Research with Polarized Slow Neutrons" (1987).

We will cite other books and review articles on more special topics in the respective chapters of this review. When existing literature is too extensive to be cited without omissions, we refer to recent reviews or papers and the references therein, with no focus on historical priorities. Particle data quoted without a reference are from the listings and review articles of the Particle Data Group, Nakamura *et al.* (2010), shortly PDG 2010. Astrophysical data are mostly from the review Bartelmann (2010), and fundamental constants from the CODATA compilation Mohr *et al.* (2008). Instrumental data quoted without a reference are from the website of the respective instrument.

The good existing coverage of past experiments allows us to focus on the relevance of neutron studies in cosmology and astrophysics, where lately many new and exciting topics have come to the forefront. On the experimental side, we concentrate on recent developments, and in addition discuss ongoing or planned experiments with no published results yet. The present arXiv article is an extended version of the Reviews of Modern Physics article, with more information on upcoming experiments, and derivations of some formulae that may be difficult to find in the literature.

TABLE I. The successive transitions of the universe; $M_{Pl}$, $T_{Pl}$, and $t_{Pl}$ are Planck mass, temperature, and time.

| Transition | Time | Temperature | |
|---|---|---|---|
| | $t_{Pl} = 5 \times 10^{-44}$ s | $kT_{Pl} = M_{Pl}c^2 = 1.2 \times 10^{19}$ GeV | |
| Inflation, GUT transitions | $t \geq 10^{-38...-36}$ s | $kT \leq 10^{15-16}$ GeV | |
| Electroweak → Electromagn. and weak | $10^{-11}$ s | 246 GeV | ↓ Standard |
| Quark-gluon plasma → Nucleons | $3 \times 10^{-5}$ s | 170 MeV | Model |
| Nucleons → Nuclei | 150 s | 78 keV | |
| Plasma of nuclei + electrons → Atoms | 374,000 yr | 0.3 eV | |
| Atoms → Stars and galaxies | ~ $5 \times 10^8$ yr | $T$ ~ 30 K | |
| Today | $13.7(1) \times 10^9$ yr | $T = 2.728(4)$ K | |

The present review is organised in such a way as to follow the cosmological evolution of the universe (as outlined in Table I), and to look at what role neutrons play at a given cosmological time/energy scale and in the sector of particle physics relevant at this scale. In Sec. II, we first give an unsophisticated outline of Standard Big Bang theory. The chapters then following are rather self-contained and can be read in any order. This reflects the fact that with each new chapter, the universe enters a new phase, and after such a transition, the new phase usually does not remember much of the old phase. In Sec. III, we discuss what neutron experiments can tell us about baryogenesis and the accompanying symmetry violations. Here the discussion of a neutron EDM plays a particular role both because it might signal new sources of *CP* violation beyond the Standard Model (SM) of particle physics and because this is a most important ingredient of baryogenesis in cosmology. Section IV is a short résumé on neutron oscillations. In Sec. V, we present new results on neutron searches for new forces, and, in particular, on searches for deviations from Newton's law of gravitation, due to, for instance, extra spatial dimension. Going down the timeline in Table I we reach the electroweak scale of the SM in Sec. VI and then discuss results and relevance of weak interaction parameters derived from neutron decay. Section VII closes our review with a description of the neutron's role in creation of the elements, both during the first minutes after the big bang and in the course of stellar evolution.

Although much of scientific progress is based on progress in instruments and methods, in this review we cannot adequately cover all the important developments made in recent years in the field of sources, transport, and detection of neutrons. For references to technical papers, see Abele (2008) and Dubbers (1991a).

Furthermore, scientific topics where there was little progress in the past decade will be treated rather shortly. This is the case for most of the neutron's electromagnetic properties listed in Sec. III.D, for neutron-antineutron oscillations shortly covered in Sec. IV.A, and for the neutron-nuclear weak interactions mentioned at the end of Sec. VI.A.3. Neutron optics, well covered in the book by Utsuro and Ignatovich (2010), will be treated here mainly in a technical context, and the beautiful experiments on neutron interferometry, which are covered in the book by Rauch and Werner (2000), and which had paved the way for much of modern atom optics, are not part of this review.

Below we give a list, in alphabetical order, of some research centers that have neutron beam stations dedicated to neutron-particle physics.

ESS: European Spallation Source Project (under design), Lund, Sweden (Vettier *et al.*, 2009).

FRM-II: National neutron source Heinz-Maier-Leibnitz, Technical University Munich, Germany.

ILL: European neutron source Institute Max von Laue - Paul Langevin, Grenoble, France.

J-PARC: Japan Proton Accelerator Research Complex, with Japanese Spallation Neutron Source (JSNS, under construction), Tokai, Ibaraki, Japan.

LANSCE: Los Alamos Neutron Science Center at Los Alamos National Laboratory (LANL), USA.

NIST: National Institute of Science and Technology, with NCNR neutron source, Gaithersburg, USA.

PNPI: Petersburg Nuclear Physics Institute, with WWR-M neutron source, Gatchina, Russia,

PSI: Paul Scherrer Institut, with SINQ quasi-continuous spallation source, Villigen, Switzerland.

SNS: Spallation Neutron Source at Oak Ridge National Laboratory (ORNL), USA.

TRIUMF: National Laboratory for Particle and Nuclear Physics, with ultracold neutron (UCN) Source Project (design phase), Vancouver, Canada (Martin *et al.*, 2008).

Tables II and III give, where available, some basic data for these reactor and spallation neutron sources, taken from the facilities' web sites, unless cited otherwise. Of course, other beam quantities like neutron brightness, i.e., neutrons/cm$^2$/s/nm/sr, neutron temperature, beam profile, divergence, time structure, or degree of spin-polarization are also important parameters that may vary from place to place. Most facilities listed have projects on upgrades, either by an increase in source power, or by the installation of better neutron guides, see for instance Cook (2009). We shall discuss neutrons fluxes and the specific properties of cold and ultracold neutrons in the context of the individual experiments.

TABLE II. Main neutron reactor sources having cold neutron beamline(s) for particle physics. Given are the reactor power, the in-pile neutron flux density, and the capture flux density available at the cold neutron beam station.

| Reactor facility | Start/ upgrade | Power (MW) | $n$-capture flux (cm$^{-2}$ s$^{-1}$) in-source | $n$-capture flux (cm$^{-2}$ s$^{-1}$) in-beam | References on $n$-flux |
|---|---|---|---|---|---|
| ILL | 1971/94 | 58 | $1.5\times10^{15}$ | $2\times10^{10}$ | Häse *et al.*, 2002 (upgrade 2009) |
| FRM-II | 2004 | 20 | $8\times10^{14}$ | $1\times10^{10}$ | |
| NIST | 1967/97 | 20 | $4\times10^{14}$ | $2\times10^{9}$ | Nico and Snow, 2005 |
| PNPI | 1959 | 16 | $1\times10^{14}$ | $6\times10^{8}$ | |

TABLE III. Main neutron spallation sources having cold neutron beamline(s) for particle physics. Listed are the power of the proton beam (a + sign indicates: in the process of ramping up), the neutron flux density within the source (time-average and peak, where applicable), and the capture flux density available at the cold neutron beam station.

| Spallation facility | Start/ upgrade | Power (MW) | $n$-capture flux (cm$^{-2}$ s$^{-1}$) in-source peak | $n$-capture flux (cm$^{-2}$ s$^{-1}$) in-source average | $n$-capture flux (cm$^{-2}$ s$^{-1}$) in-beam average | References on $n$-flux |
|---|---|---|---|---|---|---|
| SNS | 2008 | ~1+ | $5\times10^{16}$ | $1\times10^{14}$ | ~$1.3\times10^{10}$ @1.4 MW | Ferguson, Greene, 2010, private communication |
| PSI | 1998 | 1.3+ | n.a. | $1\times10^{14}$ | $10^{9}$ | |
| LANSCE | 1985 | 0.1 | $1.7\times10^{16}$ | -- | $1\times10^{8}$ | Nico and Snow, 2005 |
| J-PARC | 2008 | 0.4+ | -- | -- | ~$10^{9}$ | Mishima *et al.*, 2009 |

# II. HISTORY OF THE EARLY UNIVERSE

Particle physics, the physics of the very small, and cosmology, the physics of the very large, are intimately interwoven in standard big bang theory (or variants thereof). Hence, for each phenomenon seen in particle physics at some high-energy scale, we find an epoch in the evolution of the universe when temperature and density were so high that the phenomenon in question played a dominant role. To set the scene, we shall therefore first give a short history of the early universe, and then discuss the various experiments done in neutron physics that help shed light on the corresponding epoch of the universe. Even when a scale is not accessible with neutron experiments (the inflation scale, for example), it must still be shortly discussed in order to understand the subsequent processes of baryogenesis, etc.

We start at the earliest age of the universe, at which our present understanding of the laws of physics may be sufficient to describe the evolution of the universe, using the standard quantum field theory leading to the Standard Model (SM) of elementary particle physics, and general relativity leading to the Standard Model of cosmology. This starting time is definitely later than the Planck time $t_{Pl} = 5 \times 10^{-44}$ s, which latter corresponds to an energy scale of $E_{Pl} = 1.2 \times 10^{19}$ GeV. At this scale, quantum effects of gravity are dominant, and today's theories are not developed enough to cope with in any detail.

After this time, the history of the young universe becomes a succession of phase transitions (either in the strict sense, or less so), starting with an inflationary period, explained below, and (perhaps simultaneously) with a breaking of grand unification. Above the scale of grand unification, the electroweak and the strong interactions are unified, below this scale, they describe separate processes. Later on, we reach the scale of the present SM, at which electroweak unification is broken, and below which electromagnetic and weak interactions are separated. These transitions occur at specific transition temperatures during the cool-down of the expanding universe, as listed in Table I, see, for example, the books by Kolb and Turner (1990) or by Mukhanov (2005), and later on are followed by a series of condensations (nucleons to nuclei, etc.).

In each of the early transitions new properties of the system pop up spontaneously, which were not present in the system before the transition. In this way, the universe successively acquired more and more structure, which is a precondition for filling it with life; the initial high symmetry, although of great beauty, in the end would be rather sterile. Such a transition could be, for example, from a single type of particle field possessing high internal symmetry to multiplets of fields of lower symmetry. In this way, instead of one "grand-unified" interaction of fields and particles, there appear various interactions with different properties. In case of a phase transition, the critical temperature sets the energy scale of the process of spontaneous symmetry breaking. We call the hot phase above the critical temperature with vanishing order parameter the unbroken or symmetric phase, and the phase below the critical temperature broken or unsymmetrical.

We start with a look on the "old" big bang theory (Weinberg's "first three minutes"), whose overall evolution is very simple so we retrace it in the following few lines. For a modern view see Bartelmann (2010) and Mukhanov (2005), and the books recommended there. Let $R(t)$ be the cosmological scale factor at a given time $t$ (the "radius" of the universe), then the expansion rate of the universe at time $t$ is the Hubble function defined as $\mathcal{H}(t) = \dot{R}/R$. The evolution of $R(t)$, $\mathcal{H}(t)$, and of temperature $T(t)$ (see Table I) are derived from the Friedmann equation, which reads, for a curvature parameter $k = 0$ of the flat universe, and a cosmological constant $\Lambda = 0$,

$$\mathcal{H} = \frac{\dot{R}}{R} = \frac{1}{c}\left(\frac{8\pi}{3}G\rho\right)^{1/2}. \tag{2.1}$$

The gravitational constant $G$ is related to the Planck mass in Table I as $G = \hbar c / M_{Pl}^2$.

The energy density of the universe is divided into the mass density $\rho_m$ and the relativistic radiative energy density $\rho_r$, such that $\rho = \rho_r + \rho_m$. In the radiation-dominated era, which lasted for the first about 400 000 years of the universe, the energy density varies with scale $R$ as $\rho \approx \rho_r \propto 1/R^4$. The extra factor of $1/R$ as compared to the mass density $\rho_m \propto 1/R^3$ stems from the red-shift of all wavelengths $\lambda \propto R$ in the expanding universe, such that radiative energy of relativistic particles varies as $E_r \approx pc = hc/\lambda \propto 1/R$, where we have used the de Broglie relation. With $\rho_r \propto 1/R^4$ entered in Eq. (2.1) we obtain $\dot{R} \propto 1/R$ and hence, as verified by insertion,

$$R(t) \propto \sqrt{t}, \tag{2.2}$$

$$\mathcal{H}(t) = \dot{R}/R = 1/2t, \tag{2.3}$$

$$T(t) \propto 1/\sqrt{t}. \tag{2.4}$$

The last equation follows from the Stefan-Boltzmann law

$$\rho_r = \frac{\pi^2}{30}\frac{N}{(\hbar c)^3}(kT)^4, \tag{2.5}$$

when we compare $\rho_r \propto T^4$ with $\rho_r \propto 1/R^4$, which gives $T \propto 1/R$, and insert Eq. (2.2). The numerical relation is $kT/\text{MeV} = 1.55 N^{-1/4} (t/\text{s})^{-1/2}$, by insertion of Eq. (2.5) into Eq. (2.1), and use of Eq. (2.3). To take into account the transitions between the different stages of the universe, we note that the number of degrees of freedom of highly-relativistic particles $N$ in Eq. (2.5) decreases with every freeze-out process in Table 1, so $N = N(T)$ as shown in Fig. 19.3 in Olive and Peacock (2009), which leads to a small correction of Eq. (2.4). The number of degrees of freedom finally reached the stable value $N = 7\frac{1}{4}$ before nucleosynthesis set in, see Sec. VII.

In the later matter-dominated universe, the energy density becomes $\rho \approx \rho_m \propto 1/R^3$, which, when inserted into Eq. (2.1), changes the solutions from Eqs. (2.2) and (2.3) to $R \propto t^{2/3}$, and $\mathcal{H} = 2/3t$. At all times, temperature changes and scale changes are related as $\dot{T}/T = -\dot{R}/R = -\mathcal{H}$. Both rates diverge at the "origin of time", the unphysical big bang singularity, and so need to be substituted by a forthcoming theory of gravity.

However, to our present understanding, for a certain period of time in the very early universe, the expansion rate of the universe is determined by an effective cosmological constant $\Lambda > 0$ representing the energy density of an "inflaton field". Instead of Eq. (2.1), we then have $\dot{R}/R \approx (\Lambda/3)^{1/2}$, which is the equation for exponential growth with the "inflationary" solution

$$R(t) \propto \exp\sqrt{\Lambda/3}\, t. \qquad (2.6)$$

If inflation lasts for a time such that the exponent reaches at least 50 to 60, it can explain the observation that the universe is both homogeneous and isotropic, which, without inflation, would lead to problems with causality. Furthermore, inflation explains how the universe is flat on the one percent level, which would otherwise require an immense fine-tuning, but is now, instead, substituted by a set of initial conditions for the inflaton field $\varphi$, a scalar field with a self-interaction potential $V(\varphi)$. For access to the vast literature on this subject, we again refer to Bartelmann (2010) and Mukhanov (2005).

It is also a wonderful effect of inflation that it automatically creates fluctuations, first as quantum fluctuations, which then turn into classical fluctuations. This effect is absolutely necessary for the structure formation (galaxies, stars) in the late history of our universe. In recent years the inflationary scenario has become accessible to observation via the precision measurements of the cosmic microwave background (CMB), offering a snapshot of the universe when it became transparent at temperature $T \approx 3000$ K. The directional fluctuations of this background reflect the quantum fluctuations of the inflationary epoch. They are found to be nearly scale invariant, the relevant measure for this being the "spectral index" $n_s$, expected to be equal to one for everlasting inflation, and slightly less for a finite inflation period. Indeed, CMB satellite measurements find $n_s = 0.963 \pm 0.015$. The parameter describing the amplitudes of these quantum fluctuations can be measured precisely, too, so inflation is very accessible to observation today. These data are compatible with the simplest inflationary model, while some other more complicated models are in trouble, see Hinshaw *et al.* (2009), and references therein. In spite of these successes, we have to admit that we still do not understand the real nature of inflation.

When inflation ends, all ordinary matter and radiation fields will have their densities diluted by up to 100 orders of magnitude, so the universe ends up extremely cold and essentially empty. The inflaton field $\phi$, however, is about the same as before, but its kinetic energy then dominates over the potential $V(\phi)$, so the field may oscillate rapidly and decay into ordinary matter and radiation. By this refilling or "reheating" process, all the particles are created from the inflaton energy liberated in the transition.

Due to the continuing expansion of the universe, the temperature reached in the reheating process after inflation is much lower than the about $10^{14-16}$ GeV involved in inflation. Grand unification, however, must take place at about $10^{16}$ GeV, and therefore cannot occur after the end of inflation. Thus, Grand Unified Theory (GUT) can only play a role if GUT breaking comes together with inflation as done in "hybrid" inflation models. Therefore inflation and GUTs form one entry in Table I.

The cosmological constant after inflation and GUT transitions is near zero, but evaluation of recent observations on the CMB and on supernovae showed that, besides "dark" unknown matter, again, like in inflation, a kind of (now "small" compared to the previous value!) cosmological constant $\Lambda$ is required, leading to an accelerated expansion of the universe today. If, in particular, this "constant $\Lambda$" is a time-dependent field, it is called dark energy ("quintessence", "cosmon"), see Wetterich (1988), and Ratra and Peebles (1988). During the past few years, new astrophysical data from different sources (structure formation, redshifts, and others) confirmed consistently the presence of large amounts of dark matter and of dark energy, see, for example, Mukhanov (2005).

To conclude, cosmology and astrophysics have become accessible to experimental tests up to the very first instances of the big bang, and we shall see what neutron physics has to offer in this endeavor.

## III. BARYOGENESIS, *CP* VIOLATION, AND THE NEUTRON ELECTRIC DIPOLE MOMENT

In cosmology, the most obvious fact still lacking explanation is the existence of large amounts of matter that form the galaxies, the stars, and interstellar matter, with essentially no antimatter, in a universe that started with no particle content right after inflation. This problem of baryogenesis turns out to be closely linked to the possible existence of electric dipole moments of particles. All explanations of baryogenesis in the early universe necessarily require a significant violation of what is called "*CP* symmetry". In this case, particles like the neutron or the electron should carry a *CP* violating electric dipole moment, where specific models of baryogenesis may be tested in ongoing EDM experiments. To substantiate this claim we first discuss the models of baryogenesis in some detail. We then give a survey of ongoing and planned neutron EDM experiments, and, finally, give a review of the theoretical implications of the neutron and atomic EDM limits.

### A. Baryogenesis

#### 1. The dominance of matter over antimatter

The matter content of the universe usually is quantified by the ratio of the number density of baryons (i.e., hadrons composed of three quarks, which today are essentially the proton and the neutron) to the number density of photons, which is

$$n_b / n_\gamma = (6.08 \pm 0.14) \times 10^{-10}, \qquad (3.1)$$

as derived from the recent CMB satellite data and from primordial nucleosynthesis data, see Sec. VII.A.4. At the present temperature $T = (2.728 \pm 0.004)$ K of the universe the photon number density is $n_\gamma = 405 \text{ cm}^{-3}$. There is strong evidence that the universe consists entirely of matter, as opposed to antimatter, see Cohen *et al.* (1998), and the few antiprotons $n_{\bar{p}}/n_p \sim 10^{-4}$ seen in cosmic rays can all be explained by secondary pair production processes. This observed dominance of matter over antimatter is not what we expect from the standard big bang model discussed above, in combination with the known laws of physics as subsumed in the SM of particle physics.

A cornerstone of particle physics is the *CPT* theorem. Here, *C* stands for the operation of charge conjugation, which carries a particle state into its corresponding antiparticle state, *P* stands for the parity operation that inverts the spatial coordinates $\mathbf{r} \to -\mathbf{r}$ of the particle's state, and *T* stands for the time reversal operation that inverts the time coordinate $t \to -t$ of the state, and, simultaneously, interchanges the ingoing and outgoing states. A system is called *C*, *P*, or *T*-symmetric, when it is invariant under these operations, respectively. In nature, each of these discrete symmetries is violated in some system or other. However, in standard local quantum field theory, our basis, the interactions are invariant under the combined operations *CPT*, with the consequence that particles and antiparticles have identical properties, except for the signs of their various charges, and have equal distributions at thermal equilibrium. *CPT* invariance implies that *CP* and *T* are either both good symmetries, or are both violated.

Given *C* and *CP* invariance, equal numbers of particles and antiparticles were created in the early universe after inflation. Later on, particles and antiparticles should have almost completely annihilated each other, falling out of equilibrium only at a time when the density of the expanding and cooling universe had become so low that the remaining particles and antiparticles in question did not find each other anymore. For the nucleons and antinucleons this happened when the universe had an age of about 2 ms, with the corresponding freeze-out temperature of $kT \approx 20$ MeV. We shall discuss in more detail the mechanism of decoupling from thermal equilibrium in another context in Sec. VII.A.

The number of baryons and antibaryons left over after this "carnage" should be a minuscule $n_b / n_\gamma = n_{\bar{b}} / n_\gamma \sim 10^{-18}$, which, for the baryons, is more than eight orders of magnitude below the observed value, Eq. (3.1). Hence, one would expect the so-called baryon asymmetry $\eta = (n_b - n_{\bar{b}})/n_\gamma$ to vanish, whereas observation gives $n_{\bar{b}} \approx 0$, so $\eta \approx n_b/n_\gamma \approx 6 \times 10^{-10}$. The process responsible for this unexpectedly large baryon asymmetry of the universe (acronym BAU), generated from an initially symmetric configuration, is called baryogenesis. Baryogenesis must have taken place after inflation, otherwise particle density would have been diluted to zero.

There are three necessary conditions for baryogenesis called the three Sakharov criteria:

(1) Baryon number *B*, which is the number of baryons minus the number of antibaryons, should not be a conserved quantity, that is, there must be elementary processes that violate baryon number, in order that baryogenesis can proceed from an initial total $B = 0$ to a universe with $B > 0$.
(2) Charge conjugation symmetry *C* and the combined symmetry *CP* should both be violated, the latter of

which, under the *CPT* theorem, is equivalent to a violation of time reversal symmetry *T*. Indeed, one can show that if *C* or *CP* were exact symmetries, then the reactions that generate an excess of baryons would occur at the same rate as the conjugate reactions that generate an excess of antibaryons, such that the baryon asymmetry would retain its initial value $\eta = 0$.

(3) Baryon-asymmetry generating processes must take place far from thermal equilibrium, because otherwise, even for processes violating baryon number *B* as well as *C* and *CP* symmetry, both particles and antiparticles, due to *CPT* symmetry, would still have the same thermal distribution.

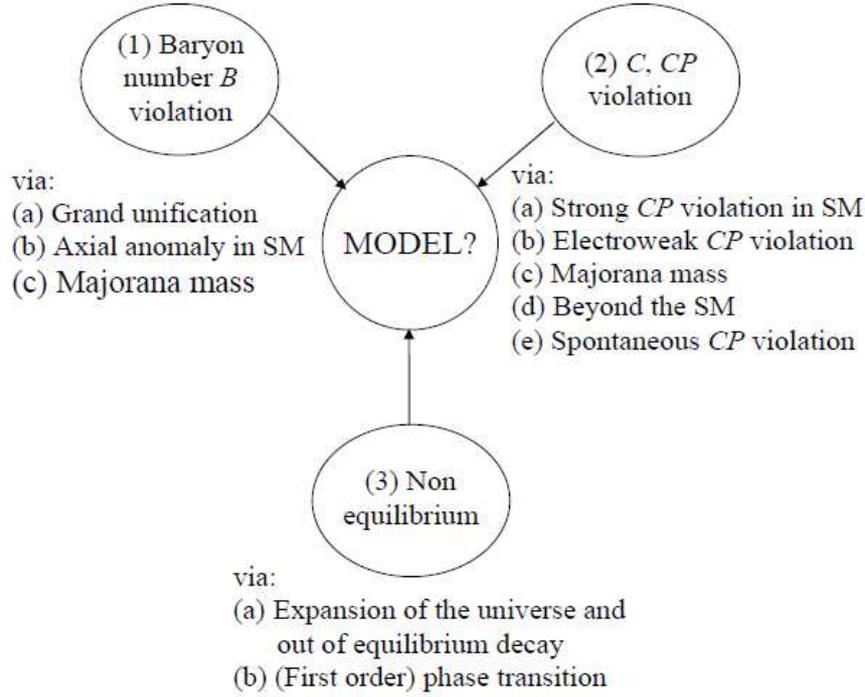

FIG. 1. The three Sakharov criteria, necessary for baryogenesis, and their possible realizations (SM = Standard Model).

Our mere existence requires that all of these three necessary conditions for generating a baryon asymmetry have been satisfied at some time in the early universe. As is visualized in Fig. 1, there are various ways to satisfy these conditions, as we shall discuss in the following. However, even for a given set of realizations, it will turn out that the construction of models is certainly restricted but is in no way unique.

## 2. Some quantities of the Standard Model

Let us first recall some basic quantities of the electroweak Standard Model that will be needed in the discussions to follow. In the SM, all particles obtain their mass through the Higgs process. The Landau "quartic" effective potential of the Higgs field $\Phi$

$$V(\Phi) = \mu^2|\Phi|^2 + \tfrac{1}{4}\lambda|\Phi|^4, \qquad (3.2)$$

acquires its characteristic symmetry breaking "Mexican hat" shape when $\mu^2(T)$ turns negative at the critical temperature $T = T_c$, see the dashed line in Fig. 4 below.

In Landau theory, for $T < T_c$ the classical nonzero expectation value of the Higgs boson $\upsilon = \sqrt{2}\langle\Phi\rangle = 2(-\mu^2/\lambda)^{1/2}$ appears as an order parameter. Near the new potential minimum, the still unobserved Higgs boson moves with a mass $m_H = (-2\mu^2)^{1/2} = \upsilon(\lambda/2)^{1/2}$. The *W* boson, which mediates the electroweak interaction, obtains its mass $m_W = \tfrac{1}{2}g_w\upsilon$ from the Higgs process, with the observed value $m_W \approx 80 \text{ GeV}$ at $T = 0$, where $\upsilon = 246 \text{ GeV}$. The range of the electroweak interaction is the Compton wavelength of the *W* boson $\lambda_C = 2\pi\hbar/m_W c = 1.5\times 10^{-17}$ m. The weak $SU(2)$ gauge coupling constant in the SM is

$$g_w = e / \sin\theta_W \approx 0.65, \qquad (3.3)$$

where $e = (4\pi\alpha)^{1/2} \approx 0.31$ is the (dimensionless) elementary electric charge. The fine structure constant therein is taken at the scale of $M_Z$, the mass of the neutral electroweak $Z^0$ exchange boson ($\alpha \approx 1/128$ instead of its usual value $1/137$). At scale $M_Z$, the weak or Weinberg angle has been measured to $\sin\theta_W \approx 0.48$, and $M_Z = M_W / \cos\theta_W \approx 91$ MeV.

## 3. First Sakharov criterion: Violation of baryon number

Violation of baryon number can have several sources.

a. *Grand unification*: GUT models violate baryon number $B$ because in their "unbroken" phase they unify quark, lepton, and gauge interactions, as we shall exemplify in the simplest GUT, based on $SU(5)$ symmetry. In this model, quarks (baryons) and leptons occupy the same particle multiplet such that they can transform into each other, which changes both baryon number $B$ and lepton number $L$. The proton ($B=1$), for instance, may decay, $p^+ \to \pi^0 + e^+$, by changing a quark into a positron (lepton number $L=-1$), with $\Delta B = 1 - 0 = 1$ equal to $\Delta L = 0 - (-1) = 1$, so $\Delta(B+L) = 2$ and $\Delta(B-L) = 0$, thus violating $B+L$ but conserving $B-L$.

*CP*-violating GUT breaking must have taken place after inflation, however, there are the problems mentioned to return, after inflation, to the unbroken GUT-symmetry (see end of the preceding Sec. II). Therefore GUT-baryogenesis, which was most important in the early days of cosmology, is not attractive anymore, unless combined with inflation in hybrid inflation models. Furthermore, in the case of $B-L$ conserving $SU(5)$, even if a baryon asymmetry had been generated, it would have been washed out later on in an equilibrium epoch before the electroweak phase transition due to ("hot"!) sphaleron transitions, outlined next.

b. \**Axial anomaly*: The SM permits baryon number violation and, in principle, baryogenesis, too (Kuzmin *et al.*, 1985). Baryon number violation then is due to the so-called axial anomaly in the electroweak sector. In this effect of quantum field theory, the total number $B+L$ of quarks and leptons is violated if the topological winding number of a certain weak gauge field configuration that they are located in does not equal zero ("change in the Chern-Simons number"), while $B-L$ remains conserved in this process. (Example: The creation of a hydrogen atom as found in large amounts in the present universe changes an initial $B+L=0$ to $B+L=2$ while leaving $B-L=0$ unchanged.)

For temperature $T=0$ this is a tiny effect, presumably unmeasurable in laboratory experiments. This is because it is induced by "instanton" gauge field configurations, quantum mechanical tunneling solutions in Euclidean space with imaginary time that mediate tunneling between the different topologically inequivalent vacua. They have an exponential transition probability, typical for tunneling,

$$\Gamma_{instanton} \sim \exp(-8\pi^2 / g_w^2), \qquad (3.4)$$

with the electroweak gauge coupling $g_w$ from Eq. (3.3) With a "Gamow factor" of $8\pi^2 / g_w^2 \approx 190$, this transition probability is extremely small.

Figure 2 gives a typical graph of the effective action for chiral left-handed quarks and leptons with initially $B+L=2$ for the simplified case of just one generation. This demonstrates that $\Delta(B+L)=2$ and $\Delta(B-L)=0$.

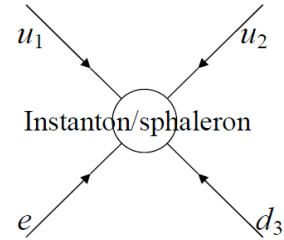

FIG. 2. In instanton or sphaleron transition between two vacuum states of Fig. 3, a baryon $B$ number violating effective instanton interaction (1, 2, 3 = color indices) is induced between $u$, $d$ quarks and the electron (one generation).

If, however, the temperature $T$ is high, then there are similar transitions, but now we have thermal transitions over an unstable three-dimensional sphaleron (gauge-Higgs) configuration, see Fig. 3. This transition again is exponentially suppressed, but now only by a Boltzmann factor

$$\Gamma_{sphaleron} \sim \exp(-E_{sph} / kT), \qquad (3.5)$$

where $E_{sph}(T) = f \, m_W(T) \, (8\pi / g_w^2)$ is the sphaleron energy (setting $c = \hbar = 1$), with $m_W$ the mass of the W-boson, and a numerical factor $f$ (Klinkhammer and Manton, 1984).

As already mentioned in Sec. 2, the value of $v$, and with it the W mass $m_W = \tfrac{1}{2} g_W v$, is temperature dependent, $v = v(T)$: For increasing temperature $T$, the Higgs field, besides a negative (mass)$^2$, acquires an additional positive mass term that can be shown to be $\sim T^2|\Phi|^2$, and its classical value $v(T)$ is reduced, being zero in the hot phase above the electroweak transition temperature. The sphaleron energy $E_{sph}$ of Eq. (3.5) is due to a classical solution of the gauge-Higgs system. In the hot phase, with the Higgs expectation value vanishing, one has a much more complicated three-dimensional nonperturbative gauge configuration, which leads to unsuppressed $B+L$ violation (Bödeker *et al.*, 2000), the "hot sphaleron effect".

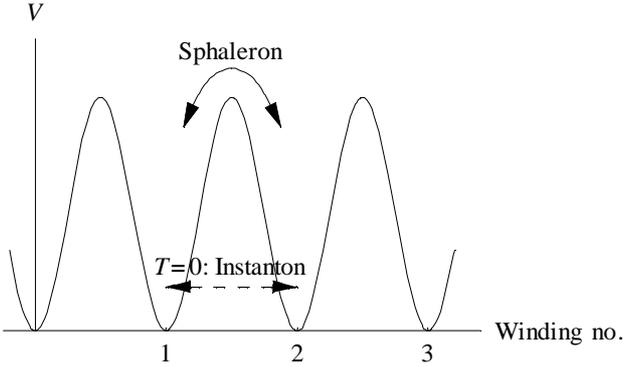

FIG. 3. In the $SU(2)_w$ gauge theory, there are degenerate ground states with nonzero topological winding number. Baryon number violating transitions between two such SM vacua are possible, either by tunnelling (instanton transition at $T=0$), or by thermal excitation over the barrier (sphaleron transition).

c. *Majorana mass*: Very interesting for baryogenesis would be the existence of very massive Majorana neutrinos giving a mass to ordinary neutrinos via the so-called seesaw mechanism. A massive Majorana lepton (i.e., a lepton that is identical to its antiparticle) has a mass term in the Lagrangian that violates lepton number $L$ conservation. The out-of-equilibrium decay of massive Majorana neutrinos ("leptogenesis", see Buchmüller *et al.*, 2005, and references therein) can violate both *CP* and lepton number. A lepton asymmetry then is partly transformed into a baryon $B$-asymmetry via a hot sphaleron process in thermal equilibrium above the electroweak scale. Starting with a lepton asymmetry, $B_0=0$ and $L_0 \neq 0$, one ends up with a baryon asymmetry $B = (8N+4)/(22N+13)$ $\times |B_0 - L_0| \neq 0$, that is, $B = \tfrac{28}{79}|L_0|$ for $N=3$ particle generations.

If instead one starts with $B_0 - L_0 = 0$, as in $SU(5)$ GUTs, then one finds that $B_0 + L_0 = 2L_0 \neq 0$ is washed out to $B+L \approx 0$, and with conserved $B-L=0$ this leads to $B=0$. Thus, the hot sphaleron transition has two faces: In thermal equilibrium, it leads to $B=0$ if originally $B_0 - L_0 = 0$, and in thermal nonequilibrium (as in a strong electroweak phase transition to be discussed later) it can produce a baryon asymmetry via a hot sphaleron process.

## 4. Second criterion: *CP* Violation

The violation of *C* (and *P*) is no problem as it is large for the electroweak theory, so the violation of *CP* is the real problem. In the SM there are two kinds of possible *CP* violations, and we will discuss others beyond the SM.

a. *Strong CP violation*: In the strong sector of the SM, one expects *CP* violation by an additional gauge kinetic term

$$L^{CPV} = -\theta \frac{g_s^2}{32\pi^2} \sum_{a=1...8} \tilde{G}_{\mu\nu}^a G^{\mu\nu a}, \quad (3.6)$$

the so-called $\theta$-term in the quantum chromodynamic (QCD) Lagrangian, with the strong coupling constant $g_s$, field tensors $G^{\mu\nu a}$ and $\tilde{G}_{\mu\nu}^a = \tfrac{1}{2}\varepsilon_{\mu\nu}^{\lambda\sigma} G_{\lambda\sigma}^a$, and color index $a$. This is equivalent to a complex phase in the quark mass determinant: both are related by an anomalous axial $U(1)_A$ symmetry transformation. This "strong *CP* violation", see Peccei (2008) and references therein, could be large in principle, which is called the strong *CP* problem, but up to now, there is no experimental indication for this. Even small $\theta$-values would produce large EDMs of the neutron. Introducing $\theta$ as an "axion" field and a $U(1)_{PQ}$ - Peccei-Quinn symmetry, as in, for example, the Minimal Supersymmetric Standard Model (MSSM, see below), $\theta \equiv 0$ is a minimum of the classical action. Equally, perturbative effects mixing the strong *CP* problem (present case a) and *CP* violation in the Cabibbo-Kobayashi-Maskawa quark mixing matrix (or CKM-matrix, next case b) only lead to tiny values of $\theta$.

b. *Electroweak CP violation*: In the electroweak sector of the three-generation SM, there is the experimentally well confirmed *CP* violation in *K* and *B* decays. This *CP* violation is naturally integrated into one complex phase $\delta$ of the left-handed down quarks in the CKM matrix (described in more detail in Sec. VI.A). As we shall see below, the Jarlskog determinant as a measure of the strength of electroweak *CP* violation would indicate that this effect is much too small for baryogenesis, but one has to inspect that very carefully. Both sources (a) and (b) of *CP* violations are genuine in the SM.

c. *Massive neutrinos*: If we admit massive neutrinos in the SM, as dictated by the recent discovery of neutrino oscillations, a Majorana type mass leading to the seesaw mechanism generates not only lepton number violation, as mentioned in Sec. 3.b above, but also free complex phases in the mass matrix that violate *CP*. Such *CP* violation in the lepton sector may be interesting for baryogenesis (then called leptogenesis, see Buchmüller *et al.*, 2005, and for the "νMSM" extension of the SM with three right-handed neutrinos see Shaposhnikov and Tkachev, 2006), but it would have very tiny effects on the neutron EDM.

d. *In models beyond the SM* there are more *CP* violating field phases that cannot be transformed away, and which could be large. This already happens in models with two Higgs doublets, see Bernreuther (2002), but is most prominent in supersymmetric models (SSM). Supersymmetry (SUSY) is a symmetry between bosons and fermions. In the minimal SUSY model (MSSM), each particle of the SM has a superpartner with equal mass and with equal couplings – the quarks share a supermultiplet with the squarks and the gauge bosons with the gauginos. As these superpartners are not observed in experiment, the supersymmetry has to be broken, explicitly by soft breakings in an effective theory, or spontaneously in a more fundamental theory. Furthermore, instead of one Higgs doublet, two doublets are required, $H_u$ coupling to the up and $H_d$ to the down quarks. In the MSSM, free phases in the Higgsino-gaugino and the $stop_R - stop_L$ mass matrices violate *CP* ($stop_R$ and $stop_L$ are SUSY-partners of the right and left-handed top quark). Extensions of the (minimal!) MSSM have an additional SM-neutral singlet chiral superfield (NMSSM, nMSSM, see also Sec. 7 below).

e. *Spontaneous breaking of CP*: Whereas in the model(s) with two Higgs doublets we have spontaneous *CP* violation in the Higgs sector, the breaking of *CP* symmetry is explicit in most SUSY models. There is also the exciting possibility of a spontaneous breaking of *CP* symmetry just around the phase transition. In the MSSM, unfortunately, this does not happen (Huber *et al.*, 2000, Laine and Rummukainen, 1999), but might happen in the NMSSM (Huber and Schmidt, 2001).

**5. Third criterion: Nonequilibrium**

Thermal nonequilibrium in the early cosmos, in principle, poses no problem:

a. *Expansion of the universe*: Some reactions or decays fall out of equilibrium if they become too slow as compared to the Hubble expansion rate, $1/t_{\text{reaction}} < \mathcal{H}$ or $\Gamma_{\text{decay}} < \mathcal{H}$, respectively. Still, this is a quantitative question if it comes to a successful model of baryogenesis. This source of nonequilibrium is particularly effective at the very early times soon after inflation, for leptogenesis, for example, with a slow out of equilibrium decay of Majorana neutrinos (Buchmüller *et al.*, 2005).

b. *Phase transitions* occurring either after the inflationary phase or at the electroweak scale are more vigorous as sources of nonequilibrium. In the first case (high scale), as mentioned before, one can also construct "hybrid inflation models" where inflation and the breaking of a GUT symmetry come together. By introducing a *CP* violating term in the effective interaction of such hybrid models, one can study an asymmetry in the production of particles after inflation using (quantum) transport equations (Garbrecht *et al.*, 2006). This asymmetry then is transformed into a lepton asymmetry via Majorana fermions, which appear quite naturally in, for example, *SO*(10) GUTs, and into a baryon asymmetry by the (hot) sphaleron transition.

The second case of a (rather low scale) electroweak transition is most interesting because present experiments at CERN (the European center for particle physics in Geneva, Switzerland) approach the TeV energy scale, and variants of the SM can be tested. In contrast, the succeeding QCD transition for the quark gluon plasma to the hadronic phase (Table I) is very mild ("crossover" transition), judging from QCD lattice calculations (unless there are strong effects of the baryonic chemical potential, see Boeckel and Schaffner-Bielich, 2010).

**6. Baryogenesis in the Standard Model (SM)**

Indeed, at first look the SM seems to fulfill all three conditions (Kuzmin *et al.*, 1985):

(1) There is baryon number *B* violation from the sphaleron transition.
(2) There is *CP* violation in the electroweak sector (CKM matrix), and, possibly, in the strong sector ($\theta$-term).
(3) Far off equilibrium, there could be a strong electroweak phase transition, with $\langle H \rangle = 0$ above the electroweak scale, and a classical neutral Higgs-field $H$ value $v(T) = 2^{1/2} \langle H(T) \rangle$ below the electroweak scale going to $v = 246$ GeV at $T = 0$.

Naively, within the SM, in a Higgs effective potential, Eq. (3.2), a temperature dependent additional positive Higgs mass term proportional to $T^2$ (mentioned in Sec. 3.b) turns around the unstable situation at $\Phi=0$ for $T=0$ (see dashed and continuous lines in Fig. 4). If one adds to this a one-loop contribution (integrating out the gauge bosons in the loop) proportional to $-\Phi^3$, these effects together produce the typical action (free energy) of a first order phase transition. The two minima corresponding to the two phases, with a critical temperature $T_c$, at which both phases coexist, give the typical picture of condensing droplets or boiling bubbles of a first-order phase transition.

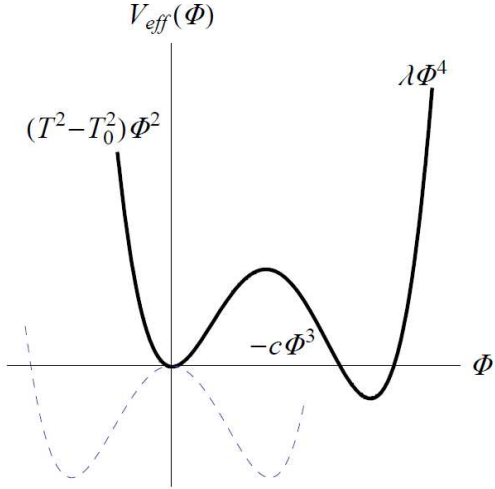

FIG. 4. The Higgs potential at $T<T_c$ (dashed line) has the usual "Mexican hat shape". With additional $T^2$ and $\Phi^3$ terms, the Higgs potential resembles the free energy of a first-order phase transition.

*CP* violation can then take place far off equilibrium in the walls of the expanding bubbles. This would produce a chiral asymmetry between left-handed and right-handed quarks, which, in the hot plasma in front of the expanding bubble, is transformed into a baryon asymmetry by the hot sphaleron transition. Due to the exponential sphaleron suppression Eq. (3.5), this asymmetry then has to freeze out rapidly in the Higgs phase when the expanding bubble takes over. This leads to the postulate $\upsilon(T^*)\geq T^*$, which makes the Boltzmann factor Eq. (3.5)(with exponent $\sim -\upsilon/T^*$) sufficiently small, where $T^*$ is the transition temperature (with $T^*<T_c$ because there is a delay transition time *vs.* expansion time).

Later on it turned out in lattice simulations (Kajantie *et al.*, 1996) that at Higgs-masses $m_H$ above the *W*-mass $m_W \approx 80$ GeV (current Higgs limit $m_H > 114$ GeV, 95% C.L. = confidence level) the transition is very smooth, a crossover rather than a strong first order phase transition, such that the third criterion would not be fulfilled. Looking at the SM as an effective theory, one can speculate that further (nonrenomalizable) terms like $\Phi^6$ could appear in the SM Lagrangian. Their origin would need explanation, but in this way one can enforce a strong first order phase transition (see Sec. C.3).

The *CP* violation through the CKM matrix has a solid experimental basis. However, if one wants to create a *CP* violating effect in the phase transition discussed before, naively, it would be proportional to the Jarlskog determinant

$$\Delta_{CP} = \frac{J}{\upsilon^{12}}(m_u^2-m_c^2)(m_c^2-m_t^2)(m_t^2-m_d^2)$$
$$\times (m_d^2-m_s^2)(m_s^2-m_b^2)(m_b^2-m_d^2) \quad (3.7)$$

with quark masses $m_q$, and with the invariant measure of *CP* violation

$$J = \text{Im}(V_{ud}V_{dc}^\dagger V_{cb}V_{bu}^\dagger) = s_1^2 s_2 s_3 c_1 c_2 c_3 \sin\delta, \quad (3.8)$$

where $V$ is the CKM-matrix (in the KM version), the $s_i$ and $c_i$ are the generalized Cabibbo angles $\sin\theta_i$ and $\cos\theta_i$, and $\delta$ is the *CP* violating phase. Evaluating Eq. (3.8) with the experimental data gives $J=(3.05\pm 0.20)\times 10^{-5}$, and $\Delta_{CP} \leq 10^{-19}$.

Even if we substitute the electroweak scale $\upsilon$ in Eq. (3.7) by the critical temperature $T_c$ of the transition, $\Delta_{CP}$ remains very small – much too small for successful baryogenesis. This estimate is based on the assumption that the quark masses enter perturbatively in loop calculations that lead to a *CP* violation. However, in the case of strong nonequilibrium there may be effective interactions containing derivatives suppressed not by $\Delta_{CP}$, but by the Jarlskog factor only, of order $J \sim 10^{-5}$ (Tranberg and Smit, 2003). Indeed, very recently such an effective interaction was derived in the next-to-leading order of a derivative expansion of a one-loop effective action, with all quarks in the loop integrated out (Hernandez *et al.*, 2009). Still, in order to obtain strong nonequilibrium one needs a mechanism beyond the SM. Low-scale inflation with an additional inflaton field has been suggested. Indeed, one can obtain a sizable baryon asymmetry this way (Tranberg *et al.*, 2009), but this is rather speculative.

### 7. Baryogenesis beyond the Standard Model

Hence, in the SM, both *CP* violation and deviation from thermal equilibrium are too weak to permit baryogenesis. Therefore presumably we must go beyond the SM, and more general variations of the electroweak theory are mandatory if we want to stick to electroweak baryogenesis. Today, physics at the electroweak scale is under experimental investigation both at CERN and in low-energy experiments. Thus, this type of baryogenesis can be tested and therefore is particularly attractive. A simple way to enlarge the SM is to introduce two Higgs doublets, instead of the one Higgs doublet of the SM (Bernreuther, 2002, Cline, 2006, Fromme *et al*., 2006).

The most prominent model beyond the SM is the Minimal Supersymmetric Standard Model (MSSM). Besides having two Higgs doublets, the MSSM has the field content of the SM just enlarged by scalar (squarks, sleptons) and fermionic (Higgsino, gauginos) superpartners, the gaugino being the superpartner "$W$-ino" to the SM gauge particle $W$ in our case. The superpartners couple with the same strength in the case of unbroken SUSY. As no SUSY partners to the known particles have been detected so far, (soft) SUSY breaking is mandatory in a realistic model but requires further couplings.

In the MSSM, we can obtain a strong first-order phase transition if a light $stop_R$-particle in the loop strengthens the $-\Phi^3$ term in the effective Higgs action (Fig. 4) (Carena *et al*., 1996, 2003a, Laine and Rummukainen, 1998). As already seen in Bödecker *et al*. (1997), who thoroughly discuss perturbative contributions in the MSSM, the $\Phi^3$ term argument is rather sketchy and also slightly gauge dependent.

Taking into account thermal mass contributions, the conditions for such a strong transition are roughly $m_H \leq 118$ GeV for the lightest Higgs mass, just barely consistent with the present experimental limit $m_H \geq 114$ GeV, and a bound on the "light" $stop_R$-mass $110 \text{ GeV} \leq m_{stopR} \leq m_{top} = 170$ GeV (narrowing to $m_{stopR} \sim 140$ GeV at the upper limit for $m_H$). In case of a very strong mixing of the two-level Higgs boson system due to $CP$ violation, the lower limit on the Higgs can be a considerably weakened new limit ($m_H \geq 70$ GeV?) (Carena *et al*., 2003b, and references therein).

These conditions for a strong phase transition depend on the ratio $\tan\beta = v_u / v_d$ of the vacuum expectation values of the two Higgs fields $H_u$ and $H_d$ (for up and down quarks), and on the average SUSY breaking mass. They can be weakened a bit by allowing that the Higgs phase be only metastable (against decay into a colored phase), i.e., stable for the time of our universe, which leads roughly to $m_H, m_{stopR} \leq 125$ GeV (Carena *et al*., 2009). The present experimental limit for a right-handed *stop* mass is $m_{stopR} \geq 100$ GeV. The additional requirement of baryogenesis then forces the lowest lying Higgs boson to be of the SM type. Light Higgs boson scenarios are not successful (Funakubo and Senaha, 2009).

$CP$ violations in the MSSM enter preferably through a complex phase factor in the gaugino-Higgsino mass matrix

$$M = \begin{pmatrix} m_2 & g_w \langle H_d \rangle \\ g_w \langle H_u \rangle & \mu \end{pmatrix} \qquad (3.9)$$

with the $W$-ino mass $m_2$ and the $\mu$ of the two-Higgs mass term $\mu H_u H_d$ in the superpotential carrying one free $CP$ violating phase, and with the classical expectation values $\langle H_u \rangle$ and $\langle H_d \rangle$ of the two neutral Higgs fields. Substituting $m_2$ in Eq. (3.9) by the $B$-ino mass $m_1$ with a second $CP$ violating phase helps much in avoiding EDM constraints (Li *et al*., 2010). The slowly moving "thick" bubble wall ($L > 1/T$) of the first-order phase transition allows for a quasiclassical WKB type treatment of order $\hbar^2$, but close to the degeneracy $m_2 \sim \mu$ in Eq. (3.9) there is a dominating oscillating contribution of order $\hbar$. The (quantum) transport equations for the Higgsino-gaugino system in front of the expanding Higgs-phase bubble have been discussed in much detail (Carena *et al*., 2003b, Lee *et al*., 2005, Konstandin *et al*., 2006) and it turned out that one can get successful baryogenesis only after squeezing the parameters of the model.

More recently, also the diffusive processes in front of the bubble wall in the hot plasma have been reanalyzed very carefully, introducing also $b$ quarks, $\tau$ leptons, and SUSY nondegenerate gauge interactions in the diffusion chain that lead from the Higgsino-Gaugino (also $B$-ino!) system to the left-handed quarks needed for the sphaleron process (Chung *et al*., 2009, 2010, and references therein). As we will see in Sec. C.2, neutron EDM constraints are very restrictive.

Baryogenesis is much easier to achieve in models that introduce, besides the MSSM superfields, a further scalar gauge singlet field $N$ (sometimes called $S$), which also can obtain a classical expectation value $\langle N \rangle$ of the order of $\langle H \rangle$ in an enlarged Higgs potential. These models are called NMSSM (next to MSSM) (Huber and Schmidt, 2001) and nMSSM (nearly MSSM) (Menon *et al*., 2004, Huber *et al*., 2006). In these models, a $\Phi^3$ type term is already present in the effective potential at the tree level.

In Profumo *et al*. (2007a), and in a recent paper by Patel and Ramsey-Musolf (2011), also a reduction of the effective scalar quartic self-coupling at the transition temperature can do the job. This alleviates much the search for a strong first-order phase transition, and this remains true in more general models (Noble and Perlstein, 2008). This allows the conditions on the Higgs mass $m_H \geq 114$ GeV and on the "light" $stop_R$-mass $110 \text{ GeV} \leq m_{stopR} \leq m_{top}$

=170 GeV quoted above to be relaxed, furthermore, a "low"-mass $stop_R$ is not needed. There still is a low lying Higgs, but now Higgs masses up to $m_H \leq 125$ GeV are allowed for successful baryogenesis, causing a better overlap with the experimental lower limit.

Recently it was proposed to use the three MSSM superpartners of the right-handed neutrinos as singlets mentioned just before ("$\mu\nu$MSSM", Chung and Long 2010). This offers new interesting possibilities. Models without SUSY going beyond the SM mentioned at the beginning of the present section and discussed in Sec. 4.d above offer more freedom but are less motivated.

In conclusion, the necessary conditions for baryogenesis certainly limit our choice of models. When we go through the possible realizations listed in Fig. 1, baryogenesis probably does *not* happen in the cases 1a): *B*-violation from the GUT transitions, and 2a) and 2b): *CP* violation from the SM electroweak transition. Still, this allows for quite a variety of different models. The electroweak baryogenesis – if realized in models beyond the SM, in particular, in supersymmetric models – is very attractive because its ingredients may be tested experimentally in the near future in high energy physics at TeV energies and in EDM experiments at neV energies. *CP* violation in these models being in the gaugino-Higgsino sector is completely different from the CKM-matrix *CP* violation of the SM, and might have a big *CP* violating phase. This raises hopes that a large EDM of the neutron could be present at the current experimental limit – different from the situation for the SM where the neutron EDM is tiny. We shall discuss this in Sec. C. One also can face another possibility: *CP* is broken in some model spontaneously (Huber and Schmidt, 2001) just at the high temperatures of the electroweak phase transition but not at $T = 0$. In this case, there would be no trace in the EDMs.

One has to admit that baryogenesis via leptogenesis (Buchmüller *et al.*, 2005, and references therein) with just heavy Majorana neutrinos added to the content of the SM is also highly interesting and convincing in view of its connection to neutrino physics. The physics of the decay of the presumptive Majorana neutrinos is at much higher scales (~ $10^{10}$ GeV) and much harder to test. The relevant *CP* violation in this case would have to be in the Majorana neutrino mass matrix. Being in the lepton sector it may influence the EDM of the electron but practically not that of the neutron.

## B. Electric dipole moments (EDMs)

Next, we discuss EDMs and their symmetries, describe experimental searches for the neutron EDM, give current limits for EDMs of the neutron and other particles, and give a survey of upcoming neutron EDM experiments.

### 1. EDMs and symmetry violations

Before discussing the electric dipole moment, we recall that the *magnetic* dipole moment operator of a particle is $\boldsymbol{\mu} = g\mu_N \mathbf{j}/\hbar$, with spin operator *j*. Its size is defined as $\mu = g\mu_N j$, measured in units of the nuclear magneton $\mu_N = e\hbar/2m_p = 3.15 \times 10^{-8}$ eV/T, with the Landé factor *g*, and proton mass $m_p$. Magnetic moments can be measured with high precision by observing the spin precession of the polarization vector $\mathbf{P} = \langle \mathbf{j} \rangle / \hbar j$ about an external magnetic field **B**,

$$\dot{\mathbf{P}} = \boldsymbol{\omega}_0 \times \mathbf{P}, \qquad (3.10)$$

with the precession frequency vector $\boldsymbol{\omega}_0 = \mu \mathbf{B}/j\hbar = \gamma \mathbf{B}$, and gyromagnetic ratio $\gamma = \mu/j\hbar = g\mu_N/\hbar$. N.B.: We do not have to apologize, as is frequently done, for using the "semiclassical" Larmor precession Eq. (3.10), which is just the Heisenberg equation $i\hbar d\langle \mathbf{j}\rangle/dt = H\langle \mathbf{j}\rangle - \langle \mathbf{j}\rangle H$ for the interaction of a particle at rest with any classical field **B**, with Hamiltonian $H = -\gamma \mathbf{j} \cdot \mathbf{B}$. (Semiclassical here only means that field **B** is unaffected by the quantum transitions.) More generally, Eq. (3.10) is the irreducible representation of the Liouville equation for the density operator, and can, for $j > \frac{1}{2}$, be generalized to higher multipole interactions, see Fano (1964).

For the neutron with $j = \frac{1}{2}$ one usually writes $\boldsymbol{\mu} = \mu \boldsymbol{\sigma}$ and $\mathbf{P} = \langle \boldsymbol{\sigma} \rangle$, with the Pauli spin operator $\boldsymbol{\sigma} = 2\mathbf{j}/\hbar$. The neutron's g-factor $g_n = -3.826$ gives the magnetic moment $\mu_n = -0.603 \times 10^{-7}$ eV/T and the gyromagnetic ratio $\gamma_n/2\pi = g_n\mu_N/2\pi\hbar = 29.16$ MHz/T.

If the particle also has an *electric* dipole moment **d** then, like any vector operator in quantum mechanics, it is connected to the spin operator as $\mathbf{d} = d\,\mathbf{j}/j\hbar$, or, for $j = \frac{1}{2}$, as $\mathbf{d} = d\boldsymbol{\sigma}$, where *d* gives the size of the EDM, usually in units of *e* cm. (N.B.: In quantum physics, isotropy of space requires that, in a rotationally symmetric problem, any operator is a multiple of a so-called spherical tensor; that is, all tensors of the same rank are proportional to each other, and the vector operators **μ**, **d**, and **σ** all have rank one.) If we expose the particle also to an electric field $E$, then we have to replace in Eq. (3.10) the precession frequency $\omega_0 = 2\mu\mathbf{B}/\hbar$ by

$$\omega_+ = (\mu\mathbf{B} + d\mathbf{E})/j\hbar. \quad (3.11)$$

The electric field is a polar vector, which, under parity transformation $P: \mathbf{r} \to -\mathbf{r}$, transforms as $P: \mathbf{E} \to -\mathbf{E}$, whereas the magnetic field **B** is an axial vector, which transforms as $P: \mathbf{B} \to \mathbf{B}$. Therefore the precession frequency Eq. (3.11) transforms under parity as

$$P: \omega_+ \to \omega_- = 2(\mu\mathbf{B} - d\mathbf{E})/\hbar. \quad (3.12)$$

Hence, for a nonvanishing EDM the frequency of spin precession changes under parity operation $P$, so $P$ symmetry is violated. Under $P$, the axial vector **j** changes sign on both sides of Eq. (3.10), so the precession equation remains the same, but with a shifted frequency.

An EDM $d \neq 0$ would also violate time reversal $T: t \to -t$, under which axial vectors like **B** and **j** change their "sense of rotation", $T: \mathbf{B} \to -\mathbf{B}$ and $T: \mathbf{j} \to -\mathbf{j}$, whereas $E$ does not change, $T: \mathbf{E} \to \mathbf{E}$, nor does $d\langle\mathbf{j}\rangle/dt$ on the left of Eq. (3.10). Therefore

$$T: \omega_+ \to \omega_- = 2(\mu\mathbf{B} - d\mathbf{E})/\hbar, \quad (3.13)$$

hence, time reversal symmetry and $CP$ symmetry (under the $CPT$ theorem) are violated for a particle with nonzero EDM.

A riddle: The neutron is a composite particle with quark content *dud*, and we showed that a nonzero EDM requires a violation of $P$ and $T$ symmetry. The water molecule $H_2O$ is a composite particle, too, with atom content HOH, and has a huge EDM. So why search an EDM on the neutron *dud* when we have found it already on HOH? For a solution, see Golub and Pendlebury (1972) or Golub and Lamoreaux (1994).

### 2. Neutron EDM experiments

One can measure particle EDMs with the highest precision by applying Ramsey's method of separate oscillatory fields (as used in all atomic clocks) to an ensemble of spin-polarized particles. The Ramsey method compares the internal "Larmor clock" of the particle with the external clock of a radiofrequency (RF) generator. We first discuss free neutron EDM results, and at the end quote EDM results for other particles.

a. *History and present status*: The first experiment on the neutron EDM started many decades ago (Purcell and Ramsey, 1950) with a beam of thermal neutrons at the reactor of ORNL, USA. The result $d_n < 5 \times 10^{-20} e$ cm at 95% C.L. was published by Smith *et al.* only in 1957, the year of the discovery of parity violation (we write $d_n$ for $|d_n|$, as long as no nonzero EDM is found). Since then the sensitivity of these experiments has improved by more than six orders of magnitude to today's limit

$$d_n < 2.9 \times 10^{-26} e \text{ cm (90\% C.L., neutron)} \quad (3.14)$$

obtained at ILL by a Sussex-RAL-ILL collaboration (Baker *et al.*, 2006), using ultracold neutrons (UCN) stored in a "neutron bottle". The energy sensitivity of the apparatus required to reach this level of accuracy is $\sim 10^{-26}$ eV per 1 V/cm electric field unit, or $10^{-22}$ eV for an electric field of $E \sim 10^4$ V/cm.

This EDM limit was preceded by results $d_n < 6.3 \times 10^{-26} e$ cm at 90% C.L. from ILL (Harris *et al.*, 1999), $d_n < 9.7 \times 10^{-26} e$ cm at 90% C.L. from PNPI (Altarev *et al.*, 1992, 1996), and $d_n < 12 \times 10^{-26} e$ cm at 95% C.L. again from ILL (Smith *et al.*, 1990). There exist several review articles on the experimental searches for a neutron EDM. For the early experiments see Ramsey (1978, 1990), and for more recent developments, Golub and Lamoreaux (1994), and Lamoreaux and Golub (2009). UCN physics in general is well described in the books by Golub *et al.* (1991), and by Ignatovitch (1990).

TABLE IV. Some useful wall materials for UCN traps or neutron guides, their calculated Fermi potential $V$, and critical neutron velocity $v_c$. The critical angle of total reflection $\theta_c$ for cold neutrons is also given, in mm per m.

| Material | $V$ (neV) | $v_c$ (ms$^{-1}$) | $\theta_c$ ($10^{-3}$) |
|---|---|---|---|
| $^{58}$Ni | 343 | 8.1 | 11.6 |
| Diamond-like carbon | 260 | 7.1 | 10.1 |
| Ni, Be, $^{65}$Cu | ~245 | 6.8 | 9.8 |
| Stainless steel | 188 | 6.0 | 8.6 |
| Cu | 169 | 5.7 | 8.1 |
| Al$_2$O$_3$ sapphire | 149 | 5.3 | 7.6 |
| SiO$_2$ (quartz), Fomblin | ~108 | 4.6 | 6.5 |
| SiO$_2$ (glass) | 91 | 4.2 | 6.0 |

b. *Principles of nEDM experiments*: Since 1984, all neutron-EDM searches in external electric fields practised the storage of ultracold neutrons. UCN are defined as neutrons with velocities so low that they are totally reflected by many materials under all angles of incidence. The average nuclear "Fermi" pseudopotential of materials seen by neutrons typically is of order $V \sim 100$ neV, see Table I; for thin coatings with reduced density, the values of $V$ can be 10 to 20% lower. (The low value of $V$ can be understood by considering that the long-wavelength UCN see the nuclear potential of some 10 MeV diluted in volume by the ration of atomic to nuclear volume $(1 \text{ Å} / 1 \text{ fm})^3 = 10^{15}$.) Total reflection of UCN, and hence their storage in closed vessels, occurs up to a material dependent critical neutron velocity $v_c = (2V/m)^{1/2}$, which is also listed in Table IV. N.B.: The neutron guides mentioned on several occasions rely on the same principle and will be discussed in Sec. VI.B.1.

In the laboratory, UCN can be manipulated equally well by gravitational, magnetic, and average nuclear forces, as the neutron experiences a gravitational force $m_n g = 102$ neV/m, and has a magnetic moment of $\mu_n = 62$ neV/T. Hence, UCN of kinetic energy $E = 100$ neV rise in the earth's gravitational field to a maximum height of $h = E/m_n g = 1.0$ m, and a magnetic field totally repels neutrons of one spin-component if the field surpasses the value $B = E/\mu_n = 1.6$ T.

Figure 5 schematically shows the EDM apparatus used at ILL. The neutron bottle of volume $V = 20 \ \ell$, consisting of two diamond-like-carbon covered electrodes separated by a cylindric mantle of quartz, is filled with $N \approx 10^4$ polarized UCN from ILL's UCN turbine source. Originally, this source had been developed for the FRM reactor in Munich, see Steyerl *et al*. (1986). After closure of port A in Fig. 5, two short Ramsey $\pi/2$-resonance flips are applied to the UCN at the beginning and end of the storage period $T$ of typically two minutes duration. The first RF pulse flips the initial "longitudinal" neutron polarization $P_{z0} \approx 1$ into the *x-y* plane, causing it to lie at right angles to both **B** and **B**$_1$. Between the two flips, while the now "transversally" polarized neutrons are freely precessing about the magnetic field $B = 1$ μT, the radiofrequency field (with rotating component $B_1$) is gated off, such that its phase coherence is maintained between the two spin-flips.

If the frequency $\omega$ of the RF field exactly equals the Larmor frequency $\omega_0 = \gamma B \approx 2\pi \times 30$ Hz of the neutron, the second flip will turn the polarization by another angle $\pi/2$ into $-P_{z0}$. The neutron bottle is then emptied onto the UCN detector, with ~ 2 m of free fall to accelerate the UCN sufficiently so they can overcome the repulsive potential of the aluminium window of the gas detector for UCN. On their way down, the neutrons pass the magnetized polarizer foil (see Fig. 5) a second time, the foil then acts as a spin analyser. The additional spin-flipper shown in Fig. 5 can induce a $\pi$-flip so one can start the measurement cycle also with an inverted initial polarization $-P_{z0}$, which helps to eliminate systematic errors.

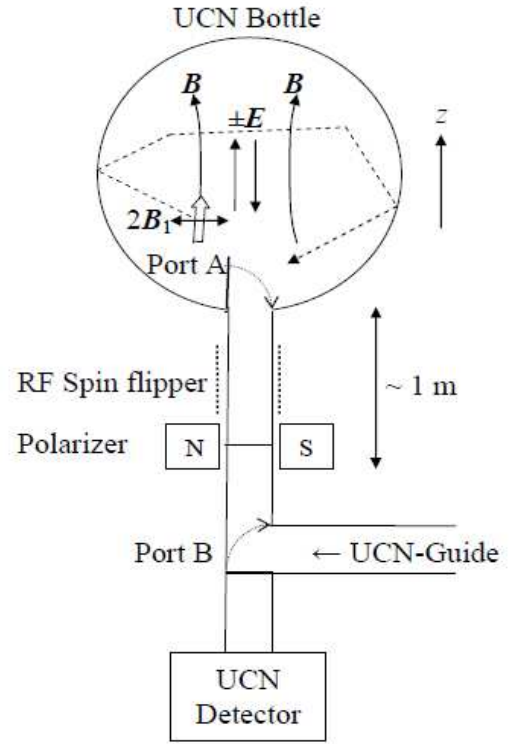

FIG. 5. Sketch of the Ramsey apparatus used at ILL to search for a neutron electric dipole moment (*n*EDM). The block arrow shows the neutron polarization right after filling of the UCN trap (Port A and B open). The dashed line represents a UCN trajectory in the UCN bottle. The sources of the **E** and **B** fields are not shown.

To monitor the magnetic field, atoms of the spin-½ isotope $^{199}$Hg were stored together with the UCN as a "comagnetometer" at about $10^{-6}$ millibar pressure. $^{199}$Hg nuclei are known to have an EDM $<10^{-28}$ *e* cm, see Sec. 4 below. The nuclei were polarized by atomic optical pumping, and their precession frequency was monitored optically.

c. *Sensitivity of EDM experiments*: The Ramsey method under nonperfect conditions (neutron polarization $P < 1$ and/or nonzero background $B \neq 0$) is most sensitive to an external perturbation when $\omega$ is slightly detuned from $\omega_0$ such that, at the end of the free precession period $T$, and before the second RF pulse is applied, the spins have accumulated an additional phase $(\omega - \omega_0)T = \pi/2$ in the *x-y* plane. (N.B.: For $P = 1$ and $B = 0$, this sensitivity does no longer dependent on such detuning.) The neutron polarization $\mathbf{P}_\perp$ in this case no longer lies at a right angle to $\mathbf{B}_1$, but is parallel to $\mathbf{B}_1$, such that the second $\pi/2$-flip is no longer effective. However, if an electric field is superimposed (in the ILL experiment of size $E_z = 10$ kV/cm), for $d_n \neq 0$ an additional phase shift $\varphi = 2d_n E_z T/\hbar$ will be induced in the $x$-$y$ plane, after which the second $\pi/2$ RF pulse will then produce a longitudinal polarization $P_z = -\sin\varphi \approx -\varphi$ (for $P_{z0} = 1$).

How well can such a small rotation out of the $x$-$y$ plane through $\varphi \ll 1$ be measured, in principle? Let the direction of $\mathbf{B}_1$ define the quantization axis. For a single particle, the polarization $P_z$ ("transverse" with respect to the quantization axis) remains completely uncertain. Let, for $N \gg 1$ particles, a spin analyzer be aligned along $P_z \approx -\varphi$. The analyzer transmits the number $N_+ \approx \frac{1}{2}N(1+\varphi)$ or $N_- \approx \frac{1}{2}N(1-\varphi)$ of particles when oriented parallel or antiparallel to axis $z$, respectively. With $N = N_+ + N_- \gg 1$, the statistical error of $\varphi = (N_+ - N_-)/N$ is then calculated as $\Delta\varphi \approx 1/N^{1/2}$, or $N^{1/2}\Delta\varphi \approx 1$, which connects the number of neutrons under study to the uncertainty of their spin phase.

When the EDM-induced phase is measured with the $\mathbf{E}$-field successively set parallel and antiparallel to the $\mathbf{B}$-field, the difference of both measurements gives the overall phase shift of $4d_n E_z T/\hbar$, with $|\mathbf{E}| = E_z$. The measured EDM and its statistical error then are

$$d_n \pm \Delta d_n = \frac{(N_+ - N_-)\hbar}{2NE_z T} \pm \frac{\hbar}{2\sqrt{N}E_z T}, \quad (3.15)$$

where the $+$ and $-$ subscripts refer to measurements with $+E_z$ and $-E_z$, and $N$ is the total number of UCN stored during a measurement campaign.

The figure of merit for a given value of $d_n$ then is $\propto N^{1/2} E_z T$. The free-precession period $T$ is naturally limited by the neutron lifetime $\tau_n \approx 880$ s. The electric field strength $E_z$ is limited to several tens of kV/cm by the requirement that leakage currents and their magnetic fields must be strictly avoided, because, like the EDM interaction, they change sign when $E_z$ is inverted to $-E_z$. Therefore for both $T$ and $E_z$ one can expect improvements of at most a factor of two or three. UCN density $\rho$ and bottle volume $V$ then are the only parameters that can bring significant statistical improvement in future EDM experiments via the total number $N = V\rho$ of bottled UCN.

d. *False effects from geometric phases*: In addition to the various systematic errors since long known in neutron EDM searches, a rather dangerous systematic error has surfaced lately (Pendlebury *et al.*, 2004), which was already known in earlier atomic EDM work (Commins, 1991). Once recognized as such, this source of error is not difficult to understand. It is due to unavoidable residual field inhomogeneities. The UCN in their rest frame see these inhomogeneities as a time dependent magnetic field $\mathbf{B}$ that continuously changes direction.

We know from the work of Berry (1984) that the exposure of a spin-magnetic moment to a magnetic field of slowly varying direction during time $T$ leads to a geometric phase $\gamma$, in addition to the usual dynamic Larmor phase $\gamma_n BT$ (with the neutron's gyromagnetic ratio $\gamma_n$). This extra phase develops continuously and reaches the size $\gamma = \frac{1}{2}\Omega$ after the field has come back to its original orientation, where $\Omega$ is the solid angle of the cone formed when the $\mathbf{B}$-field vector as seen by the moving neutron sweeps through its changing orientations. In contrast to the dynamical phase, this geometric phase is independent of the magnetic properties of the neutron. Berry's formula $\gamma = \frac{1}{2}\Omega$ was checked experimentally with a beam of polarized cold neutrons by Bitter and Dubbers (1987), and with stored polarized UCN by Richardson *et al.* (1988).

To quantify this geometric effect we approximate the UCN trajectory in the trap by a horizontal circle of radius $r$, so the neutrons move with angular velocity $\omega_r = v/r$ at right angles to the local field $\mathbf{B}$ oriented vertically as shown in Fig. 5. If $\mathbf{B}$ has a small radial component $\mathbf{B}_\perp$ of constant amplitude everywhere on this circle, the solid angle of the magnetic-field sweep is $\Omega \approx \pi B_\perp^2/B^2$. After $n = \omega_r T/2\pi$ turns, the extra phase of the UCN accumulates to $n\gamma = (\omega_r T/2\pi) \times \frac{1}{2}\Omega$. If solely the transversal magnetic field $\mathbf{B}_\perp$ entered our argument, this geometric phase would cancel in an EDM search, with signals taken for $\pm \mathbf{E}$, because it is independent of size and direction of the electric field $\mathbf{E}$. However, besides $\mathbf{B}_\perp$, there is the velocity-induced magnetic field $\mathbf{B}_v = \mathbf{v} \times \mathbf{E}/c^2$, which in our simple example is collinear with $\mathbf{B}_\perp$, so the geometric phase becomes $\gamma = \frac{1}{2}\Omega \approx \frac{1}{2}\pi(\mathbf{B}_\perp + \mathbf{B}_v)^2/B^2$. Now the cross term $2\mathbf{B}_\perp \cdot \mathbf{B}_v$, which is linear in the electric field $\mathbf{E}$, no longer cancels, and the extra geometric phase $n\gamma' = (B_\perp B_v/2B^2)\omega_r T$ cannot be distinguished from a true EDM signal.

In a real experiment, one must average over all neutron trajectories in the neutron bottle, although one finds from numerical calculations (Pendlebury *et al.*, 2004) that the rate of accumulation of the geometric phase is nearly independent of the specific trajectory of the trapped neutron. In the ILL experiment, the dominant geometric phase was that of the $^{199}$Hg comagnetometer. Future EDM searches aim at an EDM down to $10^{-28}\,e$ cm, in which case the geometric-phase induced error may well become the leading systematic error.

e. *The crystal-EDM* experiment relies on a completely different method. It does not use trapped UCN, but a beam of monochromatic and polarized cold neutrons of velocity $v = 800$ m/s. The neutrons interact with the internal electric field $E \approx 10^8$ V/cm of a non-centrosymmetric crystal. This field is four orders of magnitude stronger than the laboratory fields used in the stored-UCN EDM experiments. On the other hand, the neutron's interaction time $T = l/v \approx 2 \cdot 10^{-4}$ s in a quartz crystal of length $l = 14$ cm is six orders of magnitude smaller than for trapped UCN. The cold-neutron count rate of 60/s is similar to the corresponding rate $N/T \approx 100/s$ reached with UCN.

The experiment uses an elegant and sophisticated two-crystal backscattering arrangement, with a temperature-induced switch from positive to negative crystal electric fields. It will be difficult for this experiment to compete once the UCN-EDM projects listed below come into operation. Still, this experiment is interesting because systematic effects seem to be well under control, and are completely different from the systematic effects in the UCN-EDM experiments. Using this method, Fedorov *et al.* (2010) reached $d_n = (2.5 \pm 6.5^{stat} \pm 5.5^{syst})$ $\times 10^{-24} e$ cm. Their EDM sensitivity could be improved by a factor of about 65 by using better neutron equipment, and so possibly could catch up with the previous ILL experiment by Baker *et al.* (2006).

f. *Other uses of nEDM instruments*: There have been other uses of the UCN-EDM instruments. Altarev *et al.* (2009) used the previous ILL-EDM apparatus to compare the spin precession of the neutron with that of the comagnetometer atoms of $^{199}$Hg, and searched for daily variations of their frequency ratio. Such variations are allowed in recent Lorentz and *CPT* symmetry violating extensions of the SM, induced by the electric term of a cosmic spin anisotropy field. Its interaction with the neutron (or rather with its component vertical to the earth's rotation axis) was constrained by this measurement to $< 2 \times 10^{-20}$ eV at 95% C.L., for an improved evaluation see Altarev *et al.* (2010a), where the energy scale for Lorentz violation is tested on the level of $10^{10}$ GeV. With polarized atoms, Brown *et al.* (2010) reached a $2\sigma$ bound of $< 3 \times 10^{-24}$ eV for the neutron, see also the Data Tables by Kostelecký and Russell (2011), in particular, their Table VII for the neutron. In addition, EDM instruments were used for searches for neutron-mirror neutron oscillations, as described in Sec. IV.B, and searches for spin-dependent "fifth forces" as reported in Sec. V.C.3.

### 3. *Upcoming neutron EDM experiments

During the past few years, several new experimental neutron EDM projects were started worldwide, each aiming at sensitivities for a neutron EDM of well below $10^{-27} e$ cm. (We do not quote individual EDM sensitivities aimed at, still too uncertain at the present stage.) To reach this aim, all projects rely on significantly improved UCN densities, estimated to lie between $10^3$ and $10^4$ UCN/cm$^3$, as compared to ILL's 40 UCN/cm$^3$ measured at the exit of the source, and ~1 UCN/cm$^3$ measured in the ILL trap. In the following, we shall give a survey on these upcoming EDM projects, in the order of the foreseen date of commissioning.

a. *CryoEDM* is a follow-up project by a Sussex-RAL-Kure-Oxford-ILL collaboration located at ILL. The "superthermal" UCN source, which is integrated in the CryoEDM apparatus, relies on a method that had been proposed by Golub and Pendlebury (1975, 1977), and that was shown to work as expected for CryoEDM by Baker *et al.* (2003). It uses a genuine cooling process, in which Liouville's law of conserved phase-space density does not limit UCN densities, as it does in ILL's turbine UCN source, which relies on purely conservative forces. This method uses the coherent excitation of superfluid helium (He-II) at low temperature, induced by the inelastic scattering of cold neutrons, which in the process lose all their energy and become ultracold. To excite superfluid helium, the neutrons must have a kinetic energy of 1.0 meV, which corresponds to a neutron temperature $T_n = 12$ K and to a neutron de Broglie wavelength $\lambda = 0.89$ nm. At a bath temperature of $T = 0.5$ K, the inverse process (energy gain of UCN) is strongly suppressed, because at this temperature there are not enough phonons or other excitations present in the fluid.

CryoEDM uses cold neutrons from one of ILL's neutron guides, spin-polarized to above 90% at 0.89 nm, with polarized capture flux of $1.5 \times 10^9$ cm$^{-2}$s$^{-1}$. The UCN produced at a rate of 1 cm$^{-3}$s$^{-1}$ in the $\approx 10\,\ell$ source volume of the He-II converter are then guided to a Ramsey precession double cell of volume $2 \times 22\,\ell$. The cells are placed in a common uniform static field of $B = 5.0\,\mu$T. All parts of the apparatus are immersed in He-II, including the EDM precession cell, the SQUID magnetometers, the superconducting magnetic shield, and the solid-state neutron detectors. This is possible because $^4$He has zero neutron-absorption cross section, it can support strong electric fields, and, as a further advantage, $^4$He has a huge thermal conductivity. In the present setup, CryoEDM will search for a neutron EDM at a sensitivity similar as in the previous ILL experiment.

The apparatus completed commissioning, for a status report see Baker et al. (2010). With an improved setup, the experiment will move to a new dedicated very-cold neutron beam-line at ILL in about 2014, where the proposers expect a considerably higher final sensitivity.

b. *A PNPI EDM* experiment is installed at ILL's turbine UCN source and uses the upgraded double cell of the previous PNPI experiment, filled with 5 UCN/cm$^3$. At present, this experiment aims at a sensitivity similar to that of the 2006 ILL experiment. It is intended to use this setup later at the existing PNPI reactor, with a new 35 $\ell$ He-II UCN source with about 10$^3$ times higher UCN density, installed in the thermal column of the reactor, see Serebrov et al. (2009a, b).

c. *The PSI nEDM* project will use the 1.3 MW proton beam (2.2 mA, 590 MeV) from the PSI quasi-continuous cyclotron in a parasitic mode, with 1% duty cycle. Every 800 s the experiment receives the full proton beam for 8 s, just the time needed to fill the UCN bottle, which is followed by a measurement period of about one neutron lifetime. The protons hit a solid lead/zirconium target, which acts as a small, dedicated pulsed spallation neutron source with 10 neutrons liberated per proton collision. This source feeds an UCN source, which consists of a 30 $\ell$ solid ortho-deuterium D$_2$ converter at 5 K, located on top of the spallation source. Both spallation and UCN sources are immersed in a common neutron moderator of 3.6 m$^3$ heavy water D$_2$O at room temperature. The UCN source is linked to a 2 m$^3$ storage reservoir, coated with diamond like carbon, from which the UCN are drawn to fill the EDM cell. Several users can be served in parallel from this UCN storage vessel.

The use of small pulsed-neutron sources with low repetition rates for UCN storage experiments had been proposed by Pokotilovski (1995), and the "parasitic" use of spallation neutron sources for UCN production in a solid D$_2$ converter by Serebrov et al. (1997).

The PSI UCN source, described in Anghel et al. (2009), was tested successfully under full power end of 2010 (though with still imperfect para to ortho-hydrogen conversion), for a status report on the whole EDM project see Altarev et al. (2010b). In a first phase up to 2012, the experiment will use the previous EDM apparatus of the Sussex-RAL-ILL collaboration, with $N = 3 \times 10^5$ UCN in the measurement cell, instead of $N = 10^4$ at ILL, with a sensitivity goal somewhat below $10^{-27} e$ cm. In a second phase from 2012 on, a new EDM apparatus, presently under development, will be used that has a double-cell with opposite **E**-fields and improved magnetometry for better systematics, with about 10 times better sensitivity aim.

d. *The TUM-nEDM* project: Technical University of Munich, Germany, is developing an *n*EDM apparatus, to be connected to an in-pile UCN source under development at the new FRM-II high-flux reactor. This source is based on a 1 cm thick solid-D$_2$ converter of 170 cm$^3$ volume ("Mini-D2") in a tangential beam tube, for a design study see Trinks et al. (2000), and Frei et al. (2009). The source is intended to become operational by 2014. PSI and FRM II collaborate, and both profit from detailed UCN studies done at the TRIGA reactor of Mainz University, Germany. In its pulsed mode, the power of this TRIGA reactor goes up to 250 MW during 30 ms every 5 minutes. This makes it well suited for trapped UCN experiments: To compare, ILL runs (continuously) at 58 MW. A number of 10$^5$ measured UCNs per reactor pulse are reported, see Frei et al. (2007), and an UCN density of 4 UCN/cm$^3$ was achieved at 8 m distance from the source, comparable to that achieved at present at ILL (T. Lauer, private communication 2009).

e. *The SNS nEDM* project at ORNL follows a radically novel concept developed by Golub (1983), and Golub and Lamoreaux (1994). The experiment will use a superthermal UCN source, though will start with a cold neutron flux lower than that of CryoEDM. The EDM precession cell is inside the UCN source, which avoids the notorious problems of UCN extraction and transport. A ~$10^{-10}$ fraction of polarized $^3$He atoms (nuclear spin $j = \frac{1}{2}$) is added to the superfluid $^4$He ($j = 0$) bath, which simultaneously serves several purposes.

(1) The $^3$He nuclei act as a comagnetometer, like the $^{199}$Hg nuclei in the ILL experiment, but with a magnetic moment similar to that of the neutron (+11%). The diffusive motion of $^3$He in the $^4$He bath in part averages out its large geometric phase effect described in Sec. 2.d.

(2) The $^3$He nuclei also act as a neutron spin analyzer: For neutron spin opposite to the $^3$He spin, the neutron capture cross section (~Megabarn for UCN) is almost 200 times larger than for parallel spins, so when neutrons and $^3$He precess at a slightly different frequency, then the neutron capture rate is modulated at the difference of these frequencies.

(3) This brings us to the third role of the $^3$He nuclei: They also act as a counting gas for neutron detection. Neutron capture in $^3$He creates a fast proton and a fast triton with combined energy of 764 keV. In He-II, these fast charged particles produce scintillation light in the far ultraviolet (~80 nm), which a wavelength shifter makes detectable in photomultiplier tubes.

The stray field of the polarized $^3$He is detected in a SQUID magnetometer, which allows to measure the $^3$He precession frequency. The difference of neutron and $^3$He precession frequencies is detected in the phototubes via the modulated UV photon signal. This difference signal is sensitive only to a neutron EDM, because a $^3$He EDM signal is suppressed by Schiff's rule mentioned in Sec. B.4. The neutron-$^3$He difference signal is only one tenth of the $^3$He frequency, and hence is 10 times less sensitive to magnetic inhomogeneities than are the separately measured precession frequencies. With another elegant trick, one can make even this difference frequency disappear, in which case the pure EDM effect with no magnetic disturbance will be left:

(4) Both neutrons and $^3$He nuclei can be forced to process at the same frequency about **B** by "dressing" them appropriately with radio frequency (RF) quanta..

Indeed, one can at will suppress the Larmor precession of a polarized ensemble by irradiating it with a strong RF field of arbitrary frequency $\omega'$. For neutrons, this was shown by Muskat *et al.* (1987), see Fig. 6, who used a second-quantization treatment of the RF field. This method was developed in atomic spectroscopy by Cohen-Tannoudji and Haroche (1969), but it is more transparent when applied to "point-particles" like neutrons, with no optical excitations involved. Of course, one can derive dressed-particle effects (i.e., dressed-energy level diagrams, dressed *g*-factors, dressed-level crossings, or multiple quantum transitions) from the Larmor precession equation (3.10) (see Dubbers, 1976), but a "dressed neutron" description gives better physical insight to these phenomena.

In any case, in the SNS experiment, the free $^3$He nuclei precess faster than the free neutrons, but dressing both with RF quanta shifts the $^3$He precession frequency more strongly than the neutron precession frequency. Therefore at a critical RF field amplitude $B_1 = 1.19 \times \omega'/\gamma_n$ (with neutron gyromagnetic ratio $\gamma_n$) both precession frequencies coincide, and the difference signal comes to a stop. For a nonzero EDM, the application of an electric field $\pm E_z$ should then make the signal process again.

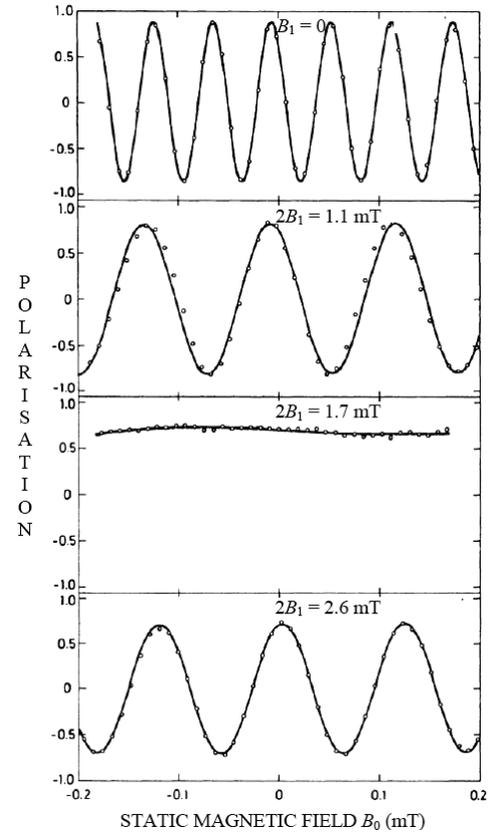

FIG. 6. The "dressed neutron" effect can be used in EDM searches to suppress unwanted magnetic effects. Spin precession about a static magnetic field $B_0$ can be suppressed when the neutrons are irradiated with a strong radio frequency field $B_1$ of arbitrary frequency. With RF amplitude increasing from the top to the bottom of the figure, spin-precession of the dressed neutrons slows down, then comes to a full stop, and reverses its sign. From Muskat *et al.* (1987).

The SNS EDM project profits from other UCN activities in the US, done at the solid $D_2$ source at LANL (Saunders *et al.*, 2004), at the solid $D_2$ source at the PULSTAR reactor of the North Carolina State University (Korobnika *et al.*, 2007), and at the Low Energy Neutron Source (LENS) at the Indiana University Cyclotron Facility (Leuschner *et al.*, 2005). The project is in its construction phase, see Alarcon *et al.* (2007) for an updated version of the EDM proposal, and Ito (2007) for a status report. In 2009, the SNS accelerator surpassed a proton beam power of 1 MW, and at the moment is heading for 1.4 MW. The current best estimate for data taking is 2017.

f. *The TRIUMF EDM* proposal foresees to transfer the UCN source developed at the Research Center for

Nuclear Physics (RCNP), Osaka, Japan (Masuda *et al.*, 2002), to TRIUMF, see Martin *et al.* (2008). At RCNP the UCN source is installed at a continuous neutron spallation source driven by a 1 μA, 390 MeV proton beam, and the He-II converter at present delivers a measured density of 15 UCN/cm$^3$. The cold-neutron moderator tank at RCNP is made of solid D$_2$O at 20 K. When installed at the TRIUMF proton beam of 40 μA and 500 MeV, and equipped with a 20 K liquid D$_2$ cold-neutron moderator and a better UCN extraction system, this source is expected to deliver up to a thousand times the UCN density measured at RCNP. First UCN experiments at TRIUMF are foreseen for 2013.

**4. Limits on atomic and particle EDMs**

Particle EDMs are available also for other neutral systems, namely, $d_e$ of electrons bound in paramagnetic atoms, and $d_N$ of heavy nuclei in diamagnetic atoms. These EDMs probe *CP* violation in different sectors, $d_e$ in the leptonic sector, $d_n$ in the hadronic sector, and $d_N$, in addition to the previous, is sensitive to nuclear effects. For a well-written account of atomic EDMs see Sandars (2001).

The interaction of atomic EDMs with external electric fields should vanish in the nonrelativistic limit for neutral atoms with point-like nuclei (according to the Schiff screening theorem, see Liu *et al.*, 2007, and references therein). However, the heavier atoms do have relativistic electrons and extended nuclei.

In heavy *paramagnetic* atoms, relativistic corrections can lead to measurable EDMs that are growing with atomic number as $Z^3$, and that are mainly sensitive to the electron EDM, plus to some small scalar *CP*-odd electron-nucleon amplitude $C_S$. For the paramagnetic thallium atom $^{205}$Tl this dependence is $d_{Tl} = -585 d_e + 5 \cdot 10^{-16} e \, \text{cm} \times C_S$, which is precise to about 10–20%, where the amplitude $C_S$ is composed of different quark-electron terms $C_{qe}$ as $C_S \approx 4C_{de} + 2C_{se} + 0.01 C_{be}$, from Huber *et al.* (2007). The $^{205}$Tl experiment gives a limit on the electron EDM of

$$d_e \approx d_{Tl}/585 < 1.6 \times 10^{-27} e \, \text{cm}$$
$$(90\% \text{ C.L., atoms}), \qquad (3.16)$$

from Regan *et al.* (2002), who neglect the contribution from $C_S$. Another promising candidate is the YbF molecule, which so far reached $d_e < (-0.2 \pm 3.2) \times 10^{-26} e \, \text{cm}$, from Hudson *et al.* (2002).

In heavy *diamagnetic* atoms, EDMs become measurable due to the finite size of the nucleus, with quark, electron, quark-quark and electron-quark amplitudes contributing. For the diamagnetic mercury atom $^{199}$Hg, Huber *et al.* (2007) find as the leading terms $d_{Hg} \approx 7 \times 10^{-3}(\tilde{d}_u - \tilde{d}_d) + 10^{-2} d_e$, combining the contributions from Schiff momentum $S(\hat{\rho}_{\pi NN}(\tilde{d}_u, \tilde{d}_d))$ and the electron EDM, where $\tilde{d}_{u,d}$ are color EDMs. There is a large expected overall uncertainty due to significant cancellations.

The experimental limit for $^{199}$Hg is $d_{Hg} < 3.1 \times 10^{-29} e \, \text{cm}$ (95% C.L.), from Griffith *et al.* (2009). These authors point out, citing Dmitriev and Sen'ov (2003), that under certain theoretical assumptions this translates into a proton EDM limit of

$$d_p < 7.9 \times 10^{-25} e \, \text{cm} \quad (95\% \text{ C.L., atoms}), \qquad (3.17)$$

and a neutron EDM limit of

$$d_n < 5.8 \times 10^{-26} e \, \text{cm} \quad (95\% \text{ C.L., atoms}). \qquad (3.18)$$

To make the picture complete we also quote EDM limits of *short-lived particles*. For the muon (we recall that all nonreferenced particle data are from PDG 2010),

$$d_\mu < 1.8 \times 10^{-19} e \, \text{cm}, \, (95\% \text{ C.L., muon}) \qquad (3.19)$$

measured with muons in a storage ring, and for the neutral $\Lambda^0$ hyperon

$$d_\Lambda < 1.5 \times 10^{-16} e \, \text{cm} \quad (95\% \text{ C.L.}, \Lambda^0). \qquad (3.20)$$

These results are by-products of magnetic moment measurements on highly relativistic particles via spin-rotation in flight. When the particles fly through a transverse magnetic field, in their rest frame, they also see a strong electric field that would act on a hypothetic EDM.

In all, there is a worldwide competition of ambitious and challenging neutron and atomic physics EDM projects, and at present it is an open question, which of the different routes chosen in the various projects will first lead to success. In the next section, we discuss the conclusions that we can draw from the constraints obtained for the EDM of neutrons and other particles.

*Note added in proof*: Very recent experimental results on an anomalous like-sign di-muon charge asymmetry, Abazov *et al.* (2010a, b) indicate a *CP* violation in the $B_{0q}$-$\bar{B}_{0q}$ mixing ($q = d$, $s$) that is 40 times larger than estimated in the SM from the CKM matrix, see Lenz and Nierste (2007), plus numerous recent papers on the subject. If confirmed, this has serious consequences for models beyond the SM.

## C. EDMs in particle physics and cosmology

In the long history of searches for particle EDMs, many theories on the origin of *CP* violation have been excluded by the increasingly tighter experimental constraints, see for instance the history plot, Fig. 1 of Pendlebury and Hinds (2000). For reviews on theories on EDMs in general, see Pospelov and Ritz (2005), on the neutron EDM, Khriplovich and Lamoreaux (1997), and on the electron EDM, Bernreuther and Suzuki (1991).

### 1. EDMs in the Standard Model

a. *Quark models*: A neutron EDM would be a clear sign of *CP* violation. As discussed before, *CP* violation in flavor changing *K* and *B* decays can be nicely explained by the phase $\delta$ in the CKM matrix of the SM. In our present understanding, the neutron is a composite object made of three "valence" quarks (*udd*) with three different colors, of gluons, and of quark-antiquark pairs ("sea quarks"), if one looks closer on varying scales. This compositeness can be taken into account using quark models in various ways for the constituent (dressed valence) quarks forming a color singlet with spin ½. In the simplest case, by using the old *SU*(6) quark model with nonrelativistic wave function of the neutron, one obtains

$$d_n = \tfrac{4}{3} d_d - \tfrac{1}{3} d_u \quad \text{(constituent quark model)} \quad (3.21)$$

for the neutron EDM $d_n$, given the EDMs of *d* and *u* quarks, see also Eq. (3.36) below for the neutron magnetic dipole moment and its derivation. More refined models for hadronic compositeness obtain roughly the same results by using a perturbative chiral quark model that includes a cloud of Goldstone bosons ($\pi$, *K*), see, for example, Borasoy (2000), Dib (2006), and Ottnad *et al*. (2010).

However, a *CP* violating $\theta$-vacuum of QCD as described in Sec. A.4.a would most strongly influence a neutron EDM. This was first worked out in the framework of current algebra by Crewther *et al*. (1979) who obtained

$$d_n = \frac{e}{4\pi^2 m_n} g_{\pi NN} \bar{g}_{\pi NN} \ln \frac{\Lambda}{m_\pi}$$
$$\text{(current algebra)} \quad (3.22)$$

as the leading log contribution with a $\Lambda$ ($\sim m_n$) cut off, and with a *CP* violating $\pi NN$ coupling

$$\bar{g}_{\pi NN} = \frac{\theta}{f_\pi} \frac{m_u m_d}{m_u + m_d} \langle p|\bar{q}\tau^3 q|p\rangle \, (1 - m_\pi^2/m_\eta^2) \quad (3.23)$$

proportional to $\theta$ and to a proton matrix element of a quark operator. Evaluation of $\bar{g}_{\pi NN} \approx 0.038 |\theta|$ tells us that from current algebra $d_n \sim 5\times 10^{-16} \theta\, e$ cm, and so $\theta$ must be really tiny $\leq 10^{-10}$ in view of the experimental neutron EDM bound $d_n < 2.9\times 10^{-26} e$ cm. This is the so-called "strong *CP* problem" mentioned before, see also Baluni (1979).

The $\theta$ term also receives a contribution from imaginary quark masses that can be calculated from the chiral anomaly. Thus, the determinant of the $6\times 6$ quark mass matrix *M* can be taken to be real with a "physical" $\bar{\theta}$ in front of the strong-interaction field tensors $G\tilde{G}$ in Eq. (3.6), $\bar{\theta} = \theta + \arg \det M_u M_d$.

One can promote $\bar{\theta}$ to become a space-time dependent field (with a $1/f_a$ factor for canonical normalization). When we enlarge the SM such that there is a Peccei-Quinn symmetry $U(1)_{PQ}$ (as mentioned before) acting on colored fields, then the spontaneous breaking of this symmetry will have this $\bar{\theta}$ field, the axion, as a Goldstone boson. Further explicit breaking by the chiral anomaly fixes the $\bar{\theta}$ to zero at a classical minimum and produces a small axion mass. Further *CP* violating terms in the effective action can lead to small "induced" contributions to $\bar{\theta}$ (Pospelov and Ritz, 2005).

b. *Effective QCD Lagrangian*: When discussing the neutron EDM today, the cleanest procedure is to start from an effective QCD Lagrangian at the 1 GeV scale that contains *CP* violating effects beyond the CKM matrix in the tree action of the SM in an integrated-out form. This can be obtained in the SM by considering loop effects (see later), but – very important – also in extensions of the SM with further massive fields that are still not observed. The additional *CP* violating part of the effective QCD Lagrangian for quarks, gluons, and photon-gluon couplings (corresponding to quark EDMs $d_i$ and quark color EDMs $\tilde{d}_i$, respectively) is added to the ordinary QCD-Lagrangian, Eq. (3.6),

$$L^{CPV} = \bar{\theta}\frac{g_s^2}{32\pi^2}G_{\mu\nu}^a \tilde{G}^{\mu\nu,a}$$
$$-\tfrac{1}{2}i\sum_{i=u,d,s,\ldots}\bar{\psi}_i(d_i F_{\mu\nu}\sigma^{\mu\nu} + \tilde{d}_i G_{\mu\nu}^a t^a \sigma^{\mu\nu})\gamma_5 \psi_i. \quad (3.24)$$

The second term actually corresponds to dimension 6 terms induced by a Higgs coupling to chiral quarks, although it looks like dimension 5. Indeed there are further dimension 6 operators not given above, a 3-gluon coupling (Weinberg operator) and 4-quark couplings, which also play a role in the extensive literature (Pospelov and Ritz, 2005) – and of course one can imagine numerous higher-dimensional terms suppressed by some power of a scale factor in the denominator.

The effective Lagrangian only covers small-distance effects. Still, in a calculation of the neutron EDM large-scale strong interaction/ confinement effects have to be included. This is a difficult problem not solved in full generality yet. Constituent quark models based on chiral symmetry and its breaking are a practical way to produce results.

c. *QCD sum rules* (Novikov, Shifman, Vainshtein, and Zakharov, or "NSVZ") have a higher ambition and have been used with high virtuosity (Pospelov and Ritz, 2005). Still they contain quite a few "technical" assumptions. Particularly for strong-interaction problems, a lattice simulation should be more reliable in the end (Shintani *et al.*, 2005, 2008, Alles *et al.*, 2007). Unfortunately, it will be rather agnostic compared to analytical calculations, but lattice data can control these analytical calculations. QCD sum rules lead to a neutron EDM in terms of quark EDMs $d_q$ and color EDMs $\tilde{d}_q$ at the 1 GeV scale (Pospelov and Ritz, 2005)

$$d_n(d_q, \tilde{d}_q) = (1 \pm 0.5) \frac{|\langle \bar{q}q \rangle|}{(225 \text{ MeV})^3}$$
$$\times [1.4(d_d - 0.25 d_u) + 1.1 e(\tilde{d}_d + 0.5 \tilde{d}_u)] \quad (3.25)$$

in the case of a Peccei-Quinn (PQ) symmetry suppressing $\bar{\theta}$.

Without PQ symmetry there is in addition the usual $\bar{\theta}$ term of size

$$d_n(\bar{\theta}) = (1 \pm 0.5) \, \bar{\theta} \, \frac{|\langle \bar{q}q \rangle|}{(225 \text{ MeV})^3} 2.5 \times 10^{-16} \, e \text{ cm} \quad (3.26)$$

and a further term due to an induced $\theta$ term proportional to the quark color-EDMs $\tilde{d}$. In these equations, $\langle \bar{q}q \rangle$ is the quark condensate, and the $d_i, \tilde{d}_i$, and $\bar{\theta}$ are defined in the effective Lagrangian Eq. (3.24). Sea quark contributions turn out to be suppressed.

d. *CP violation via CKM matrix*: In the SM, a nonvanishing quark EDM due to *CP* violation via the CKM matrix can be calculated if one goes to three-loop perturbation theory. A leading log term turns out to be most important, and the dominant EDM of the *d* quark results as (Czarnecki and Krause, 1997)

$$d_d = e m_d m_c^2 \alpha_s \, G_F^2 \, \frac{J}{108 \pi^5} \ln^2 \frac{m_b^2}{m_c^2} \ln \frac{M_W^2}{m_b^2}$$

(via CKM matrix), $\qquad\qquad\qquad$ (3.27)

where *J* is the Jarlskog measure of CKM-*CP* violation in the SM given in Eq. (3.8) (for the weak and strong coupling strengths the Fermi coupling constant $G_F \propto g_w^2$, Eq. (6.4), and $\alpha_s = g_s^2 / 4\pi$ are used here). Together with other contributions, this leads to the *d* quark EDM estimate in the SM $d_d \approx 10^{-34} \, e$ cm. This is much too small to produce (via relation Eq. (3.25)) a neutron EDM that could be measured even in the far future. Larger results can be obtained if not just one quark in the neutron is responsible for the *CP* violation. Figure 7 shows typical SM perturbation theory graphs with two or three quarks involved.

"Strong penguin" diagrams together with ordinary weak interaction (Fig. 7) turned out to be most important (Gavela *et al.*, 1982, Khriplovich and Zhitnitsky, 1982, Eeg and Picek, 1984), with typical GIM cancellation $\sim (a_s / \pi) \log (m_t^2 / m_c^2) \sim O(1)$, rather than the usual $\pi^{-2} (m_t^2 - m_c^2) / m_W^2$ $\sim O(10^{-2}...10^{-3})$. The weak interactions are short range and can be handled with perturbation theory, but the quarks in the infrared, inside the diagrams and at the outer lines, get bound into hadronic spin-$\frac{1}{2}$ states with both positive and negative parity (namely, $N^*, \Sigma, \Lambda$), and meson and baryon intermediate states play a role. The outcome of this discussion is an estimate (McKellar *et al.*, 1987)

$$d_n \approx 10^{-32} \, e \text{ cm (via CKM matrix)}, \quad (3.28)$$

still much too small to be measured.

Returning to *CP* violation by a possible $\bar{\theta}$-term, we have the opposite situation: Even a tiny $\bar{\theta} \sim 10^{-10}$ gives a neutron EDM close to the measured bounds. As we already mentioned, this EDM can be calculated in current algebra, in chiral perturbation theory, in QCD sum rules, and in constituent quark models. The values obtained are all in the same range, for a given $\bar{\theta}$,

$$d_n \sim f \bar{\theta} \times 10^{-17} \, e \text{ cm (strong } CP \text{ violation)}, \quad (3.29)$$

where $4 \leq f \leq 40$, and each of the methods has big error bars (Narison, 2008). It is also interesting to compare the hadronization procedures in a calculation of the nucleon magnetic moments, whose experimental value is well known, see Sec. D.3 below.

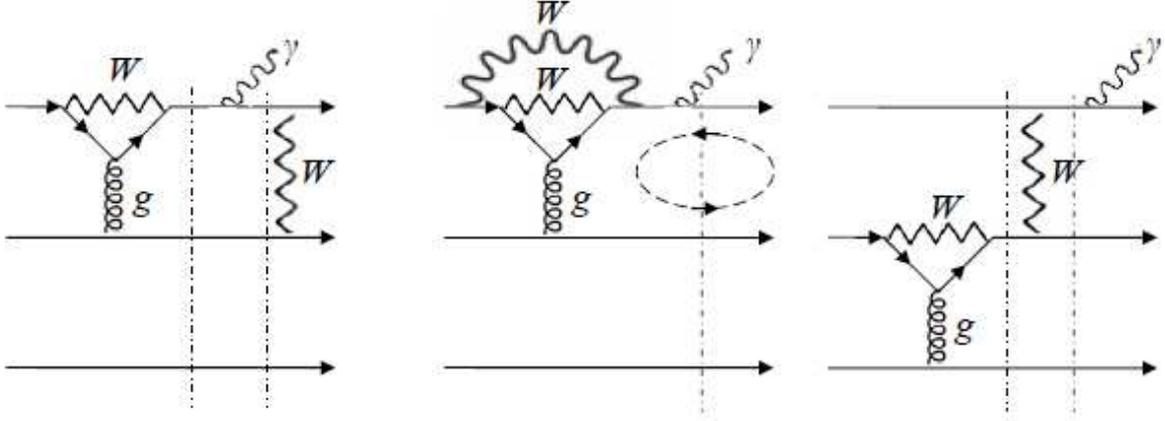

FIG. 7. Typical graphs in the Standard Model contributing to a neutron EDM. The vertical lines indicate hadronic intermediate states.

**2. EDMs in supersymmetry**

There are good reasons to believe that the SM is just an effective theory up to the scale of weak interaction of ~100 GeV, and that in the TeV range there are new particles and new couplings "beyond the SM" that will become visible in the forthcoming accelerator experiments in the TeV range. Let us just mention some of the reasons.

● the mysterious hierarchy between the weak scale and the Planck scale (see Sec. V.A.1);
● the attractive picture of a grand unification of forces;
● the very small neutrino masses presumably requiring a further large scale (for the seesaw mechanism);
● the dark matter component of our universe; and, last not least,
● the up to now experimentally unexplored Higgs sector.

a. *Dimensional analysis*: When constructing models beyond the SM one has to take care that they reduce to the SM in the well explored "low energy" region. More technically speaking, the new structure should be "integrated out" and should produce corrections to the couplings of the SM, and should have further higher dimensional terms in the interaction Lagrangian, as in Eq. (3.24). These corrections then have to carry, explicitly or implicitly, some power of a new mass scale factor $M$ in the denominator. For quark EDMs we also envisage new $CP$ violating phases that could produce large values. These EDMs also require a chiral "turner" changing quarks from left-handed to right-handed, which must be proportional to some quark mass $m_q$. Dimensional consideration then results in $d_q \sim e|f^{CPV}| m_q / M^2$ for scale $M$ of the process, where $f^{CPV}$ describes the $CP$ violating process, eventually also comprising $1/(16\pi^2)$ loop factors.

With $d_q \leq 4 \times 10^{-26} e$ cm respecting the current neutron EDM bound (using Eq. (3.25), for example), and with $m_q \sim 10$ MeV one arrives at a bound

$$M \geq |f^{CPV}|^{1/2} \times \left(\frac{m_q}{10 \text{ MeV}}\right)^{1/2} \times 70 \text{ TeV} . \qquad (3.30)$$

Thus, even without measuring new particles at the required energy scale the neutron EDM limit (or value) might tell us something about this scale $M$ in concrete "beyond" models – of course the detailed mechanism of $CP$ violation still matters.

b. *Minimal supersymmetry MSSM, one loop*: The most prominent model beyond the SM these days is the minimal supersymmetric SM (MSSM, introduced above in Sec. A.4.d). SUSY offers a solution or at least an improvement for the problems mentioned at the beginning of this section. In the MSSM with universal SUSY breaking parameters there are two $CP$ violating phases that can not be defined away: The phase $\theta_\mu$ of the $\mu$-parameter of the supersymmetric Higgs mass (appearing as $\mu H_u H_d$ in the superpotential, see Eq. (3.9)) redefining the $m_2$ phase to zero and the phase $\theta_A$ of the $A$ parameter of the SUSY-breaking coupling of three scalar fields. The phase $\theta_\mu$ induces $CP$ violation to the weak gaugino-Higgsino mass matrix and is most important for electroweak baryogenesis in the MSSM. The phase $\theta_A$ induces CP violation to the $stop_R$-$stop_L$ mass matrix.

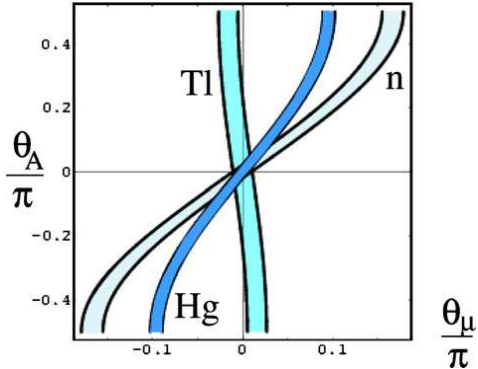

FIG. 8. (Color online) In supersymmetric (SUSY) extensions of the Standard Model, the EDMs of particles are naturally large. In the simplest version of a "minimal SUSY model" (MSSM), particle EDMs depend only on two phase angles $\theta_A$ and $\theta_\mu$. For a small SUSY breaking scale, experimental atomic and neutron EDM limits strongly constrain these phases to near the origin $\theta_A = \theta_\mu = 0$ where the three bands of the figure cross – much too small for electroweak baryogenesis. From Pospelov and Ritz (2005).

In this constrained MSSM (CMSSM) one can calculate a one-loop EDM of the $d$ quark (Ibrahim and Nath, 1998)

$$d_d = -\frac{2}{27} e g_3^2 (\sin\theta_\mu \tan\beta - \sin\theta_A) \frac{m_d}{16\pi^2 M_{SUSY}^2}$$
$$+ O(g_2^2, g_1^2) \text{ (SUSY)}. \quad (3.31)$$

Here $g_3 = g_s$ is the QCD coupling, $g_1$ and $g_2$ are the electroweak couplings, $m_d$ the quark mass at SUSY breaking mass scale $M_{SUSY}$, and $\tan\beta = v_u/v_d$ is the ratio of up/down Higgs expectation values mentioned before. This leads to an $f^{CPV}$ in (3.30) of

$$f^{CPV}_{MSSM} \sim \frac{\sin\theta_\mu \tan\beta - \sin\theta_A}{2000} (m_d/10 \text{ GeV})^{-1},$$
$$(3.32)$$

leading roughly to

$$M_{SUSY} \geq (\sin\theta_\mu \tan\beta - \sin\theta_A)^{1/2} \times 1.5 \text{ TeV}. \quad (3.33)$$

Indeed, Eq. (3.25) was derived at a scale of 1 GeV. Therefore Eq. (3.31) must be continued by the renormalization group.

A careful analysis shows (Falk and Olive, 1998, Pospelov and Ritz, 2005) that for a small SUSY breaking scale $M_{SUSY}$ (500 GeV, for example) the parameters $\theta_\mu$ and $\theta_A$ are heavily restricted. This becomes even more true if one also considers the restrictions by atomic data on the EDMs for $^{205}$Tl and $^{199}$Hg. This forces $\theta_\mu$ and $\theta_A$ to be very close to zero, see Fig. 8, much too small for electroweak baryogenesis in the MSSM.

c. *MSSM, two loops*: However, in the discussion above, the assumption of universal SUSY breaking and of a small SUSY breaking scale, and the restriction to one loop contributions in Eq. (3.31) was very (too) restrictive. If the squark masses are in the multi-TeV range, one-loop terms are suppressed, and two-loop contributions to the EDMs are dominant (Pilaftsis 2002, Chang *et al*. 2002, Li *et al.*, 2008, 2010, Ellis *et al*., 2008). In particular, restricting to *B*-ino mass-induced *CP* violation, the influence on EDM's is much suppressed (Li *et al*., 2009).

In Fig. 9, one can see that MSSM baryogenesis is seriously constrained by the present neutron (also electron) EDM data and will be conclusively tested in the near future by the LHC as well as the next generation of EDM measurements (Cirigliano *et al*., 2010b, see also Ramsey-Musolf, 2009). In the figure, $m_A$ is the mass of the *CP* odd scalar, $\tan\beta = v_u/v_d$ is the ratio of the two Higgs expectation values, and reasonable values for the *W*-ino and the *B*-ino mass parameters $m_1$ and $m_2$, and the SUSY-Higgs mass parameter $\mu$ are assumed as indicated in the figure.

d. *SUSY with additional superfield*: Data on EDMs and electroweak baryogenesis can be much easier accommodated in supersymmetric models with an additional gauge singlet superfield $S$ coupling to the $H_u$ and $H_d$ doublets as $SH_uH_d$ in the superpotential. The field value $\langle S \rangle$ then plays the role of the $\mu$-parameter in the MSSM, introduced in Eq. (3.9). In the "next to minimal" SSM (NMSSM), there is an $S^3$ coupling in the superpotential (Huber and Schmidt, 2001). In the "nearly minimal" SSM (nMSSM, Menon *et al*., 2004, Huber *et al*., 2006), the most important further term is $t_s S$ + c.c. appearing in the soft SUSY breaking potential $V_{soft}$, where $t_s$ is a complex parameter introducing *CP* violation in the *S*-system. The superfield $S$ finds its temperature dependent minimum $\langle S \rangle$ together with the two Higgs fields. There are now more parameters than in the MSSM, and one has to inspect a great variety of configurations. As indicated in Fig. 10 (Huber *et al*., 2006), it is easy to obtain a predicted baryon asymmetry higher than the observed baryon asymmetry, while simultaneously fulfilling EDM bounds (here the most restrictive is the electron EDM).

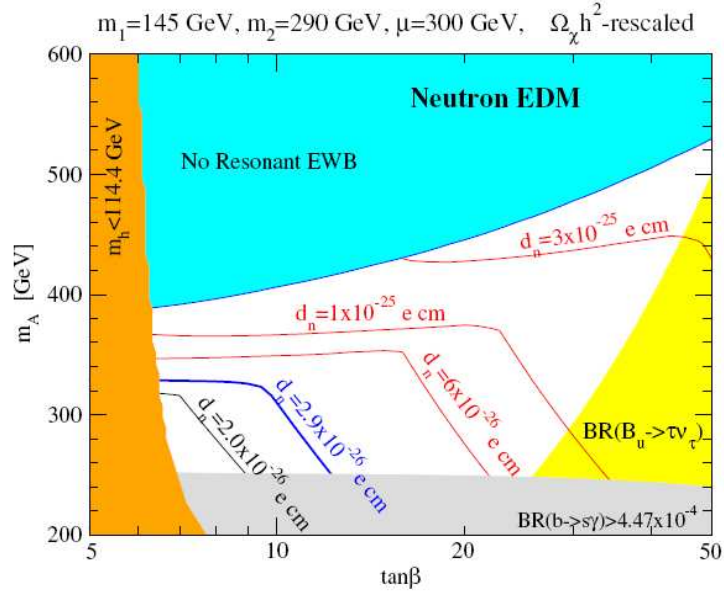

FIG. 9. (Color online) Also in a careful reanalysis taking more freedom (see text) the EDMs strongly constrain baryogenesis. The $\tan\beta$ - $m_A$ plane, with colored exclusion regions due to the Higgs-mass, electroweak baryogenesis (EWB), and branching ratios for *b*-decays. The blue line at $d_n = 2.9 \times 10^{-26}$ *e* cm indicates the current experimental limit for the *n*-EDM: the viable parameter space lies below this curve. From Cirigliano *et al*. 2010b.

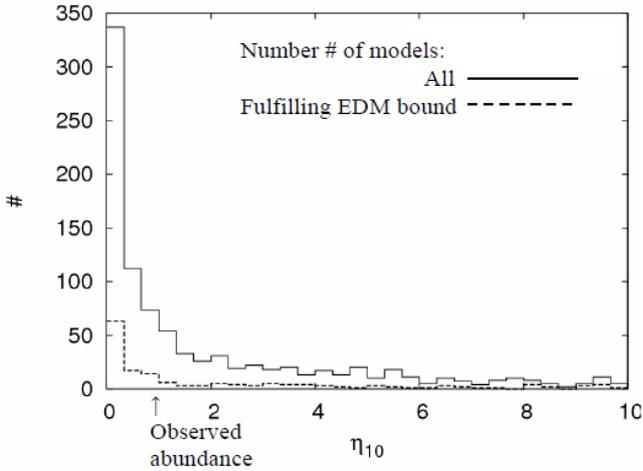

FIG. 10. Severe EDM constraints exist also when an additional scalar superfield is added to the MSSM, with more free parameters. The model is investigated by counting the (randomly chosen) points in parameter space that fulfil various constraints. The number # of results is shown for the analysis of the baryon asymmetry of the universe for large $M_2 = 1$ TeV, in dependence of the baryon-entropy abundance parameter $\eta^s_{10} = (n_B / s) \times 10^{10}$. Approximately 50% of the parameter sets predict a value of the baryon asymmetry higher than the observed value $\eta^s_{10} = 0.87$. The lower line corresponds to the small number of parameter sets (4.8%) that also fulfill current bounds on the electron EDM with 1 TeV sfermions. From Huber *et al*. (2006).

e. *Left-right symmetric SUSY*: In supersymmetric models with two Higgs doublets, the Peccei-Quinn mechanism for suppressing a QCD $\bar{\theta} \neq 0$ vacuum can be realized. Another possibility to suppress $\bar{\theta} \neq 0$ is offered by models that are parity *P* and *CP* symmetric in the ultraviolet (i.e., at very high energies), which are the left-right symmetric models with an enlarged weak gauge sector $SU(2)_L \times SU(2)_R$. The left-right symmetry in the Yukawa sector enforces Hermitian Yukawa matrices and therefore $\bar{\theta} \sim \arg \det M = 0$ (Mohapatra and Senjanovic, 1978, Ecker and Grimus, 1985, Frere *et al*., 1992). As the last possible axion window begins to close, see Sec. V.C.3, alternatives to the Peccei-Quinn mechanism are most welcome.

Of course, in order to arrive at the SM at "low energies", the *SU*(2) left-right symmetry has to be broken at some scale. This enforces spontaneous breaking of the *CP* symmetry. $\bar{\theta}$ then should stay small, but a large CKM phase is required by the present data. Supersymmetry helps to achieve such a low scale (Aulakh *et al*., 1997, Dvali *et al*., 1998): $\theta = 0$ is protected by a SUSY nonrenormalization theorem, whereas strong interactions can increase the CKM phase already below the *L–R* breaking scale, which is assumed to be much above the SUSY breaking scale (Hiller and Schmaltz, 2001, Pospelov and Ritz, 2005).

In a discussion of *CP* violating effects in SUSY theories with the focus on EDMs and baryogenesis, of course one also has to keep under control the consequences for flavor changing processes, the *CP* violating *K* and *B* decays, which at the present accuracy can already be well explained by the unitary CKM matrix of the SM. This is a very broad subject, and we refer to a few reviews (Jarlskog and Shablin, 2002, Masiero and Vives, 2003, Chung *et al.*, 2005, and Ramsey-Musolf and Su, 2008)

### 3. Other approaches

a. *Two Higgs doublets*: Without supersymmetry, there is no particular reason for two Higgs doublets. However, in view of our ignorance about the Higgs sector of the SM and of the problems with electroweak baryogenesis in the SM, such a simple modification of the SM is phenomenologically attractive (Bochkarev *et al.*, 1990). The general form of the 2-Higgs doublet model and possible *CP* violations are well-known (Gunion and Haber, 2005). It is possible to obtain viable baryogenesis with a lightest Higgs mass as large as about 200 GeV and extra Higgs states of mass above 200 GeV. The neutron EDM is predicted to be larger than about $10^{-2}$ *e* cm close to the present bound (Fromme *et al.*, 2006).

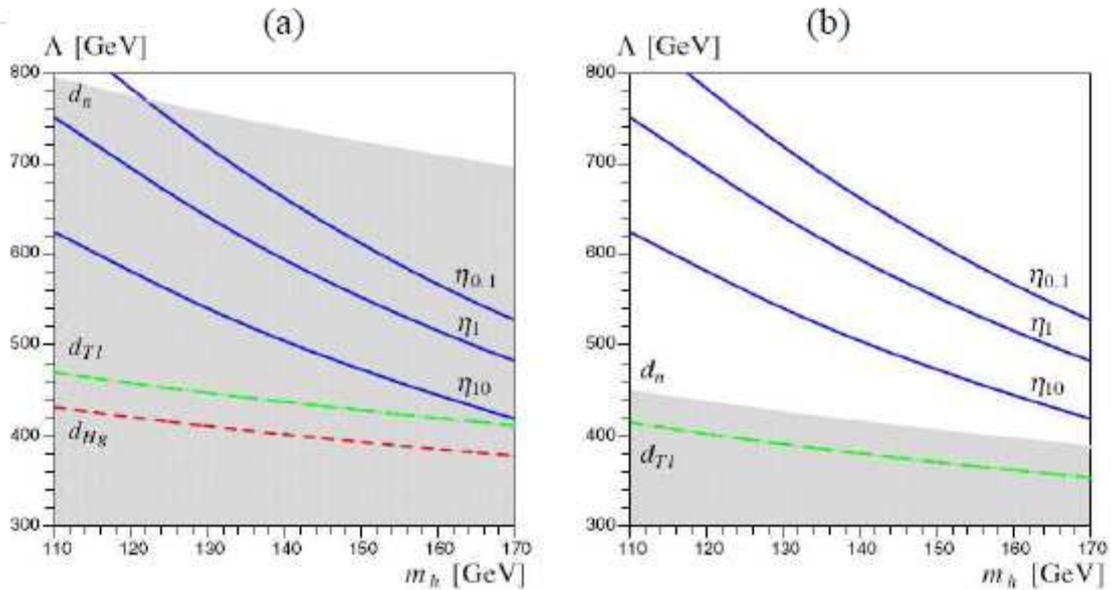

FIG. 11. (Color online) When the Standard Model is augmented not by SUSY, but by a higher dimensional term in the effective Higgs sector, strong EDM constraints exist as well. Shown are the Higgs masses $m_h$ and scales $\Lambda$ of new physics excluded (shaded areas) by the present EDM limits for neutron $d_n$, thallium $d_{Tl}$, and mercury $d_{Hg}$. (a) In the minimal scenario, the observed baryon asymmetry $\eta_1 = 6\times10^{-10}$ is excluded by the present limit on the neutron EDM $d_n$ (and so is one tenth and ten times $\eta_1$). (b) Therefore additional *CP*-odd sources must be introduced to lower the exclusion limits. From Huber *et al.* (2007).

b. *Higher-dimensional operators*: Trying to stay as close to the SM as possible but still to obtain a strong first order phase transition it has been proposed (Bödeker *et al.*, 2005, Grojean *et al.*, 2005, Huber *et al.*, 2007, Grinstein and Trott, 2008) to augment the SM with dimension 6 terms in the Higgs Lagrangian. In particular, one can have a term $(H^{\dagger}H)^3/\Lambda^2$, but also further *CP* violating terms of the type (usual Higgs coupling)$\times(H^{\dagger}H)/\Lambda^2_{CP}$ in the simplest case. In order to understand such nonrenormalizable terms one can imagine that they are obtained by integrating out a further heavy Higgs doublet (as discussed before). The role of the $\Phi^4$ term in the Higgs potential Eq. (3.2) is now played by the $\Phi^6/\Lambda^2$ term, and $\lambda\Phi^4$ plays the role of the perturbative $-\Phi^3$ term discussed in Sec. A.6 above, see Fig 4. If the two cut offs $\Lambda$ and $\Lambda_{CP}$ are identified (and taken to be quite low), and if only a single *CP* violating phase in the top-Higgs vertex is retained, then Huber *et al.* (2007) obtain successful baryogenesis.

However, this causes, in particular, the neutron EDM bound to exclude almost all of the $m_h - \Lambda$ parameter space, as shown in Fig. 11. On the other hand, one can easily fulfill the requirement that the full set of dimension-6 Higgs-quark couplings with SM flavor structure as indicated above does not violate the EDM bounds – a nice example how the EDMs constrain model building. Higher-dimensional operators in an effective theory can also be considered in SUSY theories (B(beyond)MSSM, Blum and Nir, 2008, Blum *et al.*, 2010).

## 4. Conclusions on EDM constraints

In particle physics, a neutron EDM is a clear signal of *CP* and *T* violation in a flavor conserving system. In the SM, *CP* violation induced via the electroweak CKM matrix is discussed thoroughly in theoretical papers, but it turns out to be extremely small, nothing to be tested by experiments in foreseeable future. On the contrary, strong *CP* violation via a QCD $\theta$-term is severely restricted to $\theta \leq 10^{-10}$ by present EDM measurements. Here, theoretical uncertainties are related to hadronization of matter. In theories beyond the SM, desirable for various reasons listed at the beginning of the preceding section, there are further sources of *CP* violation, and the measured bound on the neutron EDM constitutes a serious limitation on such violations, in particular, if it is combined with other EDM measurements. This is an extremely interesting connection between neutron physics and "beyond" models. It would be exciting to finally measure a concrete value and thus to be able to further restrict the parameters of such models in particle theory.

Concerning cosmology, we found that already present EDM results strongly constrain possible routes for baryogenesis as well as models beyond the SM, and we presented prominent examples:

● The experimental neutron and atomic EDM limits rule out electroweak baryogenesis in the MSSM with a low SUSY breaking scale and in a most simple setting, Fig. 8.
● The present *n*EDM limit also strongly constrains more refined MSSM results on baryogenesis, Fig.. 9.
● For the nearly minimal SUSY model nMSSM, which has an additional scalar superfield and further free parameters, the measured EDM limit reduces heavily the parameter space viable for baryogenesis, see Fig. 10.
● When the SM is extended by additional higher-dimensional terms, a strong first-order phase transition can be achieved, but baryogenesis is excluded by the neutron EDM in some versions, see Fig. 11.

If in future EDM experiments with sensitivities down to $10^{-28}$ *e* cm a nonzero EDM is found, then this will be a strong hint that the creation of the baryon asymmetry has occurred near the electroweak scale. If no EDM is seen, other ways of baryogenesis like leptogenesis or other models at intermediate energy scales or close to the inflationary scale are more probable.

## D. The other electromagnetic properties of the neutron

A multipole ($2^L$-pole) operator $M^{(L)}$ of a quantum systems with angular momentum quantum number *j* is determined by its reduced matrix element $\langle j || M^{(L)} || j \rangle$, with $\mathbf{L} = \mathbf{j} + \mathbf{j}$, so it obeys the triangle rule $L \leq 2j$. For the neutron with $j = \frac{1}{2}$, there are no multipole moment observables beyond the dipole with $L = 1$. We keep the following discussion on the neutron's multipole moments rather short, as there have been few new experimental results during the past decade.

### 1. Electric charge

The electric charge of the neutron is known to be zero to a high precision, but the Standard Model containing an abelian $U(1)_Y$ gauge symmetry does not require this, even after taking into account quantum anomaly cancellations (Foot *et al.*, 1993). Majorana masses of the neutrinos and, of course, a nonabelian gauge group of grand unification support charge neutrality of atoms and neutrons, but there remain problems (Witten, 1979).

Overall, only few hints exist for physics beyond the Standard Model, and the neutrality of neutrons (Baumann *et al.*, 1988) and of atoms (Dylla and King, 1973), both of order $10^{-21}$ *e*, is such a hint, since in the SM this would require an incredible fine-tuning. Some models beyond the SM that violate $B - L$ symmetry could accommodate a nonzero neutron charge $q_n = \varepsilon(B - L) \neq 0$, too, with the interesting signature that the charge of the hydrogen atom (which has $B = L$) would remain zero (Foot *et al*, 1993).

The elegant experiments quoted above were done decades ago, and, at the time, were rather one-man shows. This is a pity in view of the efforts invested in other searches beyond the SM. See Unnikrishnan and Gillies (2004) for a review on neutron and atom charges and on astrophysical limits, and Fedorov *et al.* (2008) and Plonka-Spehr (2010) for recent proposals for tests of neutron and Arvanitaki *et al.* (2008) for tests of atom neutrality.

### 2. Magnetic monopole moment

The neutron was also tested for a magnetic monopole moment, that is, for a free magnetic charge $g_n$ residing on the neutron. Magnetic monopoles play an important role in grand unified theories. When such a magnetic monopole $g_n$ flies through a magnetic field **B**, it follows a flight parabola due to the acceleration $\mathbf{a} = (g_n/m_n)\mathbf{B}$. Using neutrons flying through a crystal with "zero effective mass" $m_{n\,\text{eff}} \ll m_n$, see Sec. V.D.4, in a superimposed strong magnetic field, Finkelstein *et al.* (1986) obtained for the neutron magnetic monopole moment

$$g_n = (0.85 \pm 2.2) \times 10^{-20} \, g_D, \quad (3.34)$$

where $g_D$ is the Dirac magnetic charge $g_D = ce/\alpha = \varepsilon_0 c^2 (2\pi\hbar/e)$. This is the best limit for a free particle, although a tighter limit $g_n < 10^{-26} g_D$ is obtained from SQUID measurements on bulk matter (Vant-Hull, 1968).

### 3. Magnetic dipole moment

The proton and the neutron magnetic moments are known very precisely, the latter as a by-product of a neutron EDM search (Greene *et al.*, 1979) with

$$\mu_n = -1.913\,042\,73(45)\mu_N, \text{ and}$$
$$\mu_n/\mu_p = -0.684\,979\,35(16), \quad (3.35)$$

where the numbers in parentheses are the standard error in the last two digits, with the nuclear magneton $\mu_N = 3.152\,451\,2326(45) \times 10^{-8}$ eV/Tesla. In the simplified quark model, the neutron magnetic moment is given by

$$\mu_n = \tfrac{4}{3}\mu_d - \tfrac{1}{3}\mu_u \text{ (constituent quark model)}, \quad (3.36)$$

see also Eq. (3.21). In lowest order of perturbation theory, fundamental particles like the quarks (or the electron), have the Dirac magnetic moment $\mu_q = e_q/(2m_q^{\text{con}}) \approx 3e_q/(2m_N)$ for a constituent quark mass $m_q^{\text{con}} \approx \tfrac{1}{3}m_N$. The up quark with charge $+\tfrac{2}{3}e$ has $\mu = 2\mu_N$, the down quark with charge $-\tfrac{1}{3}e$ has $\mu = -\mu_N$, and from this

$$\mu_n = \tfrac{4}{3}(-\mu_N) - \tfrac{1}{3}(2\mu_N) = -2\mu_N, \quad (3.37)$$

which is in reasonable agreement with the measured value in Eq. (3.35).

As we already discussed in the context of the neutron EDM, the detailed description of a neutron in terms of quarks and gluons is much more involved. One can use chiral perturbation theory (Puglia and Ramsey-Musolf, 2000, Berger *et al.*, 2004, and references therein), QCD sum rules (Ioffe and Smilga, 1984, Balitzky and Young, 1983, Aliev *et al.*, 2002) or lattice gauge theory (Pascalutsa and Vanderhaeghen, 2005). There is also the interesting related question of form factors (Alexandrou *et al.*, 2006).

Generally, the accuracy of theoretical predictions for the magnetic moments of hadrons is very limited and cannot compete at all with the high precision from experimental result. Given some uncertainties/free parameters, these model calculations hardly allow one to draw definitive and precise conclusions on the hadronization in the infrared for the neutron and intermediate states. The test of nonstandard contributions then is in much worse shape than for the neutron EDM, where *CP* violation is filtered out.

### 4. Mean square charge radius

The internal structure of the neutron is mostly studied in high-energy experiments. In recent years, the field has seen much progress, see Grabmayr (2008) for a review. At low energy, the neutron mean-square charge radius and with it the initial slope of the neutron's electric form factor can be measured in neutron-electron scattering on heavy atoms. High and low-energy data on this quantity are now consistent with each other, see also the discussion in Sec. V.C.2.

### 5. Electromagnetic polarizability

The neutron *electric* polarizability was measured by the energy dependent neutron transmission though lead (Schmiedmayer *et al.*, 1991), with an error of same size as for the result obtained from deuteron break-up $\gamma d \to np$. The PDG 2010 average is $(11.6 \pm 1.5) \times 10^{-4}$ fm$^3$. The neutron electric polarizability can be thought of as measuring the slope of the confining strong potential acting between the quarks. The neutron *magnetic* polarizability was not measured with free neutrons, the value obtained from deuteron break-up is $(3.7 \pm 2.0) \times 10^{-4}$ fm$^3$. A more thorough discussion of these quantities is beyond the scope of this article.

In heavy nuclei, one also finds a *P* violating amplitude due to what is called an anapole moment. We shall leave this topic aside, because in the SM the anapole moment of the elementary fermion is gauge dependent and is not truly a physical observable.

## IV. BARYON NUMBER VIOLATION, MIRROR UNIVERSES, AND NEUTRON OSCILLATIONS

In the preceding Sec. III we had seen that in the GUT era of the universe, see Table I, baryon number $B$ is violated. This chapter is on baryon number violating processes of the neutron. We have seen that the proton can be unstable in grand unified theories, for example, $p^+ \to e^+ + \pi^0$, a process that has been searched for intensively, with a limit on the partial lifetime of $\tau_{pe\pi} > 8.2 \times 10^{33}$ years (90% C.L.), obtained in Super-Kamiokande by watching 50 000 m$^3$ of water over many years. The neutron can undergo similar $B - L$ conserving $\Delta B = \Delta L = 1$ processes, e.g., $n \to e^+ + \pi^-$, which has a limit of

$\tau_{ne\pi} > 1.6 \times 10^{32}$ years
(90% C.L., bound nuclei, $\Delta B = 1$). (4.1)

### A. Neutron-antineutron oscillations

In some models beyond the SM also $\Delta B = 2$ processes are allowed, in which case a single baryon with $B = 1$ changes into an antibaryon $B = -1$. Mesons indeed do "oscillate" into antimesons, as seen in the $CP$ violating $K^0 \leftrightarrow \bar{K}^0$ and $B^0 \leftrightarrow \bar{B}_0$ processes, but mesons and antimesons are quark-antiquark pairs both with $B = 0$. What oscillates is the meson's strangeness quantum number $S = +1 \leftrightarrow S = -1$, with $\Delta S = 2$.

Protons $p^+$ cannot oscillate into antiprotons $p^-$ because this violates charge conservation, but nothing in principle forbids neutron-antineutron oscillations $n \leftrightarrow \bar{n}$. A search for such oscillations was done years ago at ILL, see Baldo-Ceolin et al. (1994). A cold neutron beam of intensity $10^{11}$ s$^{-1}$ freely propagated in vacuum over a length of 80 m with a mean flight time of $t \approx 0.1$ s. During this time, a fraction $P_{n\bar{n}} \approx (t/\tau_{n\bar{n}})^2$ of antineutrons would develop and annihilate in a carbon foil spanned across the beam at the end of the flight path. This annihilation would liberate $2m_n c^2 \approx 2$ GeV, mostly in form of pions that would have been detected in a large-volume tracking detector. The limit of

$\tau_{n\bar{n}} > 0.86 \times 10^8$ s
(90% C.L., free neutron, $\Delta B = 2$) (4.2)

was obtained for the oscillation period after one year of running time. For a report on new plans for free $n\bar{n}$ searches, see Snow (2009).

One can also derive a limit on free-neutron-antineutron oscillations from the stability of nuclei, but to this end one has to extrapolate over 30 orders of magnitude to arrive, for instance, from the Soudan-2 result on iron from 2002 $\tau_{n\bar{n}}$(bound) $> 7.2 \times 10^{31}$ yr (90% C.L.) at

$\tau_{n\bar{n}} > 1.3 \times 10^8$ s$^{-1}$
(90% C.L., from bound nuclei, $\Delta B = 2$). (4.3)

In a conference contribution, Ganezer et al. (2008), the values $\tau_{n\bar{n}}$(bound) $> 1.77 \times 10^{32}$ yr and $\tau_{n\bar{n}}$(free) $> 2.36 \times 10^8$ s$^{-1}$ at 90% C.L. are given for the Super-Kamiokande-I measurements on $^{16}$O.

Mohapatra (2009) reviewed the theoretical rationale of $n\bar{n}$ oscillations, and emphasized that $B - L$ breaking by the seesaw mechanism in neutrino oscillations ($\Delta L = 2$) may be related to a seesaw $B - L$ breaking in $n\bar{n}$ oscillations ($\Delta B = 2$), which could bring $n\bar{n}$ oscillations into experimental reach. In particular, a mechanism was proposed, in which baryogenesis occurs via a baryon-number carrying scalar, whose asymmetric decays into $6q$ and $6\bar{q}$ could provide a source for baryon asymmetry and baryogenesis, see Babu et al. (2009). This scenario easily satisfies the three Sakharov conditions, and has the remarkable feature that it does not need the sphaleron mechanism, as do many of the models discussed before, because the decays are allowed well below the sphaleron decoupling temperature of about 100 GeV.

Since the review of Dubbers (1991a), no new data have been taken on free neutron oscillations, and following the criteria given at the end of the introductory Sec. I.B, we would not have taken up this subject again, had there not been new results on another type of neutron oscillations, not from neutrons to antineutrons, but from neutrons to so-called mirror neutrons, treated in the following.

### B. Neutron-mirrorneutron oscillations

The existence of a mirror world that would compensate for the $P$ violating mirror asymmetry of the weak interaction is an old idea that resurfaced in recent years in the context of searches for dark-matter candidates, see Berezhiani and Bento (2006), and Mohapatra et al. (2005). Mirror particles could be present in the universe and would not interact with ordinary particles except for their gravitational interaction, and except for some mixing between the two worlds via appropriate gauge particles.

Due to such mixing, neutrons $n$ could possibly oscillate into mirror neutrons $n'$, and thereby cease to interact with ordinary matter, that is they simply would disappear from our world. Pokotilovski (2006a) discussed various schemes for experiments on neutron-mirror neutron oscillations (both in "appearance" and "disappearance" mode, in the language of neutrino physicists).

One would think that the disappearance of our cherished neutrons into nowhere would not have gone unnoticed, and that experimental limits on $nn'$ oscillation times should be far longer than the measured neutron lifetime. However, this is not so, because all previous neutron lifetime experiments were done in the presence of the earth's magnetic field, invisible for the mirror neutron. If we assume that the mirror magnetic field, to which the mirror neutrons would react is negligible on earth, the requirement of energy conservation would lead to the suppression of $nn'$ oscillations. Therefore suppression of the magnetic field at the site of the experiment is required for a $nn'$ search.

For the old neutron-antineutron experiment, a 100 m³ large magnetic shield had been constructed around the free-flight volume, with residual fields $B < 10$ nT (Bitter *et al.*, 1991), small enough that the neutron's magnetic interaction $\mu B < \hbar / t$ was below the Heisenberg uncertainty limit. An equivalent requirement is that the neutron spin precession along the flight path has to be negligible, and indeed the precession angle had been measured to $\varphi = (\mu B / \hbar) t < 0.03$, Schmidt *et al.* (1992). However, in the neutron lifetime experiments, there was no such magnetic shield, and therefore $nn'$ oscillations would have been highly suppressed in these experiments.

Ban *et al.* (2007) and Serebrov *et al.* (2008a, 2009c) used their respective EDM experiments at ILL, which are well shielded magnetically, to obtain limits on neutron-mirror neutron oscillations and reached

$\tau_{nn'} > 103$ s (95% C.L.), and
$\tau_{nn'} > 448$ s (90% C.L.), (4.4)

respectively. This was an elegant low-cost experimental response to an interesting physics question.

To conclude, neutron oscillation experiments test basic symmetries of the universe, although the underlying theories are more speculative than for the other neutron tests beyond the Standard Model.

# V. EXTRA DIMENSIONS, NEW FORCES, AND THE FREE FALL OF THE NEUTRON

This chapter deals with hidden spatial dimensions of the universe, which are required in superstring theory for reasons of self-consistency, and which can lead to deviations from Newton's gravitational law. Recent experiments on neutron-state quantization in the earth's gravitational field are sensitive to such deviations at short distances. In addition, these experiments have triggered a number of other neutron studies on this subject. It seems hard to overestimate the importance of discovering these new forces, which would "provide us with a rare window into Planckean physics and the scale of supersymmetry breaking" (Dimopoulos and Giudice, 1996). We end this section with a review of some earlier experiments on the neutron's mass and gravitational interaction.

## A. Large extra dimensions

### 1. The hierarchy problem

Our universe is known to be flat to high precision, with three spatial and one temporal dimension. Is there more to it? The compatibility of general relativity and quantum mechanics/quantum field theory (QFT) today is still not well established. The most prominent version, superstring theory, requires 10 space-time dimensions for consistency. This revived the old ideas of Kaluza and Klein who introduced a fifth dimension in order to unify gravity and electromagnetism. If such extra dimensions of number *n* are compactified (that is, "curled up" to a radius *R*), the 4-dimensional description contains a whole "tower" of Kaluza-Klein particles, for each particle with $mass^2 \propto n / R^2$. This is too heavy to be seen in present experiments if $1/R$ is of the order of the Planck mass ($\hbar = c = 1$), but measurable if in the TeV range.

Forgetting about string theory for a while, theoretical ideas about the role of "extra" dimensions in gravity have been put forward in the last ten years. The extra *n* dimensions are not visible in everyday life because they are compactified to a radius so small that wavelengths of objects in these extra dimensions become very short, and the next "Kaluza-Klein level" becomes so high that it can no longer be excited in today's low-energy world.

Introduction of extra dimensions may help with another problem, namely, the vastly different energy scales existing in the universe, see Table I. The scale of the gravitational interaction is given by the Planck mass $M_{Pl} \approx 10^{19}$ GeV$/c^2$ and defines the strength of the gravitational coupling constant $G = 1/M_{Pl}^2$ in Newton's law.

The scale of the electroweak interaction is given by the Higgs expectation value $\upsilon = 246$ GeV. The ratio of gravitational to weak coupling strengths then is $\upsilon^2 / M_{Pl}^2 \approx 4 \times 10^{-34}$, and this is called the hierarchy problem of the SM: If, in a quantum system, one scale is very small compared to another scale, then unavoidable radiative corrections should make both scales of comparable size, unless a plausible reason is known for its smallness, such as an additional symmetry. Otherwise, some incredible fine-tuning is required, here of $v^2/M_{Pl}^2$ to nearly zero, which seems very unlikely and calls for some deeper explanation.

## 2. Bringing down the Planck mass

An elegant way to solve this hierarchy problem is to postulate that, in reality, there is no hierarchy because, basically, only one single energy scale exists, namely, the electroweak scale. Gravity then indeed is of the same strength as the other interactions, but only at very short distances (Arkani-Hamed *et al.*, 1998, 1999, called the ADD model). At larger distance, gravity looks weak only because it is "diluted" in the $3+n$ spatial dimensions of the "bulk", with $n \geq 1$ extra dimensions, whereas the three other forces of the SM act only in the three spatial dimensions of the "brane" and are not diluted.

Gravitational interaction has infinite range, because the exchanged graviton is massless. Its field lines are not lost but merely diluted in space, which is the essence of Gauss' law, which states that the field's flux through any spherical shell around a body is constant, from which follows the $1/r^2$ force law in the usual way. Correspondingly, in $3+n$ dimensions Gauss' law, applied to a $2+n$-dimensional shell, leads to a $1/r^{2+n}$ force law, which decays faster with distance $r$ than does Newton's $1/r^2$ law. If the $n$ extra dimensions are compactified to radius $R$, this fast fall-off occurs only for small distances $r < R$, while for $r \gg R$ the gravitational force falls off with the usual $1/r^2$. On the other hand, the electromagnetic force is assumed to act only in three dimensions, and for all distances $r$ falls off as $1/r^2$. We follow ADD and postulate that at very short distances $r \ll R$ all forces are unified and have the same strength. As gravitation initially falls off much faster than electromagnetism, in our visible world with $r \gg R$, gravitation looks much weaker than electromagnetism.

To quantify these statements we first write down Newton's gravitational potential for two masses $m$, $M$ in three spatial dimensions

$$V_{grav}(r) = -G\frac{mM}{r}, \text{ for } r \gg R. \quad (5.1)$$

At very short distances, we write the force law in $3+n$ dimensions as

$$V_{grav}(r) = -G^*\frac{mM}{r^{n+1}}, \text{ for } r \ll R, \quad (5.2)$$

as shown in Fig. 12.

The "true" gravitational constant $G^*$ in this equation for dimensional reasons must depend on the "true"  Planck mass $M_{Pl}^*$ as $G^* = 1/(M_{Pl}^*)^{2+n}$. We only want to have one single scale, and therefore we set this true Planck mass equal to the electroweak scale $M_{Pl}^* \equiv v = 246$ GeV (in SI units, $G^* = \hbar^{n+1}/(v^{n+2}c^{n-1})$). In particular, at the outer range $R$ of the compactified dimensions, we write Eq. (5.2) as

$$V_{grav}(R) = -\frac{G^*}{R^n}\frac{mM}{R}, \text{ for } r = R, \quad (5.3)$$

under the assumption that all extra dimensions of number $n$ have the same range.

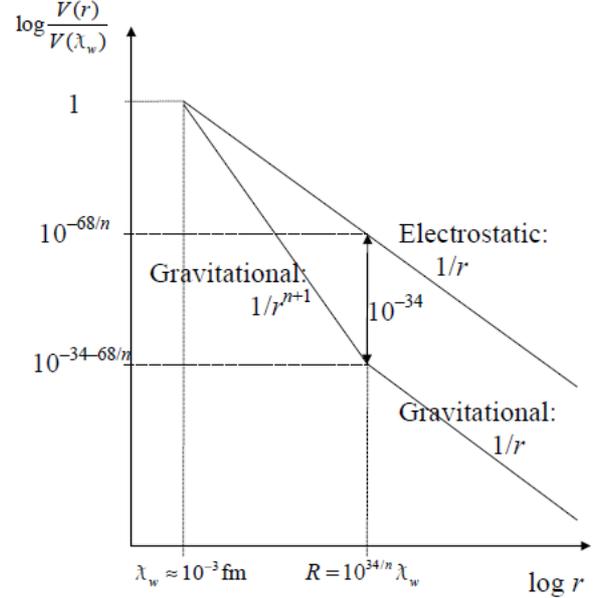

FIG. 12. The ADD model has $n$ extra dimensions, compactified to distances $r < R$, and only one single energy scale. In this model, gravitation is diluted in $3+n$ dimensions, while the other interactions are active only in three dimensions. At the electroweak scale $\lambda_w \approx 10^{-3}$ fm, the Coulomb potential $V_{el} \propto 1/r$ equals the gravitational potential $V_{grav} \propto 1/r^{n+1}$. At large distance, $r \gg R$, both have the usual $1/r$ potential, but with coupling strengths differing by a factor $\sim 10^{34}$.

At large distances $r \gg R$, we continue this with the $1/r$ potential, so

$$V_{grav}(r) = -\frac{G^*}{R^n}\frac{mM}{r}, \text{ for } r \gg R. \quad (5.4)$$

This must be identical to Newton's gravitational potential, Eq. (5.1), i.e., $G = G^*/R^n$. Insertion of $G$ and $G^*$ leads to the compactification radius of the extra dimensions

$$R = \left(\frac{G^*}{G}\right)^{1/n} = \lambdabar_w \left(\frac{M_{Pl}}{\upsilon}\right)^{2/n} \approx 10^{-18}\,\text{m} \times 10^{34/n}, \quad (5.5)$$

where $\lambdabar_w = 1/\upsilon = 0.80 \times 10^{-18}\,\text{m}$ is the reduced Compton wavelength of the electroweak scale.

The larger the number $n$ of extra dimensions, the shorter is their effective range $R$ in Eq. (5.5):

$R \approx 0.1$ m for $n = 2$;
$R \approx 0.2$ μm for $n = 3$;
$R \approx 0.3$ nm for $n = 4$;
$R \approx 0.5$ pm for $n = 6$, i.e., for $4+n = 10$ dimensions.

One extra dimension $n = 1$, with $R \approx 10^{16}$ m, is excluded because astronomical observation confirm Newton's law at this scale. For $n \geq 2$ extra dimensions and their shorter compactification radii, the exclusion plot Fig. 16 below gives present bounds, as will be discussed in Sec. C.

The Coulomb potential $V_{el}(r)$, on the other hand, falls off as $1/r$ for all distances $r$. In the ADD model of large extra dimensions, the electroweak and the gravitational interaction are unified at the electroweak scale $\upsilon$, at which point their coupling strengths coincide. This happens at $r^n = G^*/G$, or, with $G^* = 1/\upsilon^{n+2}$ and $G = 1/\upsilon^2$, at $r = 1/\upsilon = \lambdabar_w$. Hence, both potentials, Coulomb's $V_{el}(r)$ and the gravitational $V_{grav}(r)$ meet at the electroweak length scale $\lambdabar_w \approx 10^{-18}$ m. At large distances $r \gg R$, the ratio of gravitational to electrostatic interaction is $\upsilon^2/M_{Pl}^2 \approx 4 \times 10^{-34}$ as given above. Figure 12 indicates the distance dependence of both potentials. Near $r = R$, of course, a smooth transition from $1/r^{n+1}$ to $1/r$ is expected.

### 3. Competing models

The ADD model of large extra dimensions is only one of many possible models on the unification of forces. It is an attractive model as it solves deep problems of contemporary physics and, as is always welcome, predicts measurable signals both at the LHC and in low-energy physics. Randall and Sundrum (1999), for instance, in their pioneering (RS) model have only one extra spatial dimension (but with possible relations to string theory). They obtain a hierarchy of scales by introducing a 5-dimensional anti de Sitter space with one "warped" extra dimension $y$, such that the metric of our 4-dimensional subspace decreases along $y$ as $\exp(-k|y|)$. The Planck mass $M_{Pl}$ is then related to the 5-dimensional Planck mass $M$ as $M_{Pl}^2 = M^3(1 - e^{-2kr_c\pi})/k$, where the boundary $y = \pi r_c$ of the warped dimension can be extended to infinity in the special model, such that $M$ can be lowered to the TeV/$c^2$ range. The effective gravitational potential for masses $m_1$ and $m_2$ on our 4-dimensional brane then is

$$V(r) = G\frac{m_1 m_2}{r} + \int_0^\infty \frac{G}{k}\frac{m_1 m_2 e^{-mr}}{r}\frac{m}{k}\,\text{d}m, \quad (5.6)$$

where the second term comes from the exchange of massive Kaluza-Klein particles of the graviton/gravitational field.

## B. Gravitational quantum levels

### 1. Neutron states in the gravitational potential

Ultracold neutrons (UCN) moving at distance $z$ closely above the surface of a flat horizontal neutron mirror see a triangular potential, which is a superposition of the linear gravitational potential $V(z) = mgz$ for $z > 0$ and the repulsive potential $V_0 \approx 100$ neV of the neutron glass mirror for $z \leq 0$.

Neutrons can be bound in this triangular potential and form standing waves. In a simplified treatment, for $z > 0$ and $V_0 \to \infty$, the $n^{\text{th}}$ eigenfunction $\psi_n(z)$ is the (rescaled) Airy function $\text{Ai}(z - z_n)$, shifted by a distance $z_n$ such that its $n^{\text{th}}$ zero-crossing coincides with the surface of the mirror, and $\psi_n(z) \equiv 0$ for $z \leq 0$, see Fig. 13 for $n = 1$ and $n = 2$. In other words, all $\psi_n(z)$ look the same up to their $n^{\text{th}}$ node, for instance $|\psi_2(z)|^2$ in Fig. 13 looks like $|\psi_1(z)|^2$ up to its $1^{\text{st}}$ node.

These solutions of the Schrödinger equation are well known from the two-dimensional electron gas in a semi-conductor heterostructure. In the case of the electrons, eigenenergies are in the eV range, while for neutrons, the first few energy levels $E_n$ lie at (1.44, 2.53, 3.42, 4.21, …) peV, for $n = 1, 2, 3, 4, ...$, with $1\,\text{peV} = 10^{-12}$ eV, so $V_0 \gg E_n$ is fulfilled, and setting $V_0 \to \infty$ is justified. The corresponding classical turning points $z_n = E_n/mg$ are at distances (13.7, 24.1, 32.5, 40.1, …) μm above the surface of the mirror.

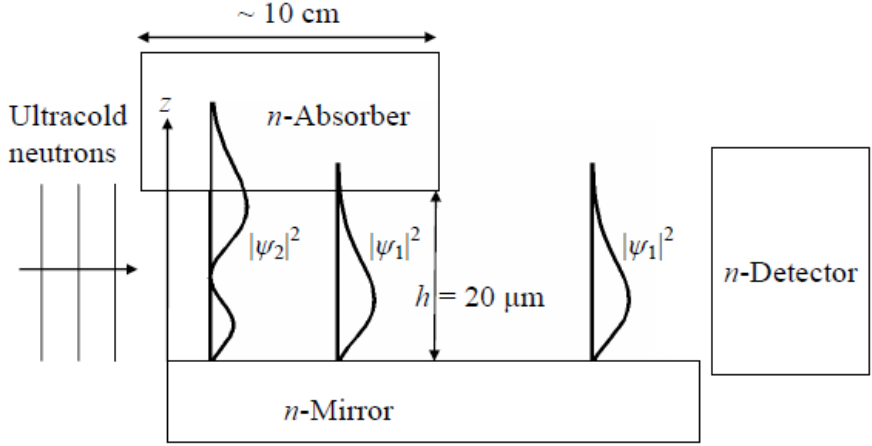

FIG. 13. Scheme of the experiment on UCN gravitational quantum levels. UCN plane waves enter from the left, and form standing waves in the triangular potential formed by the mirror potential $V \approx \infty$ and the gravitational potential $V = mgz$. For the absorber height $h = 20$ μm shown, only neutrons in the first eigenstate $\psi_1$ with energy $E_1 = 1.4$ peV pass the device and are detected, while the higher eigenstates $\psi_{n>1}$ are removed from the beam. Adapted from Luschikov and Frank (1978).

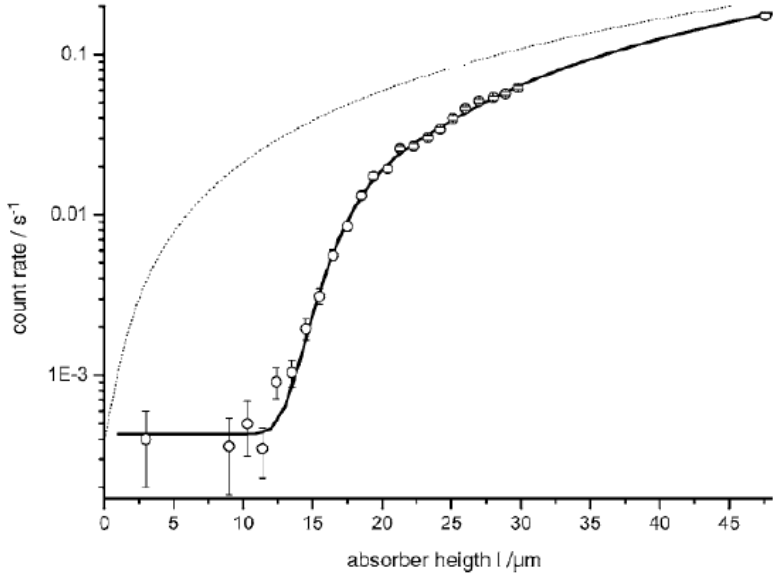

FIG. 14. Ultracold neutron transmission of the device Fig. 13, measured in dependence of the height $h$ of the absorber. Dotted line: classical expectation; solid line: quantum calculation. From Westphal et al. (2007b).

### 2. Ultracold neutron (UCN) transmission experiments

Nesvizhevsky et al. (2002, 2005) observed the quantization of gravitational neutron states in an experiment at ILL. The UCN entered a device consisting of a horizontal neutron mirror and a neutron absorber/scatterer placed at a variable height $h$ above the mirror, see Fig. 13, and were detected after leaving this slit system.

Figure 14 shows the number $N(h)$ of transmitted UCN as a function of the height $h$ of the absorber. The dotted line is the classical expectation, $N(h) \propto h^{3/2}$. The solid line is a quantum mechanical calculation with three free parameters: the overall normalization, the neutron loss rate on the absorber, and the population coefficient for the $n = 1$ ground state. For a different theoretical approach to the problem with similar results, see Adhikari et al. (2007). Up to a height $h \approx 12$ μm, no neutrons can pass the system because the $n=1$ ground state has sufficient overlap with the absorber and is blocked (although visible light with its much longer wavelength easily passes). At first sight, one would expect a stepwise increase of neutron transmission, but this is almost completely washed out. For a recent review, see Baeßler (2009).

When the roles of the absorber and the mirror were exchanged, with the mirror above the absorber, the transmission dropped to 3%, as predicted by theory, see Westphal *et al.* (2007b). This proves that the effect is really due to gravitation, and not to simple "box-state" quantization between two mirrors.

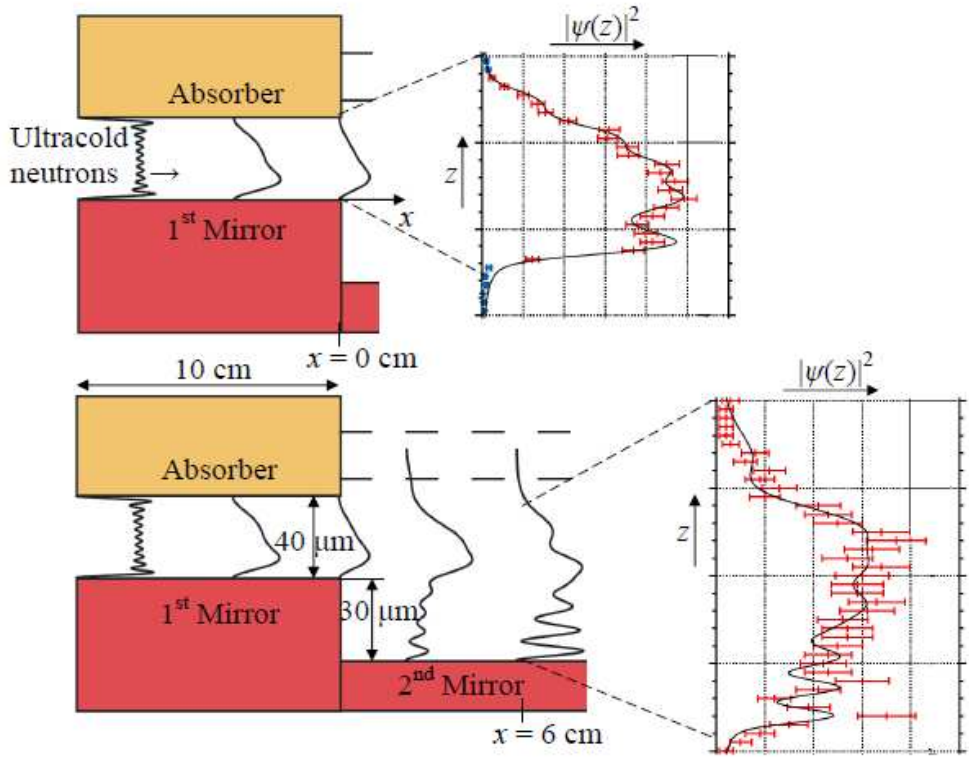

FIG. 15. (Color online) Position sensitive detection of the transmitted UCN after their "fall" over a step of 30 μm height at the exit of the slit system, measured at two different distances from the slit exit, $x=0$ cm, and $x=6$ cm. This differential phase sensitive measurement of neutron transmission in principle is much more sensitive to fifth forces than is the integral transmission curve, Fig. 14. Adapted from Jenke *et al.* (2009).

### 3. Observation of UCN gravitational wave functions and resonance transitions

In the meantime, both, the wavefunctions of UCN gravitational quantum states, and resonance transitions between such states have been measured.

The vertical distribution $|\psi(z)|^2$ of the UCN leaving the gravity spectrometer was measured in a specially developed position sensitive detector of spatial resolution 1.5 μm. After leaving the slit region, with fixed height of 40 μm, the UCN were dropped onto a second mirror, installed a distance 30 μm below the first mirror, to enhance the short wavelengths, as indicated in Fig. 15. While the UCN move along the second mirror their vertical distribution rapidly changes shape. It turns out that the detailed phase structure of $|\psi(z)|^2$ at a fixed height $h$ is more sensitive to the details of the potential $V(z)$ than is the integral transmission measurement $N(h) = \int |\psi_h(z)|^2 dz$ of the first experiment shown in Fig. 14. Figure 15 shows both the calculated and measured UCN distributions for distances $x=0$ cm and $x=6$ cm behind the exit of the slit system, see Abele *et al.* (2009).

Jenke *et al.* (2011) report the first observation of vibration-induced resonance transitions between UCN gravitational levels. In principle such resonance transitions between quantized gravitational levels can be detected in an apparatus consisting of three sections (like a Rabi resonance-in-flight apparatus). The first preparatory section singles out and transmits only the gravitational ground state $n=1$. In the second, the resonance transition section, the neutron sees a time dependent vertical vibrational perturbation that induces a resonant $\pi$-flip to the $n=3$ level. The third section is the neutron state analyser, which only transmits those UCN that remained in the $n=1$ ground state.

In the experiment quoted, the three sections are realized in a single device consisting of a mirror on the bottom and a rough scatterer on the top. The scatterer suppresses the population of higher levels by scattering and absorption, and only allows the $n=1$ ground state to pass. The state-selector on top introduces an asymmetry, because the ground state passes the system with higher probability than the excited state. Resonance transitions are induced by vibrating the entire system. The UCN detector is installed behind this device and registers the occurrence of a resonance via a sizeable (20%) reduction in UCN transmission. Meanwhile, the statistics of the resonance signals has been considerably improved over that presented in the Jenke *et al*. (2011) paper.

Nesvizhevsky *et al*. (2010a and 2010b), in what they call a "neutron whispering gallery", observed the quantization of cold neutrons that are subject to centrifugal forces near the surface of a curved neutron guide, dependent on neutron wavelength, and found results closely following theoretical expectations. This may open the road for new studies of short-range forces using cold neutrons of meV kinetic energy instead of UCN, see also Nesvizhevsky and Petukhov (2008), and Watson (2003).

The experiments on quantized gravitational levels suffer from low neutron count rates. Like the EDM experiments, they would profit a lot if one of the new UCN sources under construction would deliver the high fluxes envisaged. Furthermore, the instruments used for the searches of new interactions described above were not optimized for this purpose. Therefore with new instruments on new sources improvements for all these measurements by several orders of magnitude are expected.

For the experiment on quantized gravitational levels, a new instrument GRANIT is under construction, for details see Nesvizhevsky *et al*. (2007), Pignol (2009), and Kreuz *et al*. (2009). In the GRANIT project, UCN will be stored for long times up to the neutron lifetime. It is intended to induce radiofrequency transitions between the gravitational levels, with whatever fields couple to them. In its first version, GRANIT will run not with stored UCN, but in transmission mode, like the previous spectrometers at ILL.

Abele *et al*. (2010) discuss the use of a Ramsey separate oscillatory field apparatus with vibration-induced transitions between neutron-gravitational quantum states states. Voronin *et al*. (2011) study the possibility to extend such gravitational experiments to antineutrons.

Gudkov *et al*. (2011) investigate the possibility to study short-range interactions via Fabry-Perot interferometry between narrowly spaced plates, using parametric enhancement of the phase shifts of slow neutrons. As the resonance signals would be very narrow, one must take care that the signals are not washed out by limited resolution.

### C. Neutron constraints on new forces and on dark matter candidates

#### 1. Constraints from UCN gravitational levels

The gravitational neutron experiments described above are very sensitive to the form of the potential near the mirror. They therefore can test the existence of any short-range "fifth force" that couples to the neutron via some new charge $q$. Such new forces can be mediated by the exchange of bosons of different types (scalar, pseudoscalar, vector, etc.), originating from different sectors of models beyond the SM (supersymmetry, extra dimensions, etc.), or by the exchange of multiple bosons, see the review by Adelberger *et al*. (2003). The new interaction with coupling constant $g$ is not necessarily linked to gravitation, and one can make the general ansatz of an additional Yukawa potential as,

$$V(r) = \frac{\hbar c g^2}{4\pi} \frac{\exp(-r/\lambda)}{r}, \qquad (5.7)$$

where $\lambda = h/m_B c$ is the Compton wavelength of the exchanged boson of mass $m_B$.

This Yukawa ansatz also covers the ADD model with potential $V(r) \propto 1/r^{n+1}$ for $n$ large extra dimensions discussed above, if it is seen as an effective 4-dimensional model with towers of Kaluza-Klein particle exchanges (recurrencies), which are summed over. As a demonstration, for one extra dimension with $n+1=2$ this gives $\Sigma_{k=0}^{\infty} \exp(-kr/R)/r \approx R/r^2$ for $r \ll R$, via power series expansion. For $r \gg R$, this becomes $1/r$ multiplied by $\exp(-r/R)$ for $k=1$, which gives Eq. (5.7) again. However, the precise shape of the deviations from Newton's law are likely to depend on the details of the Kaluza-Klein spectrum (Callin and Burgess, 2006).

The range $\lambda$ is the Compton wavelength of Kaluza-Klein modes. Their strength is $g^2 \sim nG$, with the number of extra dimensions $n$, and gravitational coupling constant $G$, see Kehagias and Sfetsos (2000). However, $g^2$ may reach values six to eight orders of magnitude higher than this whenever there are additional gauge bosons freely propagating in the bulk, for instance when these are linked to a global $B-L$ symmetry, see Arkani-Hamed *et al.* (1999), and, for a more recent review, Antoniadis (2007).

The apparatus shown in Fig. 13 was built to detect neutron gravitational quantization, but is not specially optimized for detecting new forces. Still, the measurement shown in Fig. 14 gave interesting limits on such extra forces. It is expected that the upcoming dedicated instruments mentioned above will further push these limits. For scalar interactions with a Yukawa potential, Fig. 16 shows an exclusion plot in the plane of interaction strength $g^2$ versus the range $\lambda$ of the interaction. The top line $g^2=1$ of this figure corresponds to the strong interaction, and the bottom line $g^2=10^{-37}$ to the gravitational interaction. The regions above the curves in Fig. 16 are excluded by the measurements.

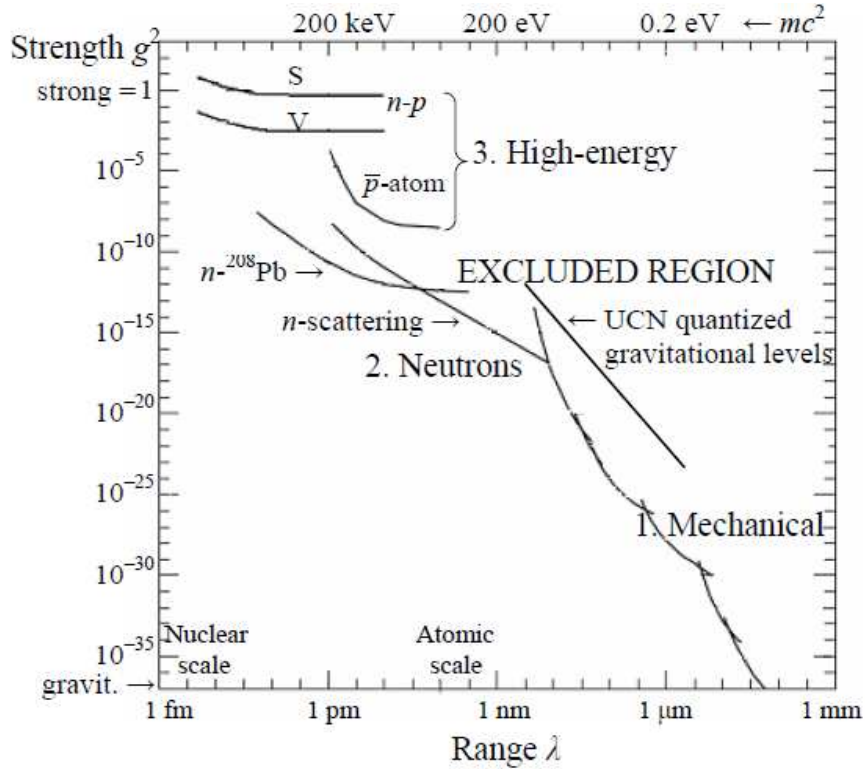

FIG. 16. Exclusion plot on new short-range interactions. Shown are the strength $g^2$ of the interaction versus its range $\lambda$. The upper scale gives the corresponding mass $m=h/\lambda c$ of the exchanged boson. Values above the curves shown are excluded by experiment. The constraints from neutron scattering in the subatomic range are combined from several different neutron measurements, see Nesvizhevsky *et al.* (2004, 2008). Most curves are adapted from Kamyshkov *et al.* (2008)

The straight line named "UCN quantized gravitational levels" in Fig. 16 gives the constraints derived from the UCN transmission curve Fig. 14 on an attractive Yukawa interaction of the neutron with the underlying mirror. The curve falls off as $g^2 \propto \lambda^{-4}$, from Nesvizhevsky and Protasov (2004), see also Abele *et al.* (2003), and Bertolami and Nunes (2003). The latter authors also discuss implications for the weak equivalence principle.

Figure 16 also shows bounds from torsion balance experiments with large objects for ranges $\lambda > 2$ μm, from Heckel *et al.* (2008), and from cantilever experiments with small objects for ranges between 2 nm and 1 μm, see Klimchitskaya *et al.* (2009) and references therein. The force on the cantilever is dominated by the Casimir force, which must be known very well in order to extract limits on new forces. The cantilever limits follow a curve $g^2 \propto \exp(d_0/\lambda)/\lambda^3$, so the bounds diverge exponentially as soon as $\lambda$ falls below the distance $d_0$ between probe and surface. Calculations of the Casimir effect between microscopic and macroscopic bodies are very demanding and not yet completely under control, see Canaguier-Ourand *et al.* (2010), and references therein. The UCN limit is not yet competitive with the published Casimir limits, but the neutron has the advantage that it has no background from Casimir forces.

Below the nanometer scale, Fig. 16 shows other limits from high-energy experiments with antiprotonic atoms at CERN ($A\bar{p}$, from Pokotilovski, 2006b), and from small-angle scattering of ~100 GeV neutrons on protons (*n-p*, from Kamyshkov and Tithof, 2008).

### 2. Constraints from neutron scattering

Also conventional neutron scattering experiments turn out to be sensitive to extra short-range forces. In Fig. 16, the new limits termed "*n*-scattering" reach far into the subatomic range. They are from a recent densely written paper by Nesvizhevsky *et al.* (2008), which presents several novel and independent methods to extract the following limits on new forces from existing neutron scattering data (see also Pignol, 2009):

● A global fit of a nuclear random-potential model to more than 200 measured neutron scattering lengths gives the limit $g^2\lambda^2 \leq 0.016$ fm$^2$. In the past, it was often asked who would need these neutron scattering lengths to such high precision, now we know.

● Scattering lengths from Bragg scattering and those from neutron interferometry have different responses to new forces. Comparison of both sets of data for 13 nuclei gives the asymptotic limit $g^2\lambda^2 \leq 0.0013$ fm$^2$.

● Comparison of scattering lengths at neutron energies of 1 eV with those at thermal neutron energies gives the asymptotic limit $g^2\lambda^2 \leq 0.0008$ fm$^2$ (all values with 95% C.L.).

In Fig. 16, the "*n*-scattering" curve shows this last, most constraining result, the linear part of the curve being the above asymptotic value. It is comforting that several independent data and evaluation methods all give very similar constraints. For completeness, we add the result of another evaluation of energy dependent total neutron scattering on $^{208}$Pb by Pokotilovski (2006b). A similar evaluation had already been made by Leeb and Schmiedmayer (1992).

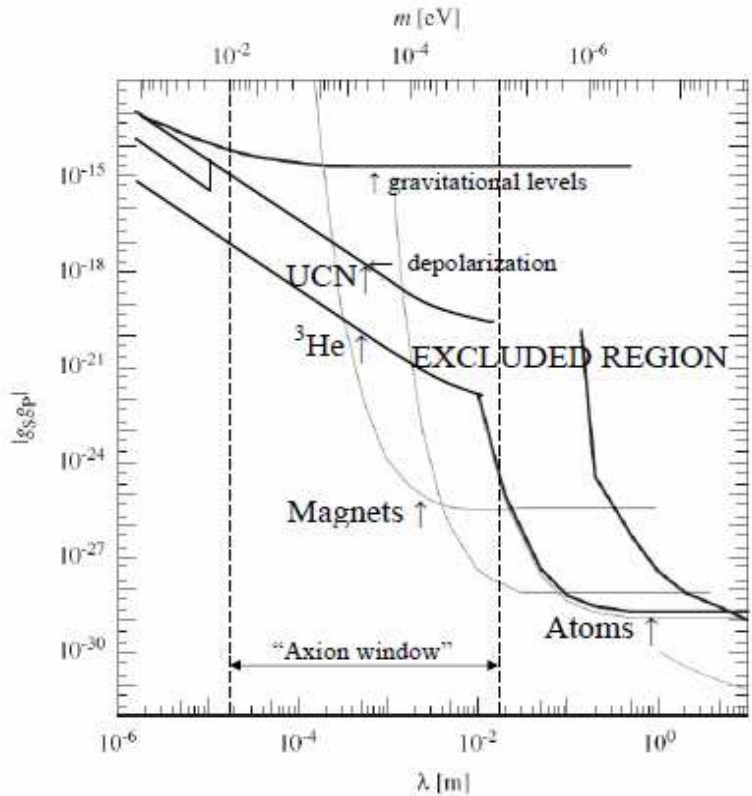

FIG. 17. Exclusion plot on new short-range spin-dependent interactions. The neutron measurements constrain axion interactions with nucleons in the axion window, left open by the astrophysical constraints (dashed vertical lines). The methods have room for improvement by several orders of magnitude. Black lines: coupling to nucleon spins; gray lines: coupling to electron spins. Most curves are adapted from Baeßler *et al.* (2009).

In the same paper, as a by-product, Nesvizhevsky et al. reconfirmed the Garching-Argonne value of the neutron-electron scattering length $b_{ne} = (-1.31 \pm 0.03) \times 10^{-3}$ fm, which significantly differed from the JINR Dubna value $b_{ne} = (-1.59 \pm 0.04) \times 10^{-3}$ fm, and which determines the neutron's mean square charge radius $\langle r_n^2 \rangle$, see Isgur (1999), Kopecky et al. (1997), and references therein. The value of $b_{ne}$ is also of high interest because it determines the initial slope of the neutron form factor $G_E(q^2)$, and for a long time was the only thing known about this form factor. Because they needed a precision value of $b_{ne}$ for their evaluation, the authors turned the problem around and extracted $b_{ne}$ from recent precise measurements of $G_E(q^2)$, finding $b_{ne} = (-1.13 \pm 0.08) \times 10^{-3}$ fm, which again excludes the Dubna value.

### 3. Constraints on spin-dependent new forces

Of high interest in modern particle theories are the limits on spin-dependent short-range interactions mediated by light pseudoscalar bosons like the axion. As discussed in Sec. III.A.4.a, we need the axion to solve the strong $CP$ problem, and, in addition, the axion is a favorite dark-matter candidate. However, the axion is experimentally excluded over a wide range of masses and coupling strengths $g^2$. As a historical remark, neutron reactions were also helpful in the hunt for the axion. Döhner et al. (1988) did an early axion search, using the PERKEO neutron decay spectrometer, and set limits in the MeV mass range (and, using the auxiliary calibration spectrometer PERKINO, the then virulent 17 keV neutrino signal could be explained as an ordinary backscattering effect, Abele et al., 1993). Today, axions are excluded except for a narrow window between $10^{-5}$ eV and $10^{-2}$ eV, which corresponds to a range $\lambda$ between 20 μm and 2 cm (dashed vertical lines in Fig. 17), with unknown interaction strength $g^2$.

Constraints on the spin-dependent axion-nucleon interaction strength within this "axion window" were derived from neutron data, as shown in Fig. 17. The potential $V(r)$ for the interaction of one fermion with the spin of another fermion is attractive for one neutron spin direction, and repulsive for the other. Its effect on the UCN transmission curve, Fig. 14, would be as shown in Fig. 18, from Baeßler et al. (2007), see also Westphal et al. (2007a). This leads to the upper exclusion curve in Fig. 17.

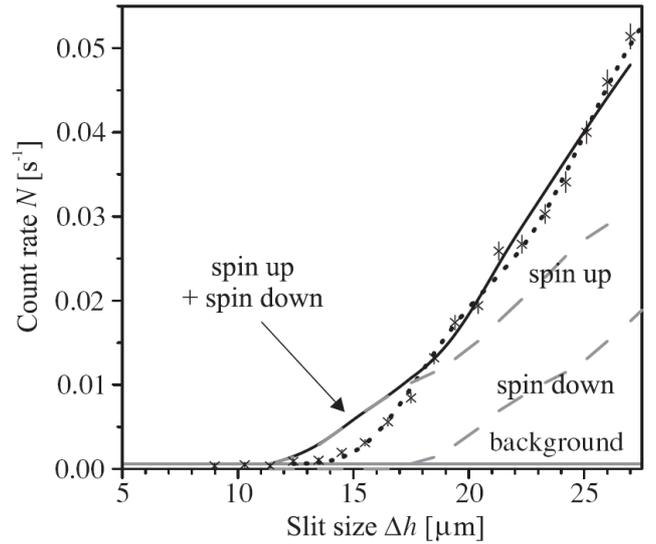

FIG. 18. Expected effect of a spin-dependent extra interaction on the UCN transmission curve from the gravitational level experiment. One neutron spin-component would be attracted, the other repelled at short distances (dashed lines). Even unpolarized neutrons are sensitivity to such spin-dependent interactions (solid line). From Baeßler (2009).

Recently, Zimmer (2008, 2010) proposed that UCN trapped in rather large neutron bottles could also be used to obtain tight constraints on short-range spin-dependent interactions of type $V(\mathbf{r}) = \boldsymbol{\sigma}_n \cdot \mathbf{B}_{eff}(\mathbf{r})$, with a suitable pseudo-magnetic field $\mathbf{B}_{eff}$. Every wall encounter of the UCN would slightly shift the spin phase of neutron polarization $\langle \boldsymbol{\sigma}_n \rangle$, and would also depolarize the UCN to some degree.

Indeed, Serebrov (2009) used the existing data on the depolarization of stored UCN to derive the constraint $|g_S g_P| \lambda^2 < 2 \times 10^{-21}$ cm$^2$ for $\lambda$ between $10^{-6}$ and $10^{-2}$ m, which covers the whole "axion window", see also Serebrov et al. (2010), and further $|g_S g_P| \lambda^2 < 4 \times 10^{-22}$ cm$^2$ for $\lambda$ between $10^{-6}$ m and $10^{-5}$ m, as is also shown in Fig. 17.

Voronin et al. (2009) took their existing crystal-EDM neutron data, as discussed in Sec. III.B.2.e, and looked for a possible additional phase-shift due to extra forces. This enabled them to set new constraints $|g_S g_P| < 10^{-12}$, valid down to the atomic range $\lambda \simeq 10^{-10}$ m = 1 Å, which then turns into $|g_S g_P| \lambda^2 < 5 \times 10^{-29}$ cm$^2$, valid down to $\lambda \sim 10^{-13}$ m. These very short ranges had already been excluded by the Supernova SN 1987A neutrino events, see Hagmann et al. (2010), but such an independent check is very useful.

Petukhov *et al.* (2010) used their $^3$He neutron polarizer cell to measure $T_1$ of $^3$He as a function of an external magnetic field. By comparison of their improved relaxation theory with these data (plus $T_2$ data for $^3$He from other sources) they obtained $|g_S g_P|\lambda^2 < 3 \times 10^{-23}$ cm$^2$, shown also in Fig. 17; see also Petukhov *et al.* (2011) and Fu *et al.* (2011) for recent reevaluations. References to the other curves in Fig. 17 numbered 1. and 2. are given in Baeßler *et al.* (2009).

Hence, in recent years, it was shown that the neutron is a unique tool to search for new forces at very short ranges. Such forces are predicted by powerful new theories that advance the unification of physics and that may shed light on unsolved problems like quantum gravity or the nature of dark matter in the universe.

**D. More lessons from the neutron mass**

**1. Gravitational *vs.* inertial mass**

The neutron has no charge and so its inertial mass cannot be measured the usual way in a $e/m$-sensitive mass spectrometer. One could measure the neutron mass in nuclear recoil experiments, but not to a high precision. One could also measure the ratio of the neutron magnetic moment to neutron mass $\mu_n/m_n$ in a Stern-Gerlach experiment. Indeed, the gravitational mass of the neutron was derived from the "sagging" of an UCN beam in a magnetic storage ring (Paul, 1990) to $m_{grav} = (914 \pm 34)$ MeV/c$^2$. This agrees within the 4% error margin with the precisely known inertial mass of the neutron $m_{inertial} = 939.565\ 346(23)$ MeV/c$^2$, as is expected from the weak equivalence principle, which requires $m_{grav} = m_{inertial}$.

A better test of the weak equivalence principle was done by Schmiedmayer (1989), based on the following argument. The most precise measurements of neutron scattering lengths in condensed matter are done with gravitational spectrometers, see the compilation by Koester *et al.* (1991). In these experiments, one drops neutrons from a given height onto a horizontal surface and determines the critical height, above which the neutrons are no longer totally reflected from the material. These measurements of neutron scattering lengths depend on the gravitational neutron mass, whereas the conventional measurements via neutron scattering depend on the neutron's inertial mass. A comparison of the scattering lengths measured with both methods gives the ratio

$$m_{grav}/m_{inertial} = 1.00011 \pm 1.00017. \quad (5.8)$$

For macroscopic pieces of matter, the weak equivalence principle is tested to a $10^{-13}$ level, see Adelberger *et al.* (2009). The neutron's $10^{-4}$ limit is derived from a free matter wave, i.e., a quantum object. It is very likely, but not guaranteed that bulk matter and matter waves obey the same law of gravitation. Furthermore, the neutron is a pure $t = \frac{1}{2}$ isospin object, and gravitation possibly depends on isospin.

**2. Neutron mass, and test of $E = mc^2$**

To precisely determine the neutron's inertial mass the neutron capture reaction $p + n \rightarrow d + \gamma$ with $E_\gamma = 2.22$ MeV is used. The masses of the proton and of the deuteron in atomic mass units u are known to $10^{-10}$ or better. To be competitive, $E_\gamma$ needs to be measured with $\sim 10^{-7}$ accuracy. A precision measurement of $E_\gamma$ was done at the GAMS facility of the ILL by a NIST-ILL collaboration. At GAMS the hydrogen target (a Kapton foil) was installed in-pile and looked at by nearly perfect, flat diffraction crystals installed at about 10 m distance outside the reactor vessel. From the result for $E_\gamma$ Kessler *et al.* (1999) obtained

$$m_n = 1.008\ 664\ 916\ 37(82)\ \text{u}, \quad (5.9)$$

with $8 \times 10^{-10}$ relative error.

Neutron capture $\gamma$-ray energies were used to test the Einstein energy-mass relation $E = mc^2$ to an unprecedented precision. For a neutron capture reaction like $^{28}$Si $+ n \rightarrow ^{29}$Si $+$ multiple $\gamma$ one can write

$$\Delta Mc^2 \equiv (M_{28} + m_d - m_p - M_{29})c^2 = \sum_i E_{\gamma i}. \quad (5.10)$$

The masses $M_{28}$ and $M_{29}$ had been measured before with two silicon atoms stored simultaneously in a single Penning trap, one of the isotope $^{28}$Si, the other of $^{29}$Si. All relevant deuteron and $^{29}$Si $\gamma$-transition energies $E_i$ entering the sum on the right-hand side of Eq. (5.10) (with appropriate signs) then were measured at GAMS. With these data, one arrives at a test of the most famous formula of physics to an accuracy of $4 \times 10^{-7}$, more than 50 times better than previous tests, see Rainville *et al.* (2005), and Jentschel *et al.* (2009).

## 3. The fine structure constant

The precise value of the neutron mass was also used to determine the electromagnetic coupling constant, i.e., the fine-structure constant $\alpha = e^2/4\pi\varepsilon_0 \hbar c \approx 1/137$, to a high precision. At first sight, this is surprising, as the neutron has no electric charge. Sometimes the value of a fundamental constant can be calculated from some other constants known with higher accuracy than its directly measured value.

The fine-structure constant can be written in terms of the Rydberg constant $R_\infty = \tfrac{1}{2}\alpha^2 m_e c/h$ as $\alpha^2 = (2R_\infty/c)(h/m_e)$, where $R_\infty$ is much better known than is $\alpha$. The quantity $h/m_e$ can be determined from the de Broglie relation $h/m_e = \lambda_e v_e$ by measuring both the electron's velocity $v_e$ and de Broglie wavelength $\lambda_e$. For a charged particle like the electron, however, this cannot be done with sufficient precision, but for a neutral particle like the neutron this is possible. We use $h/m_e = (h/m_n)(m_n/m_p)(m_p/m_e)$ and write the fine-structure constant as

$$\alpha^2 = \frac{2R_\infty}{c}\frac{m_n}{m_p}\frac{m_p}{m_e}v_n\lambda_n, \qquad (5.11)$$

where $R_\infty$ is known to $7\times 10^{-12}$, the mass ratios to better than $5\times 10^{-10}$, and the velocity of light $c$ is as defined in the SI system and has no error.

For the neutron, Krüger *et al.* (1999) measured $h/m_n = v_n \lambda_n$ at ILL to $7\times 10^{-8}$, and used the precisely known value $m_n/m_p$ from the preceding section. The neutron velocity $v_n$ was measured via the beat frequency of the neutron polarization signal after double passage through a magnetic undulator (in the form of a current carrying meander coil) with 10 m flight path between the two passages. The neutron wavelength $\lambda_n$ was measured by diffraction on a perfect silicon crystal in backscattering geometry, for which the Bragg condition is nearly independent of the angle of incidence. Today the result $\alpha^{-1} = 137.036\,0114(50)$, with better knowledge of the silicon lattice constants (see Mohr *et al.*, 2008), translates to

$$\alpha^{-1} = 137.036\,0077(28) \text{ (neutron)}. \qquad (5.12)$$

In the same way, one can derive $\alpha^{-1}$ from the velocities and matter-wavelengths of free atoms, with, at the present time, three times lower error than for the neutron. The current world average of $\alpha$ is dominated by the electron $g-2$ measurement leading to $\alpha^{-1} = 137.035\,999\,070(98)$ with $7\times 10^{-10}$ accuracy (Gabrielse *et al.*, 2007). This value is based on the validity of quantum electrodynamics, whereas the composed values from $h/m$ measurements, although less accurate, rely on different assumptions.

## 4. More experiments on free falling neutrons

Many other experiments on the neutron's mass and gravitational interaction were done over the years, some of which are listed as follows.

● McReynolds (1951) at Brookhaven National Laboratory (USA) first observed the free fall of the neutron.
● Collela *et al.* (1975) at the Ford Nuclear Reactor in Michigan (USA) first studied gravitation in the quantum regime by measuring the gravitationally induced phase shift of a neutron matter wave with a neutron interferometer.
● The effective mass of a particle in a periodic potential can have either sign, and vanishes at the inflection point of the dispersion curve $\omega(k)$. The same is true for a neutron moving through the periodic potential inside a crystal. When the neutron's effective mass becomes near zero it becomes more sensitive to external perturbations by up to six orders of magnitude, as was shown experimentally by Zeilinger *et al.* (1986). This effect had been applied in a search for a magnetic monopole moment of the neutron, as was discussed in Sec. III.D.2.
● In another elegant experiment, Frank *et al.* (2009) measured the inelastic "diffraction in time" effect induced by an optical phase grating moving transversally through a beam of free-falling monochromatic UCNs (Frank *et al.*, 2003). In the experiment, parameters were chosen such that the energy loss $\hbar\Omega$ of the UCN, induced by the phase grating rotating with frequency $\Omega = 2\pi v/a$, exactly compensated the gain $m_n gh$ in UCN kinetic energy after a fall through height $h$ (grating's velocity $v$, lattice constant $a = 5\,\mu m$). Two identical neutron Fabry-Perot monochromators at distance $h$ above each other monitored this compensation. The energy loss was varied between $\hbar\Omega = 10$ neV and 40 neV by varying the grating's velocity from $v = 15$ m/s to 60 m/s. The height of fall was adjusted accordingly in the range $h = 10$ cm to 40 cm. As a by-product, the experiment tested the weak equivalence principle to $2\times 10^{-3}$.

Nowadays, atomic physics experiments have become very competitive in testing the gravitational interaction, for references see Cronin *et al.* (2009). Experiments with atoms have a number of advantages, among them high intensity, ease of manipulation, and portability. The corresponding neutron experiments have the advantage that neutrons, in this context, are structureless and "inert", can easily penetrate matter (or are totally reflected from it), and have well calculable interactions.

## VI. THE ELECTROWEAK STANDARD MODEL AND THE $\beta$-DECAY OF FREE NEUTRONS

We now turn to the epoch at a scale of $\leq 1\,\text{TeV}$, see Table I, which is very well described by the present Standard Model of particle physics, based on the $U(1)_Y \times SU(2)_L \times SU(3)_C$ gauge group. This scale is much better understood than the earlier epochs of the universe discussed in the preceding chapters. To describe the $\beta$-decay of the free neutron

$$n \to p^+ + e^- + \bar{v}_e, \qquad (6.1)$$

with half-life of ~10 minutes, and $\beta$ endpoint energy of 782.3 keV, we shall need only a small number of parameters in the SM.

There are two main reasons for our interest in precise neutron decay data. First, most "semileptonic" weak interaction processes (i.e., involving both quarks and leptons) occurring in nature must be calculated today from measured neutron decay parameters. Examples range from proton or neutron weak cross sections needed in cosmology, astrophysics, and particle physics, over pion decay and muon capture, to such mundane problems as precise neutrino detector efficiencies.

Second, in neutron decay, many tests on new physics beyond the SM can be made, relevant again both in particle physics and in studies of the very early universe. According to conventional wisdom, the SM and its vector-axialvector ($V-A$) structure are extremely well tested. However, if one looks closer then one finds that there is still much room for couplings with other symmetries, like scalar $S$ or tensor $T$ couplings, or for additional right-handed $V+A$ couplings, with incomplete parity violation. Precision bounds on such amplitudes beyond the SM all come from low-energy physics, and progress in this field slowed down years ago when low energy physics became unfashionable.

Neutron tests beyond the SM are possible because, as we shall see, there are many more observables accessible in neutron decay (more than 20) than there are SM parameters (three), which makes the problem strongly over-determined. When we go beyond the SM there are up to 10 allowed complex coupling constants. These couplings are allowed in the sense that they obey basic symmetries like Lorentz invariance (symmetries that in the age of quantum gravity and dark energy are no longer sacrosanct), and that their occurrence is only a question of observational evidence.

In the language of particle physics, neutron decays are "rare events" with a low-energy signature in a noisy environment. Therefore the experimental study of neutron decay is challenging, and error margins are easily underestimated. The precision of neutron decay data has dramatically improved over past decades. However, there are still inconsistencies between various neutron decay data that need to be resolved, but at a much higher level of precision than in the past. There is a high demand for better neutron data for physics both within and beyond the SM, which is the reason why, in recent years, more and more neutron decay "hunters" have joined the party.

References to earlier reviews on experiments and on the theory of neutron decay are given in Sections B and C, respectively.

### A. Neutron decay theory

Before we turn to the description of the weak interaction, we recall the corresponding expressions for the electromagnetic interaction, in the limit of low energy.

#### 1. The example of electromagnetism

Particles interact with each other via their currents. The interaction $\mathbf{\mu} \cdot \mathbf{B}$ between a magnetic moment and a magnetic field, for instance, is the interaction between the molecular and macroscopic electromagnetic currents in the sources of $\mathbf{\mu}$ and $\mathbf{B}$. In Dirac's relativistic theory, the electromagnetic four-current is a vector current with components $j_\mu^{elmag} = e\psi'\gamma_\mu\psi$, with $\mu = 1,...,4$. Here, $\gamma_\mu$ is a four-vector with the $4 \times 4$ Dirac matrices as elements, and $\psi$ and $\psi'$ are four-component free-particle wave functions $\psi = u(p)\exp(-ip \cdot x)$ for the initial and final states of the relativistic charged fermion.

The electromagnetic interaction between two different point-like charged fermions (denoted by the subscripts 1 and 2) proceeds via exchange of virtual photons, which are massless spin-1 (or "vector") bosons. The transition matrix element, in first-order perturbation theory, then contains the scalar product of the two vector currents $eu_1'\gamma_\mu u_1$ and $eu_2'\gamma^\mu u_2$, and a propagator $-i/q^2$ for the massless photon,

$$\mathcal{M}^{elmag} = -ie^2(u_1'\gamma_\mu u_1)q^{-2}(u_2'\gamma^\mu u_2), \qquad (6.2)$$

with four-momentum transfer $q = p_2 - p_1$. One calls this a neutral-current interaction, because the photon exchanged in the process has no electric charge.

If an electron interacts with an extended object like a proton, then there is an additional induced-tensor interaction with the proton's anomalous magnetic moment $\kappa_p \mu_N = (\frac{1}{2} g_p - 1)\mu_N = 1.793 \mu_N$, given in units of the nuclear magneton $\mu_N = e\hbar/2m_p$ with g-factor $g_p = 5.586$. This interaction involves currents of the form $j_\mu^{anom} = -i\kappa_p \mu_N (p\sigma_{\mu\nu} q^\nu p)$, where $p$ stands for the proton wavefunction, and $\sigma_{\mu\nu} = \frac{1}{2} i(\gamma_\mu \gamma_\nu - \gamma_\nu \gamma_\mu)$.

## 2. Weak interactions of leptons and quarks

We first regard the weak interaction of structureless quarks $q$ and leptons $l$ before we turn to the weak interactions of composite neutrons and protons.

a. *Charged-current weak interactions*, on the other hand, like neutron decay in Eq. (6.1), are mediated by the exchange of massive vector bosons $W^\pm$. Their mass is $M_W \approx 80$ GeV, and they carry both electric and weak charges. In most nuclear and particle $\beta$-decays, the exchanged momenta are low, $(pc)^2 \ll (M_W c^2)^2$, and the corresponding propagator becomes $\propto 1/M_W^2$, see Eq. (1.2). As the range of the interaction is given by the Compton wavelength of the exchanged boson $\lambdabar = \hbar/M_W c$, this leads to a point-like interaction between two weak currents.

The weak current in the "$V-A$" electroweak SM is composed of a vector part $V_\mu = g_w \psi' \gamma_\mu \psi$, the same as for the electromagnetic interaction, and an additional axial-vector part $A_\mu = g_w \psi' \gamma_\mu \gamma_5 \psi$ of equal amplitude, but of opposite sign, so $j_\mu = V_\mu - A_\mu = g_w \psi' \gamma_\mu (1-\gamma_5)\psi$, with weak coupling constant $g_w$. The factor $(1-\gamma_5)$ (with $\gamma_5 \equiv i\gamma_1 \gamma_2 \gamma_3 \gamma_4$) projects out the left-handed part of the spinor $\psi$, so the weak current $j_\mu$ is completely left-handed. As vectors $V_\mu$ and axial-vectors $A_\mu$ have different signs under a parity transformation $P$, the parity-transformed current $Pj_\mu = -g_w \psi' \gamma_\mu (1+\gamma_5)\psi = -(V_\mu + A_\mu)$ is completely right-handed.

In nature so far, only left-handed weak currents have been observed, and therefore parity symmetry is said to be maximally violated in the weak interaction. This means that if one looked into a mirror, one would see processes that are not present in nature without the mirror. 50 years after its discovery, the origin of this symmetry violation is still an open question, as will be discussed in Sec. D.2.

For $\psi$ and $\psi'$ one can insert any of the six known quark fields $q = (u,d,s,c,b,t)$ or lepton fields $l = (e, \nu_e, \mu, \nu_\mu, \tau, \nu_\tau)$ (or the antiparticle fields $\bar{q}$, $\bar{l}$). If two lepton currents interact, then one calls this a purely leptonic interaction; if a lepton and a quark current interact, a semileptonic interaction; and if two quark currents interact, a purely hadronic interaction.

b. *For purely leptonic muon decay*, $\mu \to e + \bar{\nu}_e + \nu_\mu$, for instance, the transition matrix element is (we omit the "weak" superscript)

$$\mathcal{M}_{muon} = (G_F/\sqrt{2})[\nu_\mu \gamma_\mu (1-\gamma_5)\mu][e\gamma^\mu (1-\gamma_5)\nu_e], \tag{6.3}$$

where $e, \nu_e$, etc., are the single particle Dirac spinors for the electron, electron neutrino, etc., and summation over repeated indices $\mu = 1,...,4$ is implied (the upright $\mu$ in this formula stands for the muon).

The strength of the weak interaction is given by the Fermi constant $G_F/(\hbar c)^3 = 1.166\ 378\ 8(7) \times 10^{-5}$ GeV$^{-2}$, newly determined by Webber *et al.* (2011), which is related to the weak coupling constant $g_w$, Eq. (3.3), as

$$G_F/\sqrt{2} = \frac{1}{8} g_w^2 / (m_W c^2)^2. \tag{6.4}$$

c. *Semileptonic neutron decay* on the quark level reads $d \to u + e^- + \bar{\nu}_e$. The corresponding matrix element for the point-like quarks is

$$\mathcal{M}_{quark} = (G_F/\sqrt{2})[u\gamma_\mu (1-\gamma_5)d][e\gamma^\mu (1-\gamma_5)\nu_e]. \tag{6.5}$$

Here two complications arise. The first is due to "quark mixing". In the early 1960's, it was observed that, while purely leptonic muon decay proceeds with a strength given by $G_F^2$, semileptonic neutron or nuclear decays proceed only with $0.95 \times G_F^2$, while strangeness-changing decays of strange particles (like $\Sigma, \Lambda, \Xi$-baryons or $K$-mesons) proceed with $0.05 \times G_F^2$. Cabibbo then postulated that the down quark state $d'$ that participates in the weak interaction is not the ordinary mass eigenstate $d$ but has a small admixture of strange quark state $s$, and the strange quark has a small admixture of $d$, such that $d' = d\cos\theta_C + s\sin\theta_C$, $s' = -d\sin\theta_C + s\cos\theta_C$. The Cabibbo angle $\theta_C$ has $\cos^2\theta_C \approx 0.95$ and $\sin^2\theta_C \approx 0.05$, so the probabilities of strangeness conserving and strangeness changing transitions sum up to unity, and together have the same strength as muon decay ("universality of the weak interaction").

Hence, the hadronic weak interaction acts on quark states that are obtained by a rotation in flavor space (the properties down, strange, etc., being called the flavors of the quarks). Later on, this concept was extended to three families of particles (even before the detection of the third family in experiment), in order to take into account time-reversal $T$ violating amplitudes, via an additional complex phase factor $\exp(i\delta)$ in the CKM quark-mixing matrix

$$\begin{pmatrix} d' \\ s' \\ b' \end{pmatrix} = \begin{pmatrix} V_{ud} & V_{us} & V_{ub} \\ V_{cd} & V_{cs} & V_{sb} \\ V_{td} & V_{ts} & V_{tb} \end{pmatrix} \begin{pmatrix} d \\ s \\ b \end{pmatrix}. \qquad (6.6)$$

The CKM matrix $V$ depends on three angles and one complex phase $\delta$, which are free parameters of the SM, though there are also other ways of parameterizing the matrix (after the detection of neutrino flavor oscillations, a similar mixing matrix is postulated for weak interactions of the leptons). Rotations in complex spaces are described by unitary rotation matrices, hence the CKM matrix $V$ acting in a Hilbert space should obey $VV^{\dagger} = 1$, where $V^{\dagger}$ is the conjugate transpose of $V$. This means that quark mixing is a zero-sum game: every quark gives as much as it takes. More on this will be discussed in Sec. C.3.

**3. Weak interactions of nucleons**

Nucleon structure introduces a second complication: In semileptonic decays, here neutron decay, it is not free quark currents that interact with the leptonic current, but the currents of quarks bound within a neutron. As already mentioned, the neutron is a complicated object made up not only of three valence quarks but also of so-called sea quarks (virtual $q\bar{q}$ pairs of all flavors) and gluons, all coupled to each other by the exchange of strongly self-interacting gluons. This makes a neutron state difficult to calculate.

a. *Form factors*: To account for the internal structure of the nucleons one introduces empirical form factors $f_i(q^2)$ and $g_i(q^2)$, with $i = 1, 2, 3$, the requirement of Lorentz invariance limiting the number of form-factors to six. Furthermore, transition energies in neutron decay are so low compared to the nucleon mass that these form factors only need to be taken at zero momentum transfer $q^2 \to 0$. They therefore reduce to pure numbers $f_i(0)$ and $g_i(0)$, whose quotients are real in the limit of $T$-invariance. We adopt a notation that is consistent with Holstein, 1974.

b. *Forbidden induced terms*: The SM excludes "second class" currents, which involve the terms $+f_3(0)q_\mu / 2M$ and $-ig_2(0)\sigma_{\mu\nu}q^\nu\gamma_5 / 2M$ (with $2M = m_p + m_n$), because of their weird properties under combined charge conjugation and isospin transformations, though, of course, these parameters should be tested if possible. Second class currents are forbidden in the SM, but can be induced by isospin-violating effects due to the differences in mass and charge of the $u$ and $d$ quarks. In neutron decay, such isospin breaking effects are expected within the SM only on the $4\times10^{-5}$ scale, see Kaiser (2001), and their detection at the present level of accuracy is unlikely and would require exotic processes beyond the SM.

We can furthermore safely neglect the "induced pseudoscalar" term $-g_3(0)\gamma_5 q_\mu$ at the low energies of neutron decay. In the nuclear current we therefore only retain $f_1$, $f_2$, and $g_1$, which amounts to replacing $u\gamma_\mu(1-\gamma_5)d$ in Eq. (6.5) by $p(f_1(0)\gamma_\mu + g_1(0)\gamma_\mu\gamma_5 -if_2(0)\sigma_{\mu\nu}q^\nu / 2M)n$, where $p$ is the proton and $n$ is the neutron wave function.

c. *Conserved Vector Current*: We can simplify this further. The electromagnetic vector current of hadrons (i.e., their isospin current) is known to be conserved (i.e., it is divergence-free), which means that the electric charge of the proton, for instance, is simply the sum of electric charges of its constituents, and is not renormalized by "dressed nucleon" strong-interaction effects. The SM can unify the weak and the electromagnetic interactions only if the weak hadronic vector current has the same property, i.e., if the weak charges remain unaltered by the strong interaction. Therefore conservation of the electroweak vector current ("CVC") requires that the vector coupling be unaffected by the intrinsic environment of the nucleon, i.e., $f_1(0) = 1$.

d. *Weak magnetism*: With the same reasoning we can attack $f_2(0)$. Under CVC, not only the charges but also the higher multipoles of the electromagnetic and weak hadronic couplings should remain unaffected by dressed-nucleon effects. Therefore under CVC the hadronic vector current in an electromagnetic transition and the hadronic vector currents in weak $\beta^-$ or $\beta^+$ decays should form the $(+1, 0, -1)$-components of an isovector triplet, linked to each other by the Wigner-Eckart theorem. Hence, there should be a simple link between the ordinary magnetism of the nucleons and the "weak magnetism" term $f_2$ in $\beta$- decay. This hypothesis on the connection between weak and electromagnetic multipole interaction parameters, made in the early 1960's, was an important precursor of electroweak unification.

The relation between the weak magnetism amplitude $f_2(0)$ in neutron-to-proton $\beta$-decay and ordinary magnetism of protons and neutrons is derived from the general relation between the spin-½ matrix elements of a vector operator $V = (V^+, V^0, V^-)$, namely, $\langle +\frac{1}{2}|V^+|-\frac{1}{2}\rangle = \langle +\frac{1}{2}|V^0|+\frac{1}{2}\rangle - \langle -\frac{1}{2}|V^0|-\frac{1}{2}\rangle$. In our case of isospin-½ and the electroweak hadronic vector current $V_\mu$, this reads

$$\langle p|V_\mu^+|n\rangle = \langle p|V_\mu^z|p\rangle - \langle n|V_\mu^z|n\rangle . \qquad (6.7)$$

The anomalous magnetic moments of the proton and the neutron are described by the currents $-i\kappa_p(e/2m_p)(p\sigma_{\mu\nu}q^\nu p)$, $-i\kappa_n(e/2m_p)(n\sigma_{\mu\nu}q^\nu n)$, respectively, with nuclear magneton $e/2m_p$, and with the weak charge $g_w = e/\sin\theta_W$ absorbed in the Fermi coupling constant. The "weak magnetism" term $-i(f_2(0)/2M)(p\sigma_{\mu\nu}q^\nu n)$ is given by the difference between proton and neutron anomalous magnetic moments, $f_2(0) = \kappa_p - \kappa_n = 3.706$, with $\kappa_p = \frac{1}{2}g_p - 1 = +1.793$ and $\kappa_n = \frac{1}{2}g_n = -1.913$.

e. *Neutron decay matrix element*: Hence, in $V - A$ electroweak SM the $\beta$-decay matrix element is

$$\mathcal{M}_{neutron} = \frac{G_F}{\sqrt{2}} V_{ud} [p(\gamma_\mu(1+\lambda\gamma_5) + \frac{\kappa_p - \kappa_n}{2M}\sigma_{\mu\nu}q^\nu)n] \times [e\gamma_\mu(1-\gamma_5)\nu_e] . \qquad (6.8)$$

In the end, the complicated interior of the neutron is taken care of by one free parameter

$$\lambda = g_1(0) = |g_1(0)|e^{i\varphi} , \qquad (6.9)$$

which becomes $\lambda = -|g_1(0)|$ in the case of time reversal invariance. Indeed, from searches for $T$ violation in neutron decay we know that $\varphi \approx \pi$, within error, see Sec. B.2.f below.

Conventionally one writes $\lambda = g_A/g_V$, with $g_A/g_V = g_1(0)/f_1(0)$, though one should not confuse these zero-momentum form factors $g_A$ and $g_V$ with the elementary coupling constants $g_V \approx 0$ and $g_A = -1/2$ of the electroweak neutral leptonic current in the SM. As stated above, the Fermi constant $G_F$ is known very precisely from the measured muon lifetime. Therefore when we neglect $T$ violation and take $G_F$ from muon decay, the matrix element Eq. (6.8) has only two free parameters: the first element $V_{ud}$ of the CKM matrix and the ratio $\lambda$ of axial-vector to vector amplitudes.

f. *Partially Conserved Axialvector Current*: We saw that the electroweak hadronic vector current is conserved (CVC), which led to $f_1(0) = 1$. For the axial-vector coupling, things are different. The pion is a pseudoscalar particle that can weakly decay into two leptons that are vector particles, hence the hadronic axial (= pseudovector) current cannot be conserved under the weak interaction. Without a conservation law for the axial current one would expect that strong-interaction radiative effects would make the electroweak axial coupling of hadrons comparable in size to the strong interaction, i.e., very large as compared to the weak coupling.

However, quantum chromodynamics, the theory of the strong interaction of the SM, is based on a symmetry between left-handed and right-handed hadronic currents called chiral symmetry. This symmetry would also require the conservation of the axial vector current, i.e. $|g_1(0)| = 1$, and $\lambda = -1$. Though, at low energies, chiral symmetry is spontaneously broken, i.e., the symmetry of the dynamic system is broken, while the symmetry of the underlying effective Lagrangian remains intact. Therefore the axial current is expected to be at least partially conserved (PCAC), specifically in the limit of vanishing pion mass (connected to vanishing quark masses, which break the symmetry explicitly), with $|g_1(0)|$ near 1, that is, $\lambda = g_1(0)/f_1(0)$ near –1.

Historically, the value of $\lambda$ measured in neutron decay, now at $\lambda \approx -1.27$, gave a first hint to chiral symmetry and its spontaneous breaking. As, within PCAC, $\lambda$ is linked to the axial coupling of the pseudoscalar pion, one can make a connection between $\lambda$ and the measured strong and weak interaction parameters of the pion, called the Goldberger-Treiman relation. In this way, the untractable parts in the calculation of $\lambda$ are shifted into measured parameters, though this procedure has an accuracy of only several percent. The deviation of $\lambda$ from –1, due to the strong interaction, can also be derived from QCD calculations on the lattice, with the result $\lambda = -1.26 \pm 0.11$, for references see Abele (2008).

g. *Neutron-nuclear weak interactions*: The weak interaction between neutrons and nuclei is of high interest, exploring otherwise inaccessible features of the strong interaction, though this topic is beyond the scope of this review. Experiments on slow-neutron transmission through bulk matter, for instance, produce the most blatant manifestations of parity violation. In one version of these beautiful experiments, unpolarized neutrons, upon transmission through an "isotropic" crystal (nonmagnetic, cubic, etc.), become longitudinally polarized. Even nonscientists feel that something is wrong. The solution is that all bulk matter is left-handed with respect to its weak interaction with the neutron. See Snow *et al.* (2011) for an experiment on *P* violating spin rotation in $^4$He, and Gericke *et al.* (2011) for an experiment on *P* violation in the $n + p \to d + \gamma$ reaction.

There exist several reviews on the theoretical implications of these measurements, in particular, in the context of effective field theory, see Holstein (2009b), Ramsey-Musolf and Page (2006), and references therein. Both experiment and theory on neutron-nuclear weak interactions are very challenging, in particular for the more interesting low-mass nuclei.

## 4. The 20 and more observables accessible from neutron decay

In free neutron $\beta^-$-decay, there are astonishingly many observables accessible to experiment. Precise values for all these observables are needed in order to obtain all the coupling constants $g_i$ and $g'_i$, with $i = S, V, T, A, P$ ($S$ for scalar, $T$ for tensor, and $P$ for pseudoscalar interaction, see Sec. D.1). In the SM with purely left-handed currents, one has $g_i = g'_i$ for $i = V, A$, and $g_i = g'_i = 0$ for $i = S, T$, and $g_P$ and $g'_P$ are negligible.

With the matrix element of Eq. (6.8) one arrives at a neutron decay rate

$$\tau_n^{-1} = \frac{c}{2\pi^3} \frac{(m_e c^2)^5}{(\hbar c)^7} G_F^2 |V_{ud}|^2 (1 + 3\lambda^2) f, \qquad (6.10)$$

with the phase space factor $f = 1.6887(2)$ (from Towner and Hardy, 2010), which is the integral over the $\beta$-spectrum (for the precise shape of the electron and proton spectra see, for instance, Glück, 1993). After corrections for radiative effects and weak magnetism, the lifetime becomes (Marciano and Sirlin, 2006)

$$\tau_n = \frac{(4908.7 \pm 1.9)\text{ s}}{|V_{ud}|^2 (1 + 3\lambda^2)}, \qquad (6.11)$$

where the error in the numerator reflects the uncertainty of electroweak radiative corrections due to hadronic loop effects.

The ratio $\lambda = g_A / g_V$, on the other hand, can be obtained from the measurements of one of the many neutron-decay correlation coefficients discussed next. In the SM, these coefficients are insensitive to Coulomb corrections of order $\alpha$.

a. *Electron-antineutrino coefficient a*: The $e$-$\bar{\nu}_e$ correlation parameter $a$ describes the angular correlation between electron and antineutrino three-momenta $\mathbf{p}_e$ and $\mathbf{p}_\nu$. This correlation is parity $P$ conserving and, for a given $\lambda$, is determined by momentum conservation of the three spin-½ particles emitted in unpolarized neutron decay. It leads to an angular distribution of $\bar{\nu}_e$ emission with respect to the direction of electron emission

$$d^2\Gamma \propto (1 + a \frac{c\mathbf{p}_e \cdot c\mathbf{p}_\nu}{W_e W_\nu}) \, d\Omega_e = (1 + a \frac{v_e}{c} \cos\theta) \, d\Omega_e, \qquad (6.12)$$

with total electron energy $W_e = (p_e^2 c^2 + m_e^2 c^4)^{1/2} = \gamma m_e c^2$, electron velocity $v_e / c = c p_e / W_e$, neutrino energy $W_\nu = p_\nu c$, and the angle $\theta$ of electron emission with respect to the direction of neutrino emission. In the $V - A$ model, the coefficient $a$ depends on $\lambda$ as

$$a = \frac{1 - \lambda^2}{1 + 3\lambda^2}. \qquad (6.13)$$

A simple scheme for deriving the dependence of this and other correlation coefficients on $\lambda$ was given in Dubbers (1991a). As $a$ measures the deviation of $\lambda^2$ from 1, it is highly sensitive to the violation of axial vector current conservation (PCAC), with $\partial_\lambda a / a = -2.8$ at $a \approx -0.10$ (for $\lambda \approx -1.27$, where $\partial_\lambda$ stands for $\partial / \partial \lambda$).

The coefficient $a$ is $P$ conserving because it involves the product of two polar vectors, and was already searched for before the discovery of parity violation. The neutrino momentum cannot be measured in experiment, but must be reconstructed as $\mathbf{p}_\nu = -(\mathbf{p}_e + \mathbf{p}_p)$ from the measured electron and proton momenta. Though, up to now, coefficient $a$ has always been determined from the shape of the proton spectrum, see Sec. B.2.a, with the electron remaining undetected. The decay protons have an endpoint energy of 750 eV and for detection must be accelerated to typically 30 keV.

b. *$\beta$-decay asymmetry A*: The $\beta$-asymmetry parameter $A$ is the correlation coefficient between neutron spin $\mathbf{\sigma}_n$ and electron momentum $\mathbf{p}_e$. This asymmetry is due to the parity violating helicity of the emitted electrons (of helicity $-v_e / c$, with spins pointing preferentially against the direction of flight), in conjunction with angular momentum conservation. The correlation leads to an angular distribution of the $e^-$ emitted in the decay of spin-polarized neutrons

$$d^2\Gamma \propto (1 + A\langle\sigma_n\rangle \cdot \frac{c\mathbf{p}_e}{W_e})\, d\Omega_e = (1 + AP_n \frac{v_e}{c}\cos\theta)\, d\Omega_e$$
(6.14)

where $\theta$ is the angle of electron emission with respect to the direction of neutron polarization $\mathbf{P}_n = \langle\sigma_n\rangle$.

The $\beta$-asymmetry is $P$ violating because it contains the scalar product of a vector $\mathbf{p}_e$ and an axialvector $\sigma_n$. Within the $V-A$ model it depends on $\lambda$ as

$$A = -2\frac{\lambda(\lambda+1)}{1+3\lambda^2}.$$
(6.15)

Like coefficient $a$, the $\beta$-asymmetry $A$ directly measures the deviation of $\lambda$ from $-1$ and is very sensitive to it, with $\partial_\lambda A / A = -3.2$ at $A \approx -0.12$.

c. *Antineutrino asymmetry B*: The $\bar{\nu}_e$-asymmetry parameter $B$ is the correlation coefficient between neutron spin $\sigma_n$ and antineutrino momentum $\mathbf{p}_\nu$. Nonzero $B$ leads to a $P$ violating angular distribution of the $\bar{\nu}_e$ (of helicity one, with spins all pointing into the direction of flight) emitted from polarized neutrons

$$d^2\Gamma \propto (1 + B\langle\sigma_n\rangle \cdot \frac{c\mathbf{p}_\nu}{W_\nu})\, d\Omega_e = (1 + BP_n \cos\theta)\, d\Omega_e$$
(6.16)

with neutrino energy $W_\nu$, and

$$B = 2\frac{\lambda(\lambda-1)}{1+3\lambda^2}.$$
(6.17)

With $\partial_\lambda B / B = 0.077$ at $B \approx 1.0$, the parameter $B$ is about 40 times less sensitive to variations of $\lambda$ than are the parameters $A$ and $a$. This makes $B$ valuable for searches of decay amplitudes beyond the SM, as we shall see in Sec. D.

Correlation $B$ is related to correlation $A$ by a rule given by Weinberg (1959): If in $V-A$ leptonic decays one interchanges the roles of electrons and neutrinos, for instance by measuring the neutrino asymmetry instead of the $\beta$-asymmetry, then even powers of $\lambda$ in the numerator of the coefficient change sign, while the odd interference terms do not.

d. *Proton asymmetry C*: The proton asymmetry parameter $C$ is the correlation coefficient between neutron spin $\sigma_n$ and proton momentum $\mathbf{p}_p$. It leads to a $P$ violating angular distribution of the protons emitted from polarized neutrons $W(\theta) = 1 + 2CP_n \cos\theta$, with

$$C = x_C \frac{4\lambda}{1+3\lambda^2} = -x_C(A+B).$$
(6.18)

The kinematic factor $x_C$ depends on the total energy release $W_0$, and for the neutron is $x_C = 0.27484$, see Glück (1996), and references therein. The sensitivity to $\lambda$ is $\partial_\lambda C / C = 0.52$ at $C \approx 0.24$.

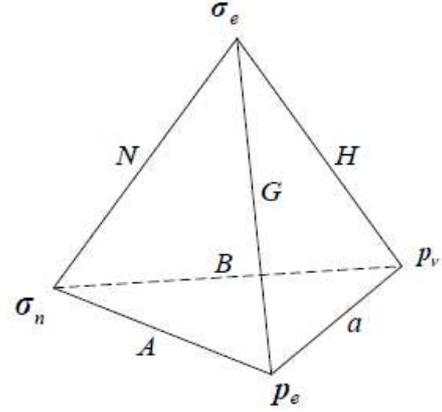

FIG. 19. Two-fold correlations between momenta $\mathbf{p}_e$, $\mathbf{p}_\nu$ and spins $\sigma_n$, $\sigma_e$ in the $\beta$-decay of slow neutrons, with the correlation coefficients: $\beta$-asymmetry $A$, neutrino-asymmetry $B$, electron-antineutrino correlation $a$, electron helicity $G$, and spin-spin and spin-momentum coefficients $N$ and $H$.

Under $V-A$, all correlation coefficients depend only on the one parameter $\lambda$, and therefore any correlation coefficient can be expressed by any other correlation coefficient, and by various combinations of such coefficients, for instance $a = 1 - B + A$, etc. Among such relations (which usually must be corrected for weak magnetism), Eq. (6.18) has the special feature of being model independent, holding as well under non-zero scalar, tensor, or right-handed amplitudes.

e. *Twofold correlations involving electron spin*: $\beta$-decay of slow neutrons involves four vector quantities accessible to experimental investigation: the momenta $\mathbf{p}_e$ and $\mathbf{p}_p$ of the electron and the proton, and the spins $\sigma_n$ and $\sigma_e$ of the neutron and the electron. In principle, proton spin $\sigma_p$ is also accessible, but in neutron decay, proton helicity $\langle\sigma_p\rangle$ is extremely small and difficult to measure.

Angular distributions and correlations are (pseudo-)scalars, and for a $\beta$-transition from an initial $j = \frac{1}{2}$ to a final $j' = \frac{1}{2}$, we can construct 17 different scalars or pseudoscalars from these four vectors. These correlation coefficients use up most of the letters of the alphabet, see Jackson *et al.* (1957a, 1957b), and Ebel and Feldman (1957).

The simplest such correlations are given by the scalar products of any two of these four vectors, which gives the six correlation coefficients shown in Fig. 19. They are all conserved under time-reversal $T$ operation ($\mathbf{p} \to -\mathbf{p}$, and simultaneously $\boldsymbol{\sigma} \to -\boldsymbol{\sigma}$). Scalar products involving two vectors (momenta) or two axial vectors (spins), i.e., $a$ and $N$, conserve parity $P$ ($\mathbf{p} \to -\mathbf{p}$, $\boldsymbol{\sigma} \to \boldsymbol{\sigma}$), while the other four, $A$, $B$, $G$, and $H$, are $P$ violating. In the SM with its maximally parity violating $V-A$ structure, one finds the electron helicity coefficient $G = -1$, and (when neglecting weak magnetism)

$$H = -\frac{m_e}{W_e} a, \text{ and } N = -\frac{m_e}{W_e} A \quad (6.19)$$

for the other bilinear correlations involving the electron spin. Beyond the SM, these relations no longer hold, as we shall see in Sec. D.

f. *Threefold correlations*: From any three different vectors out of $\mathbf{p}_e$, $\mathbf{p}_p$, $\boldsymbol{\sigma}_n$, and $\boldsymbol{\sigma}_e$ one can form four triple products, all $T$ violating:

$$D\boldsymbol{\sigma}_n \cdot (\mathbf{p}_e \times \mathbf{p}_\nu), \quad L\boldsymbol{\sigma}_e \cdot (\mathbf{p}_e \times \mathbf{p}_\nu),$$
$$R\boldsymbol{\sigma}_e \cdot (\boldsymbol{\sigma}_n \times \mathbf{p}_e), \quad V\boldsymbol{\sigma}_n \cdot (\boldsymbol{\sigma}_e \times \mathbf{p}_\nu). \quad (6.20)$$

Triple products involving two momenta are $P$ conserving, and those involving two spins are $P$ violating. Triple correlations involving neutrino momentum are measured via the proton momentum as $\boldsymbol{\sigma}_n \cdot (\mathbf{p}_e \times \mathbf{p}_p) = -\boldsymbol{\sigma}_n \cdot (\mathbf{p}_e \times \mathbf{p}_\nu)$, etc.

In the $V-A$ SM, such $T$ violating amplitudes should be immeasurably small, with

$$D = -2\frac{|\lambda|\sin\varphi}{1+3\lambda^2}. \quad (6.21)$$

Should one ever find a $D$-coefficient with nonzero $\varphi$, then one may try to corroborate it by measuring the electron energy dependence of the $V$-coefficient, which under $V-A$ is $V = -(m_e c^2/W_e)D$.

In the SM, $L = (\alpha m_e / p_e)a$ and $R = -(\alpha m_e / p_e)A$ are non-zero only from order $\alpha$ Coulomb corrections (plus imaginary parts beyond the SM that depend on scalar and tensor amplitudes only). These triple products can be used for searches for new physics beyond the SM. One must keep in mind that triple correlations can also be induced by final state effects, which, however, are far below (two and more orders of magnitude) the present experimental limits for the correlation coefficients, and can well be calculated.

g. *Four and fivefold correlations*: Next, we look for products involving four vectors. When we multiply two scalar products we obtain five more independent scalars and their respective coupling constants,

$$K(\boldsymbol{\sigma}_e \cdot \mathbf{p}_e)(\mathbf{p}_e \cdot \mathbf{p}_\nu), \quad Q(\boldsymbol{\sigma}_e \cdot \mathbf{p}_e)(\boldsymbol{\sigma}_n \cdot \mathbf{p}_e),$$
$$S(\boldsymbol{\sigma}_e \cdot \boldsymbol{\sigma}_n)(\mathbf{p}_e \cdot \mathbf{p}_\nu), \quad T(\boldsymbol{\sigma}_e \cdot \mathbf{p}_e)(\boldsymbol{\sigma}_n \cdot \mathbf{p}_\nu),$$
$$U(\boldsymbol{\sigma}_e \cdot \mathbf{p}_\nu)(\boldsymbol{\sigma}_n \cdot \mathbf{p}_e) \quad (6.22)$$

In the SM, $K = -A$, $Q = -A$, $S = 0$, $T = -B$, and $U = 0$ (the latter when we neglect a $T$ violating Coulomb correction).

Scalar products of two vector products add nothing new, as they can be decomposed into products of scalar products. Finally, we can construct a $T$ violating five-fold product $W(\boldsymbol{\sigma}_e \cdot \mathbf{p}_e)\boldsymbol{\sigma}_n \cdot (\mathbf{p}_e \times \mathbf{p}_\nu),$, with $W = -D$, corresponding to the $D$-coefficient with electron helicity analysis. This, however, will not play a role soon. Hence, indeed, within the SM, the problem is heavily over-determined, and we shall see in Sec. D how to profit from this.

### 5. Rare allowed neutron decays

There are two further allowed but rare neutron decay channels, radiative $\beta$-decay under emission of an inner bremsstrahlung photon, and bound $\beta$-decay of a neutron into a hydrogen atom.

a. *Radiative neutron decay*: The rare decay mode

$$n \to p^+ + e^- + \bar{\nu}_e + \gamma \quad (6.23)$$

has a bremsstrahlung photon in the final state. This process is overwhelmingly due to bremsstrahlung of the emitted electron, which is a well-understood process (Glück, 2002), although in effective field theory tiny deviations from the SM photon energy spectrum are possible (Bernard *et al.*, 2004).

To first order, the photon spectrum varies as $1/\omega$ and ends at the $\beta$-endpoint energy $\hbar\omega_{max} = E_0 = 782$ keV. Therefore every "octave" in the spectrum contains the same number of photons. Angular momentum conservation transforms part of the electron helicity into photon polarization. This parity violating effect could be measurable in the visible part of the spectrum, but is diluted by having four particles in the final state. For references, see Cooper *et al.* (2010).

b. *In bound $\beta$-decay* of a neutron, the emitted electron ends up in a bound state of hydrogen,

$$n \to H + \bar{\nu}_e, \qquad (6.24)$$

which must be an atomic S-state for lack of orbital angular momentum. Due to the small hydrogen binding energy of 13.6 eV, as compared to the kinetic energy release $E_0 = 782$ keV, this process has a tiny branching ratio of $4 \times 10^{-6}$, of which 10% go to the excited $n = 2$ metastable S-state of hydrogen, and another 6% to higher S-states with $n > 2$, where $n$ is the main quantum number.

The populations of the four $n = 2$ hyperfine states with total angular momentum $F = 1$, $M_F = -1, 0, +1$, and $F = 0$, $M_F = 0$ can in the $V - A$ SM be written as

$$W_1 = \tfrac{1}{2}a + 2B, \ W_2 = 2(B - A),$$
$$W_3 = 1 - W_1 - W_2, \ W_4 = 0 \qquad (6.25)$$

with the $\lambda$-dependent correlation coefficients $a$, $A$, $B$ from above, for details and references see Dollinger *et al.* (2006), and Faber *et al.* (2009). Of course, one will not measure $\lambda$ "the hard way" via bound $\beta$-decay, but one can search for interesting deviations from $V - A$ theory, as we shall see in Sec. D. This gives us, together with the branching ratio for bound beta decay, four more observables in neutron decay.

To the list of 17 correlation coefficients discussed above we add the neutron lifetime, the four quantities in bound $\beta$-decay, the radiative decay observables, plus several small but interesting terms that enter the spectral shapes (weak magnetism $f_2$, the SM-forbidden second class amplitudes $g_2$, $f_3$, and the Fierz interference terms $b$ and $b'$ discussed in Sec. D). Hence, altogether, the number of observables accessible in neutron decay is well above 20.

**B. Experiments in neutron decay**

Experimental results exist on ten of the 20 or more neutron decay parameters listed above: namely, $\tau_n$, $a$, $b$, $b'$, $A$, $B$, $C$, $N$, $D$, and $R$. For the Fierz amplitudes $b$ and $b'$ (vanishing in the SM), and for the $T$ violating triple-correlation coefficients $D$ and $R$, only upper limits have been derived. In this chapter we shall discuss the present experimental status for each of these parameters, describe the experiments that contribute significantly to the PDG 2010 world average, and add new experimental results to this average. The evaluation of these parameters within and beyond the SM will be the subject of the subsequent Sections C and D.

For reviews focussing on neutron decay experiments, see Nico (2009), Paul (2009), Byrne (1995), Yerozolimsky (1994), Schreckenbach and Mampe (1992), Dubbers (1991b), Byrne (1982), Robson (1983) for an historical review, and Wietfeldt and Greene (2011) for a discussion of the status of the neutron lifetime. For drawings of instruments used in these experiments, see the reviews by Abele (2008), Nico (2009), and Wietfeldt and Greene (2011).

**1. The neutron lifetime**

Over past decades, published neutron lifetimes have kept decreasing from $\tau_n = (1100 \pm 160)$ s at the end of the 1950's to the PDG 2010 average of $\tau_n = (885.7 \pm 0.8)$ s. Published error bars have decreased by a factor 200 over this period. At all times, however, adopted lifetime values were larger than the present PDG average by about three standard deviations. To underestimate error bars is self-destructive, because, instead of applauding the progress made, the physics community will regret the continuing lack of consistency. It now seems, though, that neutron decay data converge to consistent values, though again well below the PDG 2010 averages.

The neutron lifetime $\tau_n$ can be measured by two principally different methods: with cold neutrons "in-beam", or with UCN "in-trap".

a. *The "in-beam" method* uses electrons and/or protons emitted from a certain neutron beam volume filled with an average number $N_n$ of neutrons. The charged decay particles are counted in detectors installed near the *n*-beam at a rate of

$$n_e = n_p = N_n / \tau_n. \qquad (6.26)$$

To derive $\tau_n$ one compares the rate $n_e$ (or $n_p$) to the rate $n_n$ in a neutron detector at the end of the beam. The neutron detection efficiency depends on the neutron cross section $\sigma_n$ and the effective thickness of the detector material.

Over the years, the in-beam method has seen many improvements:

● The use of thin neutron detectors with $\sigma_n \propto 1/\upsilon_n$ compensates for variations in neutron flight time $T \propto 1/\upsilon_n$ through the decay volume.
● Magnetic guidance of the decay electrons to the detectors permits effective $4\pi$ detection of the charged decay products (Christensen *et al.*, 1972).
● The use of neutron guides permits measurements far away from the neutron source in a low-background environment (Byrne *et al.*, 1980; Last *et al.*; 1988; Nico *et al.*, 2005).

- In-beam trapping of decay protons and their sudden release onto a detector leads to considerable background reduction (Byrne *et al.*, 1980; Nico *et al.*, 2005).
- Use of a proton trap of variable length eliminates edge effects (Byrne *et al.*, 1990; Nico *et al.*, 2005).

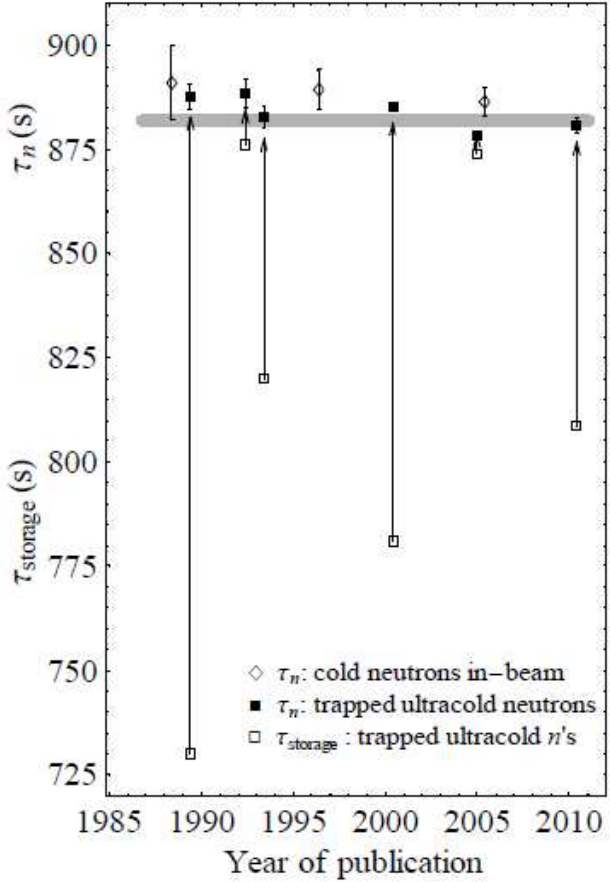

FIG. 20. The mean neutron lifetime, horizontal gray bar, as derived from cold neutron decay "in-beam" (◊), and from the decay of ultracold neutrons trapped in "bottles" (■). The width of the gray bar gives the upscaled error. For the UCN measurements, the vertical arrows show the extrapolation from the measured storage times (□) to the derived neutron life times (■), see Eq. (6.28). For details and references, see text.

The in-beam method involves absolute neutron and electron or proton counting, and for the NIST group now involved in these experiments, development of absolute neutron calibration is one of its professional duties.

A word on neutron guides, used nowadays for all such "in-beam" measurements. Neutron guides typically are rectangular glass tubes of cross section $\sim 6 \times 12$ cm$^2$ coated inside with a thin layer of totally reflecting material. They permit low-loss transport of cold or thermal neutrons over up to ~100 m distance. For neutrons with velocities above the critical velocity, $v > v_c$, see Sec. III.B.2, the critical angle of total neutron reflection is $\theta_c \approx \sin\theta_c = v_c/v$. Table IV lists $\theta_c$ for cold neutrons of effective temperature $T_n \approx 30$ K at their most probable velocity $v_0 \approx 700$ m/s. Thermal neutrons with $v_0 = 2200$ m/s have three times lower $\theta_c$.

So-called neutron supermirrors are coated with about 100 to several 1000 double layers of Ni/Ti of continuously varying thickness and have typically two to three times the critical angle of Ni coated mirrors ('m = 2' or 'm = 3'). The first neutron guide made entirely from such supermirrors was installed at ILL's fundamental physics facility as a so-called ballistic guide, whose cross section varies over its length such as to minimize losses, see Häse *et al.* (2002) for its design, and Abele *et al.* (2006) for its performance.

Figure 20 displays all neutron lifetimes that enter the PDG 2010 average, plus two other measurements discussed below. The open diamonds (◊) show the in-beam neutron lifetimes from Spivak (1988), Byrne *et al.* (1996), and Nico *et al.* (2005), see also Dewey *et al.* (2009) for future plans. The gray horizontal bar is the average $\tau_n = (881.9 \pm 1.3)$ s of these measurements. The width of the gray bar gives this error, enlarged by a scale factor $S = [\chi^2/(n-1)]^{1/2} = 2.5$, as discussed below.

b. *The UCN "in-trap" method* for neutron lifetime measurement uses stored ultracold neutrons (full squares (■) in Fig. 20). 25 years ago, sufficiently strong UCN sources were installed at the ILL (Steyerl *et al.*, 1986) and at Gatchina (Altarev *et al.*, 1986) to be used for neutron lifetime measurements with trapped UCN. In these experiments, one counts the number of UCN that survived in the neutron bottle over successive storage periods of variable lengths $T$, but constant initial neutron number $N_n(0)$,

$$N_n(T) = N_n(0)\exp(-T/\tau_{storage}), \qquad (6.27)$$

with the UCN disappearance rate

$$\frac{1}{\tau_{storage}} = \frac{1}{\tau_n} + \frac{1}{\tau_{loss}}. \qquad (6.28)$$

After each such measurement, the neutron trap is emptied and refilled with UCN. The UCN loss rate $1/\tau_{loss}$ is due to neutron interactions with the walls of the trap, whereas collisions with atoms of the rest gas usually can be neglected.

With the trapped-UCN lifetime method, no absolute particle counting is needed, because the mean residence time $\tau_{storage}$ can be obtained from a fit to the exponential neutron decay law, with $N_n(T)$ measured successively for several different storage times $T$. The important task is to eliminate $\tau_{loss}$ from Eq. (6.28). Different techniques have been developed for this purpose, as discussed below, with the results shown in Fig. 20.

In principle, the rates $n_e$, $n_p$, $n_n$ of decay electrons, protons, and of lost UCN can also be measured with suitable detectors during the whole storage interval, all obeying the same decay law (ideally, i.e., when there are no changes in the UCN spectral shape)

$$\frac{n_e(t)}{n_e(0)} = \frac{n_p(t)}{n_p(0)} = \frac{n_n(t)}{n_n(0)} = \exp(-t/\tau_{storage}). \quad (6.29)$$

Experience shows that experimenters should aim at constructing their apparatus such that large corrections to the data are avoided, because data subject to large corrections are more prone to hidden errors. Certainly, large corrections can be precise, and small corrections are no guarantee that the data are free of error. With this in mind, we shall give the size and type of the largest correction applied for each of the UCN lifetime experiments.

● The first lifetime experiment with trapped UCN by Mampe *et al*. (1989) at ILL's UCN turbine source used a rectangular glass vessel, covered inside with a special UCN-compatible oil ("Fomblin"), as a trap, see also Bates (1983). The trap had a movable back wall, such that it was possible to vary the surface-to-volume ratio of the trap, and with it, the loss rate. In UCN storage experiments, the velocity spectrum of the stored UCN changes with time, mainly because fast UCN have more wall encounters and hence higher losses than have slow UCN. A judicious choice of storage time intervals ensured that the total number of wall encounters during measurement was the same for all sizes of the UCN trap used. In this case, UCN spectral effects should cancel, except for a small correction for a vertical UCN density profile due to gravity, and for other effects unknown at the time that we shall discuss later on. The longest storage time $\tau_{storage} = 730$ s was reached for the maximum bottle volume of 140 $\ell$, see the corresponding open square in Fig. 20. To arrive at the neutron lifetime $\tau_n = (887.6 \pm 3.0)$ s (filled square), the data were extrapolated by $\Delta t \approx 150$ s (vertical arrow) to infinite trap volume. A problem common to all such UCN storage experiments is the removal of UCN with elevated energy, "marginally trapped" on quasistable closed trajectories. This was achieved by making the movable back wall of the trap from glass with a corrugated surface that made the UCN follow chaotic trajectories.

● The UCN storage experiment by Nesvizhevsky *et al*. (1992) used a neutron container with an open roof, in which the UCN were confined both by the walls and by gravity. The rate of wall encounters in the trap, and with it the UCN loss rate, was varied in two different ways. First, the UCN energy spectrum, and with it the wall encounter rate, was varied by gravitational selection. To this end, the trap was tilted under a certain angle, and the UCN with the highest energies (dependent on tilt angle) were poured out through the upper opening of the trap. Second, two different bottle volumes of about 60 $\ell$ and up to 240 $\ell$ were used, with largely different wall encounter rates. The UCN bottles were kept at temperatures of around 15 K and were coated inside with a thin layer of solid oxygen. The longest storage time reached was $\tau_{storage} = 876$ s, and to obtain the neutron lifetime of $(888.4 \pm 3.3)$ s an extrapolation over a mere $\Delta t = 12$ s was needed.

● The experiment by Arzumanov *et al*. (2000) at ILL used a trap of two concentric cylinders of slightly different diameters. UCN were stored either in the large inner volume with small surface-to-volume ratio and low losses, or in the small outer volume between the two cylinders with large surface-to-volume ratio and high losses. The surviving UCN were measured both right after the filling of the trap, of number $N_i$, and after a longer storage time, of number $N_f$. Furthermore, a "jacket" of thermal-neutron detectors covered the outside of the UCN storage vessel, which measured the integral number $J$ of inelastically upscattered UCN during the storage interval $T$. The neutron lifetime was then derived as $\tau_n = (885.4 \pm 1.0)$ s solely from the three directly measured numbers $J$, $N_i$, $N_f$, with all detector efficiencies cancelling. Among the lifetimes contributing to the PDG 2010 average, this is the one with the smallest error. The method used in this experiment is very elegant, but requires an extrapolation $\Delta t$ over more than a hundred seconds. The quoted systematic error of $\pm 0.4$ s is problematic in view of the later discovery that velocity spectra of UCN suffer minute changes during storage, even for a constant number of wall collisions, see Lamoreaux and Golub (2002), Barabanov and Belyaev (2006), and references therein. The authors Arzumanov *et al*. (2009) now acknowledge that this systematic error can optimistically only be arrived at in a future experiment after considerable additional improvements of their apparatus (now in progress).

The PDG 2010 neutron lifetime average

$$\tau_n = (885.7 \pm 0.8) \text{ s (PDG 2010)} \quad (6.30)$$

is based on the above experiments, with no scale factor needed.

• A neutron lifetime experiment installed at NIST (Huffman *et al.*, 2000) uses a superthermal $^4$He bath both for in-situ UCN production and for $e^-$ scintillation detection. The $^4$He bath is inside a magnetic UCN trap of about 1 m length. In this trap, lateral UCN confinement is assured by a long superconducting quadrupole field with transverse magnetic potential $V(\rho) \propto |\rho|$, for lateral distance $\rho$ to the field axis *z*. Axial confinement is achieved by two "humps" in the longitudinal potential $V(z)$ at the ends of the trap, produced by two solenoids with opposite windings. A prototype of this instrument had produced a value of $\tau_n = 833^{+74}_{-63}$ s, see Dzhosyuk *et al.* (2005). With a new trap of 3.1 T depth and 8 $\ell$ volume $4.5 \times 10^4$ UCN can now be stored at density 5 cm$^{-3}$. A sensitivity of $\Delta \tau_n \approx \pm 2$ s is expected after one reactor cycle (O'Shaughnessy *et al.*, 2009).

• The experiment by Serebrov *et al.* (2005, 2008b) at ILL is an improved version of the experiment by Nesvizhevsky *et al.* (1992) described above. Instead of solid oxygen, the authors used a new type of synthetic oil as a wall coating, which has lower inelastic UCN losses than Fomblin oil (used by Mampe *et al.*, 1989), and a higher coating efficiency than solid oxygen (used by Nesvizhevsky *et al.*, 1992). With a mean UCN loss time of $\tau_{loss} \approx 2$ days extrapolation over less than $\Delta t = 5$ s was required to reach the lifetime $\tau_n = (878.5 \pm 0.8)$ s in the limit of zero wall-collision rate $\gamma$, see Fig. 21. The authors do not exclude that the 2.9 $\sigma$ discrepancy to their Nesvizhevsky *et al.* (1992) result was due to some small imperfection in the oxygen coating in the earlier experiment. If this had occurred in the smaller of the two traps, this would shift up the dark points in Fig. 21 (or rather those in the corresponding 1992 figure), and would shift the $\gamma = 0$ intersection to a longer neutron lifetime. The Serebrov *et al.* (2005) experiment was excluded from the PDG 2010 average, from which it differs by 6.4 standard deviations.

Serebrov and Fomin (2010) have reanalyzed the UCN lifetime experiments of Mampe *et al.* (1989) and of Arzumanov *et al.* (2000) for effects affecting the UCN spectra that where not known at the time of measurement. They find the first lifetime shifted by $-6$ s, from $(887 \pm 3.0)$ s to $(881.6 \pm 3.0)$ s, and the second lifetime by $-5.5$ s, from $(885.4 \pm 1.0)$ s to $(879.9 \pm 2.6)$ s, which makes them both consistent with the average Eq. (6.31) below. Steyerl *et al.* (2010) point out that surface roughness effects may be another source of error in such measurements.

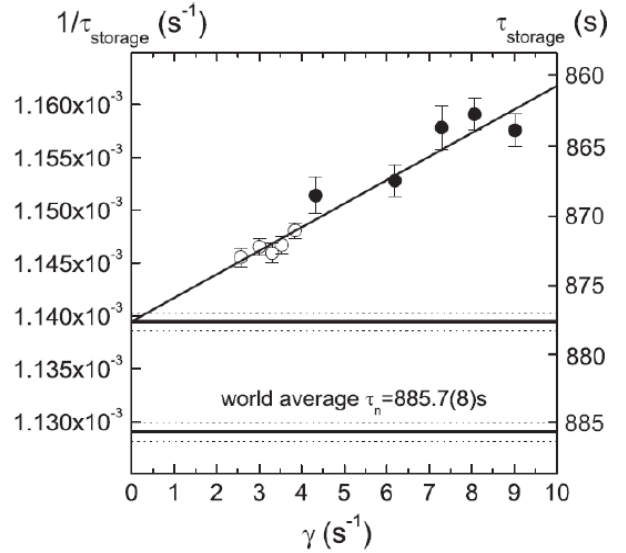

FIG. 21. Neutron lifetime from the latest UCN storage experiment. Shown is the UCN disappearance rate $1/\tau_{storage}$ versus the (normalized) loss rate $\gamma \propto 1/\tau_{loss}$, see Eq. (6.28). For $\gamma = 0$, the linear fit gives $\tau_{storage} = \tau_n = (878.5 \pm 0.8)$ s. Open dots are for the larger trap, full dots for the smaller trap, each for five different UCN energy bands. The extrapolation is over $\Delta \tau = 5$ s. Shown is also the PDG 2010 average. From Serebrov *et al.* (2005, 2008b).

• Pichlmaier *et al.* (2010) used an improved version of the first UCN lifetime apparatus by Mampe *et al.* (1989). In their experiment, the UCN spectrum was first shaped in a large prestorage volume, the UCN trap was proven to be N$_2$-gas-tight over several days, and its Fomblin oil coating was continuously being refreshed. The result $\tau_n = (880.7 \pm 1.8)$ s, with statistical and systematic errors of equal size, is 2.5 $\sigma$ below the PDG 2010 value, and 1.1 $\sigma$ above the Serebrov *et al.* (2005) value. To arrive at this number an extrapolation over rather large intervals of $\Delta t = 110$ s and more was needed. However, the dominant part (about 90%) of this extrapolation is based on geometric quantities like the length of the trap that where sufficiently well controlled.

We update the PDG 2010 average $\tau_n = (885.7 \pm 0.8)$ s (scale factor $S = 1$) with the last two results to

$$\tau_n = (881.9 \pm 1.3) \text{ s}, \text{ (our average)}, \quad (6.31)$$

with the error blown up by $S = 2.5$, following PDG procedures (see Introductory Text in Reviews, Tables, Plots, PDG 2010). Both averages differ by 2.5 $\sigma$.

A *caveat*: When applying, as we have done, scale factors to widely differing results, one must assume that all errors entering the average are underestimated by the same scale factor. *A posteriori*, this assumption holds for most past neutron decay measurements (see the neutron history plots in PDG 2010). However, it is not certain whether it also holds for the latest UCN lifetime measurements, on which the average of $\tau_n$ mostly depends.

Fortunately, the tests beyond the SM done in Sec. D below mostly depend on the correlation coefficient $B$, and less on the lifetime $\tau_n$ and the $\beta$-asymmetry $A$. Nevertheless, for quantities depending directly on $\tau_n$ like neutrino cross sections (see Sec. VII.A) we must keep this *caveat* in mind.

*Note added in proof*: In the 2011 updated web version of PDG 2010, a new lifetime average is given

$$\tau_n = (881.7 \pm 1.4) \text{ s (PDG 2011)}, \quad (6.32)$$

with a scale factor $S = 2.6$, in close agreement with our average Eq. (6.31), and with similar *caveats*.

## 2. Correlation coefficients

Next, we discuss the various neutron-decay correlation experiments. All correlation coefficients quoted are corrected for weak magnetism. Radiative corrections of order $\alpha$ are not necessary for the correlation coefficients, as they only affect the $T$ violating imaginary parts of the form factors, see Jackson *et al.* (1957b). For further radiative corrections, see Glück (1998).

a. *Electron-antineutrino coefficient a*: In principle, the correlation between electron and antineutrino momenta can be measured directly from the angular dependence of electron-proton coincidences, but such measurements so far have not been successful. Instead, the coefficient $a$ is derived from the shape of the *proton* spectrum, which is sensitive to the electron-neutrino correlation $a$, and, for polarized neutron decay, to $A$ and $B$ as well. In contrast, the shape of the *electron* spectrum from unpolarized neutron decay depends very little on the relative sizes of the coupling constants. The following experiments contribute to the PDG 2010 world average of $a$.

● Stratowa *et al.* (1978) measured $a = -0.1017 \pm 0.0051$ from the shape of the proton electron spectrum to 5%, a result that was not surpassed for 25 years. In this elegant experiment, the neutron decay volume was "in-pile" near the core of a small 7 MW Austrian research reactor. A through-going tangential beam tube was used, with no direct sight onto the reactor core or onto the moderator, and hence with rather low background. The protons emerging from this beam hole, detected at a rate of 90 s$^{-1}$, were analyzed in a spherical electrostatic spectrometer.

● Byrne *et al.* (2002) measured the proton spectrum at ILL "in-beam", using a modified version of their proton-trap lifetime apparatus mentioned above. In this experiment, the protons adiabatically emerged from a high magnetic field to a low field region. This transformed their transverse energy into longitudinal energy, which is easier to analyse with a superimposed electric field. The result was $a = -0.1054 \pm 0.0055$, with 5% error.

● *a*SPECT (Zimmer *et al.*, 2000b, Glück *et al.*, 2005) measured the energy spectrum of the decay protons, after adiabatic expansion from high into low magnetic field, too, in a retardation spectrometer of the type used for neutrino mass measurements in tritium decay. To be detected, the protons must overcome the potential of an electric counter field, which can be varied from zero to beyond the proton endpoint energy of 750 eV. The derivative of this transmission spectrum then gives the proton's energy spectrum. The apparatus was tested at FRM II and ILL at a proton count rate 460 s$^{-1}$ and a background rate of 0.2 s$^{-1}$, and reached a statistical error of $\Delta a = \pm 0.004$ (stat), but these runs had problems in quantifying the background, see Baeßler *et al.* (2008), problems likely to be avoided in a new scheduled run.

The PDG 2010 average for the electron-antineutrino correlation is

$$a = -0.103 \pm 0.004, \quad (6.33)$$

with a relative error of 4%.

b. $\beta$-*decay asymmetry A*: At present, the correlation between neutron spin and electron momentum is the main source for $\lambda = g_A / g_V$. The $\beta$-asymmetry is measured by counting the electrons from polarized neutron decay, alternately for neutrons spin up (+) and spin down (−) with respect to the holding field. The measured asymmetry is

$$A_m = \frac{N_e^+ - N_e^-}{N_e^+ + N_e^-} = A P_n \frac{v_e}{c} \langle \cos\theta \rangle, \quad (6.34)$$

with $\langle \cos\theta \rangle = \frac{1}{2}(1 + \cos\theta_0)$, where $\theta_0$ is the detector angle of admission about the axis of neutron polarization. The dependence of $A_m \propto \beta = v_e / c$ on electron energy $W_e \propto \gamma = (1 - \beta^2)^{1/2}$ has the familiar shape $\beta$ *vs.* $\gamma$ known from special relativity, plus a small (~1%) almost linear energy dependence from weak magnetism, omitted in Eq. (6.34).

Early neutron correlation experiments used neutron beam holes right outside a reactor's biological shield. To suppress background, they required multiple coincidences (between proton signals and $\Delta E_e$ plus $E_e$ electron signals), which allowed only small decay volumes. This, in turn, led to low event rates of a few counts per hour, for a review see Robson (1983).

The following experiments contribute to the present world average of $A$.

● Erozolimskii *et al.* (1990, 1991) measured $A$ on the vertical polarized cold neutron beam from the liquid hydrogen (LH$_2$) cold source that is installed in the center of the PNPI reactor core (Altarev *et al.*, 1986). The polarized beam has capture flux density $1.8 \times 10^9$ s$^{-1}$cm$^{-2}$ and polarization $P = 0.7867 \pm 0.0070$. The decay electrons emitted vertical to the neutron beam axis were counted, alternately for neutron spin up and down, in a small plastic scintillator with $\langle \cos\theta \rangle \approx 1$, in coincidence with decay protons detected under $4\pi$ solid angle. The result was $A = -0.1135 \pm 0.0014$ (Yerozolimsky *et al.*, 1997), with 1.3% relative error. The largest correction applied was +27% for neutron polarization.

● Schreckenbach *et al.* (1995) measured the $\beta$-asymmetry $A$ with a cold neutron beam passing a gas-filled time projection chamber, in which the tracks of the decay electrons were measured. This chamber was backed by plastic scintillators to measure electron energy. The experiment profited from ILL's low-background cold neutron guides and supermirror neutron polarizers with, at the time, $P_n = 0.981 \pm 0.003$ (as did the instruments discussed in the following). The result $A = 0.1160 \pm 0.0015$ had 1.3% relative error, with the largest correction of +18% on the effective solid angle $\langle \cos\theta \rangle$ of electron detection, see also Liaud *et al.* (1997).

● The PERKEO instrument in its first version (Bopp *et al.*, 1986, 1988), got rid of the need for coincident detection in the measurement of $A$, due to a drastically enlarged neutron decay volume and reduced background. With this the measured neutron decay rate increased by several orders of magnitude to 100 s$^{-1}$ per detector (for polarized neutron decay), with $P_n = 0.974 \pm 0.005$. Neutron density in-beam at ILL was 1500 cm$^{-3}$ (unpolarized) over several dm$^3$. In the instrument, a $B = 1.6$ T field from superconducting coils magnetically coupled two plastic scintillators to the neutron decay volume and to each other. The electrons, spiralling about $B$, reached the detectors with $4\pi$ effective solid angle. Electrons that backscattered on one detector deposited their remaining energy as delayed events in the other detector. In the PERKEO approach, the neutron polarization always adiabatically follows the local field $\mathbf{B}(\mathbf{x})$, hence there is a clean division between electrons emitted along $(\mathbf{p}_e \cdot \mathbf{B} > 0)$ and against $(\mathbf{p}_e \cdot \mathbf{B} < 0)$ the local field direction. In this way, electron emission with respect to the field direction is always well defined, with $\langle \cos\theta \rangle \equiv \frac{1}{2}$, and no precision alignment of the detectors relative to the decay volume is necessary. Lee and Yang (1956) had already advocated such a scheme for their proposed search for parity violation. The full energy dependence $A_m(E_e)$ was measured and $A$ was extracted on the basis of Eq. (6.34). The 1986 result was $A = -0.1146 \pm 0.0019$, with 2% relative error. The largest correction of +13% accounted for electrons counted in the wrong detector due to magnetic-mirror reversals of electron momenta induced by $\mathbf{B}$-field inhomogeneities over the neutron decay volume.

● A new version of the PERKEO instrument by Abele *et al.* (2002), used with a neutron polarizer with $P_n = 0.989 \pm 0.003$, gave $A = -0.1189 \pm 0.0007$, with a statistics-dominated 0.6% relative error. In this measurement, the correction to the magnetic mirror effect mentioned above had been suppressed to +0.09%, the largest correction then became the +1.1% correction for neutron polarization.

The PDG 2010 average derived from the above measurements is $A = -0.1173 \pm 0.0013$, where the error is increased by a scale factor of 2.3.

● A later measurement with PERKEO was done at the new fundamental physics cold neutron facility at ILL. The station's dedicated "ballistic" supermirror cold neutron guide at the time had a capture flux density of $1.3 \times 10^{10}$ cm$^{-2}$s$^{-1}$ (unpolarized). The experiment used an improved neutron polarizer with $P_n = 0.997 \pm 0.001$ (Kreuz *et al.*, 2005a). In this measurement, the error of the $\beta$-asymmetry $A$ was again cut in half, and the corrections diminished to 0.4%. The combined result covering the period 1997 to 2008 is $A = -0.11933 \pm 0.00034$, with a 0.3% relative error, as reported by Abele (2008, 2009). Publication of this result is underway, the delay being due to changes of affiliation of several of the main collaborators. This new result for $A$ differs from the PERKEO 1986 result by +2.5 standard errors.

● A new $\beta$-asymmetry result with trapped ultracold neutrons, reported by Liu *et al.* (2010), used the 5 K solid deuterium UCN source installed at the LANSCE pulsed proton beam. UCN density in the trap was 0.2 cm$^{-3}$. The initial UCN polarization, created by spin selection via a 7 T magnetic barrier, was $P_n = 1.00^{+0}_{-0.0052}$, and depolarization during storage was below 0.0065. The decay electrons were counted in a $\Delta E_e$-$E_e$ arrangement of multiwire chambers and plastic scintillators, magnetically coupled to the UCN trap volume and to each other by a $B = 1$ T field. UCN generated backgrounds were negligible. The result was $A = -0.11966 \pm 0.00089(\text{stat})^{+0.00123}_{-0.00140}(\text{syst})$.

To update the PDG 2010 average $A = -0.1173 \pm 0.0013$, which had a scale factor $S = 2.3$, we replace the Abele et al. (2002) result $A = -0.1189 \pm 0.0007$ by the Abele (2008) combined result $A = -0.11933 \pm 0.00034$, and add the Liu et al. (2010) result given above, which latter, for further analysis, we simplify to $A = -0.1197 \pm 0.0016$. This gives the new world average

$$A = -0.1188 \pm 0.0008, \quad (6.35)$$

with a scale factor $S = 2.5$ on the error. A history plot of $A$ showing the corrections applied to the raw data (similar to Fig. 20 for $\tau_n$) is given in Abele (2009). The comments made above on the scale factor of the neutron lifetime average apply equally to $A$. In particular, we note what happens when to an old measurement with error $\sigma_1$ that is a number of $m$ standard deviations away from the "true value", we add a new and more precise measurement with smaller error $\sigma_2 < \sigma_1$: The scale factor $S$ on the total error of both measurements does not decrease, but keeps increasing as $S = m / (1 + (\sigma_2 / \sigma_1)^2)$ for decreasing error $\sigma_2$ of the new measurement, to a perennial $S \to m$.

c. *Antineutrino asymmetry B*: The correlation between neutron polarization and antineutrino momentum is reconstructed from electron-proton coincidences in polarized neutron decay. The following experiments contribute to the present world average of $B$.

● Kuznetsov et al. (1995) measured $B$ at PNPI and later at ILL in a setup similar to the one used in the $A$-measurement by Erozolimskii et al. (1990) cited above. The proton time-of-flight, measured in coincidence with the energy-resolved electrons, is sensitive to the direction of antineutrino emission with respect to neutron spin direction. The neutron polarization was $P_n = 0.9752 \pm 0.0025$. The final result $B = 0.9821 \pm 0.0040$ (Serebrov et al., 1998) had a 0.4% error.

● Schumann et al. (2007), improving on Kreuz et al. (2005b), measured $B$ with PERKEO by detecting both electrons and time-delayed protons from polarized neutron decay with plastic scintillators like in the $A$-measurement, with $P_n = 0.997 \pm 0.001$. The protons were detected via secondary electrons released during their passage through a thin carbon foil at a high negative electric potential in front of the scintillator. The result $B = 0.9802 \pm 0.0050$ had an error similar to the PNPI experiment, all corrections applied were smaller than this error.

The PDG 2010 average derived from these and some earlier, statistically no longer significant measurements, is

$$B = 0.9807 \pm 0.0030, \quad (6.36)$$

with 0.3% relative error.

● An interesting measurement of the two quantities $P_n A = -0.1097 \pm 0.0016$ and $P_n B = 0.9233 \pm 0.0037$ was done by Mostovoi et al. (2000), using the same PNPI apparatus at ILL for both measurements. From these quantities $\lambda = (A - B)/(A + B) = -1.2686 \pm 0.0046$ was derived, where neutron polarization has shortened out, which often is the largest correction in such measurements. The authors then calculated both $A$ and $B$ from this value of $\lambda$, using Eqs. (6.15) and (6.17), which, of course is not an independent derivation of these quantities, so only $\lambda$ from this measurement is included in the compilations, but not $A$ and $B$. From this calculated $B$ and the measured $P_n B$ the authors derive $P_n = 0.935 \pm 0.005$.

d. *Proton asymmetry C*: Schumann et al. (2008) for the first time measured the correlation between neutron polarization and proton momentum, using the same PERKEO setup as used for the $B$-measurement. The proton asymmetry was measured in coincidence with all electrons as a function of electron energy, see Fig. 22. The result

$$C = -0.2377 \pm 0.0026 \quad (6.37)$$

with 1% relative error agrees with $C = -0.2369 \pm 0.0009$ as calculated from Eq. (6.18) with the measured $A$ and $B$ values. In principle, the proton asymmetry could also be measured with single protons as a function of proton energy. From measured energy spectra of $C$ one can derive $A$ and $B$ separately. For instance, at low electron energy, $C(E_e)$ in Fig. 22 reflects the neutrino asymmetry $B \approx +1$, and it reflects the electron asymmetry $A \approx -0.1$ at endpoint energy $E_0$.

e. *Correlation N*: This describes the correlation between neutron spin and electron spin. Kozela et al. (2009) for the first time made the electron spin variable $\sigma_e$ accessible in neutron decay experiments. Their main goal was the $T$ violating parameter $R$ described below, and the $N$-coefficient was a by-product of this research. The experiment used the fundamental physics beam-line of the continuous spallation neutron source SINQ at PSI. The apparatus consisted of multiwire proportional counters for electron track reconstruction on both sides of the cold neutron beam. These track detectors were backed by thin Pb-foils for electron spin analysis, followed by plastic scintillators for electron energy measurement.

The Pb-foil acted as a Mott polarimeter, via the spin-dependent angular distribution of the electrons backscattered by the foil. From the asymmetry of backscattered events upon reversal of the neutron spin followed

$$N = 0.056 \pm 0.011(\text{stat}) \pm 0.005(\text{syst}) . \quad (6.38)$$

Within errors, this is in agreement with the expected $N = 0.0439 \pm 0.0003$ from Eq. (6.19), $N = -(m_e/W_e)A$, taken at the electron endpoint energy $W_0$ with the measured $A$ from Eq. (6.35).

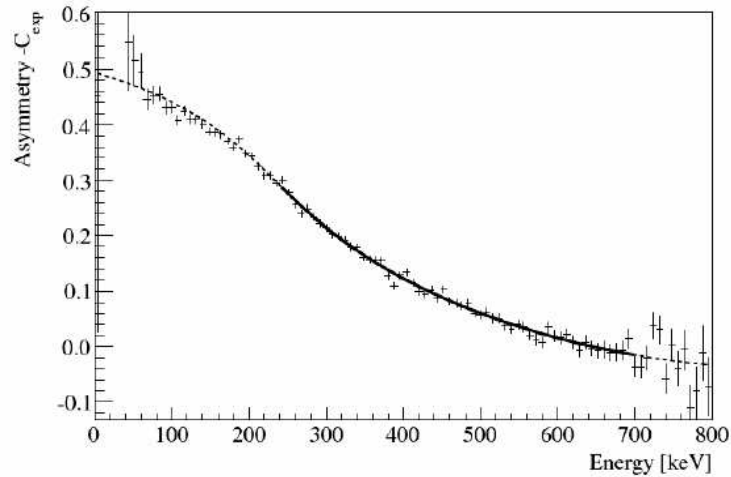

FIG. 22. The negative of the experimental proton asymmetry $-C$ as a function of electron energy $E_e$. The bold line indicates the fit region. From Schumann *et al.* (2008).

f. *Triple correlation D*: This is the coefficient for the $T$ violating triple product $\boldsymbol{\sigma}_n \cdot (\mathbf{p}_e \times \mathbf{p}_\nu)$ between neutron polarization, electron momentum, and proton momentum.

● For a long time this coefficient was dominated by the result $D = (-11 \pm 17) \times 10^{-4}$ of Steinberg *et al.* (1974, 1976), done at ILL. In this experiment, the longitudinal neutron polarization, the electron detectors and the proton detectors were arranged at right angles to each other in order to maximise the above triple product. The neutron spin was inverted in regular intervals. Erozolimskii at PNPI did similar measurements in the 1970's with a slightly larger error.

● Later on, Lising *et al.* (2000) realized that an octagonal detector arrangement with 135° between the electron and proton detector axes would give a three times higher quality factor $NS^2$, due to much higher count rate $N$ at a slightly reduced signal $S$. Their experiment, done at the fundamental physics beamline of the NIST reactor, gave $D = (-6 \pm 12(\text{stat}) \pm 5(\text{syst})) \times 10^{-4}$.

● Soldner *et al.* (2004) used an arrangement with electron track detection in a multiwire proportional chamber, which, together with the orthogonally arranged proton detectors, permitted off-line control of the delicate symmetry conditions. The experiment reached $D = (-2.8 \pm 6.4(\text{stat}) \pm 3.0(\text{syst})) \times 10^{-4}$.

The PDG 2010 average from the above measurements

$$D = (-4 \pm 6) \times 10^{-4} \quad (6.39)$$

is consistent with zero. This translates into a phase between $g_A$ and $g_V$

$$\varphi = -180.06° \pm 0.07° , \quad (6.40)$$

which means that, within error, $\lambda$ is real, and $T$ is conserved.

g. *The coefficient R* of the triple product $\boldsymbol{\sigma}_e \cdot (\boldsymbol{\sigma}_n \times \mathbf{p}_e)$ between electron spin, neutron spin, and electron momentum was measured for the first time by Kozela *et al.* (2009) with the apparatus described above for coefficient $N$. To derive $R$ the authors regarded the electron spin component perpendicular to the decay plane spanned by $\boldsymbol{\sigma}_n$ and $\mathbf{p}_e$. $R$-events were separated from $N$-events by their differing kinematic factors, which depend on the angle between the $\boldsymbol{\sigma}_n$-$\mathbf{p}_e$ decay plane and the Mott scattering plane, with the result

$$R = 0.008 \pm 0.015(\text{stat}) \pm 0.005(\text{syst}) . \quad (6.41)$$

A nonzero $R$ requires scalar and tensor couplings $g_S$ and $g_T$ not present in the $V-A$ Standard Model, as will be discussed in Sec. D.

### 3. *Upcoming neutron decay experiments

Several new neutron decay instruments are in the construction or in the test phase.

a. *Upcoming lifetime experiments*: Several lifetime experiments are underway.

● Ezhov *et al.* (2009) use a magnetic UCN trap built from rare-earth permanent magnets. The trap has the form of a bucket of 20 cm inner diameter, open at the top. Near the surface of its inner walls is a repulsive magnetic potential, produced by a magnetic field up to 1.2 T, whose direction changes sign every few centimeters in the azimuthal direction. UCN from the ILL turbine source are brought into this trap from above, via a cylindrical "elevator" box, which is filled with UCN, then slowly lowered into the trap, emptied there, and withdrawn. In the prototype apparatus 1800 UCN per filling were stored at ILL in an effective volume of 10 $\ell$. UCN lost by spin-flip were counted in the same detector as the surviving UCN, so their detection efficiencies do not enter the result. A new 90 $\ell$ version of this apparatus is under test.

● At FRM II, a superconducting trap of toroidal form called PENeLOPE is under development. It has a depth of 1.8 T, a volume of 700 $\ell$, and a proton detector above the active volume, see Zimmer (2000), and Materne *et al.* (2009). The trap will be filled from the FRM II solid $D_2$ UCN source that is under construction. It is expected to hold up to $2 \cdot 10^8$ UCN. Prototype testing is under way.

● A superthermal $^4$He UCN source placed within a magnetic Ioffe trap is proposed by Leung and Zimmer (2009), with a long octupole field with transverse magnetic potential $V(\rho) \propto |\rho|^3$ made from permanent magnets, plus two superconducting coils as magnetic end caps. After filling, the magnetic trap, and with it the trapped UCN, are withdrawn from the stationary $^4$He bath in axial direction. In the process, some UCN are lost as they must penetrate the end window of the $^4$He vessel, but this is largely compensated by the advantage that the decay protons can be detected in a low-background environment. Prototype testing has started.

● A conceptual design study for a novel magneto-gravitational UCN trap at LANSCE was presented by Walstrom *et al.* (2009).

b. *Upcoming correlation experiments*: Several correlation experiments are on the way.

● *a*CORN is built for the measurement of electron-antineutrino correlation *a* directly from electron-proton coincidences (Wietfieldt *et al.*, 2009). Decay electrons originating in a small volume of a cold neutron beam and emitted vertically to the beam axis $z$ are detected under small solid angle in an electron detector installed in +$y$ directions. Decay protons are guided electrostatically to a small proton detector installed in –$y$ direction. Precise circular apertures limit the divergence of the electron and proton fluxes. For low electron energies (80 keV $\leq E_e \leq$ 300 keV), the proton time of flight spectrum, measured in delay to the prompt electron signal, splits into two well defined groups: The early proton arrivals are those emitted in direction of the proton detector, opposite to electron emission, i.e., directionally anticorrelated with the electrons. The late proton arrivals are those first emitted in the same direction as the electrons, so they had to make a detour through the electrostatic mirror to arrive late at the proton detector. These protons are correlated with the direction of electron emission. When installed at NIST, a coincidence count rate of 0.1 s$^{-1}$ is expected in this experiment.

● The PERKEO program continued with a new instrument for neutron correlation measurements (Märkisch *et al.*, 2009). In the new apparatus, the neutron decay volume was further enlarged, with a measured decay electron rate of $5 \cdot 10^4$ s$^{-1}$ for a continuous polarized cold neutron beam. In view of the size of the new instrument, conventional magnet coils had to be used. The rather small magnetic field of $B = 0.15$ T is acceptable because the up to 2 cm radii of gyration of protons and electrons have an astonishingly small effect on the detector response function (Dubbers *et al.*, 2008b). In a recent second period of beam time, this instrument was used with a pulsed neutron beam. The decay electrons and their asymmetry were observed from a cloud of neutrons moving through the 2.5 m length of the instrument's active volume. At present the results of this run are under evaluation.

● The Nab instrument is built for the measurement of correlation *a* and Fierz interference *b* (Počanić *et al.*, 2009). It aligns the proton and electron momenta emerging from unpolarized neutron decay in a high magnetic field region by adiabatic magnetic expansion from 4 T to 0.1 T into both the +$y$ and –$y$ direction, at right angles to the neutron beam. Both protons and electrons are detected, and their relative time-of-flight is measured in two silicon detector arrays under $4\pi$ solid angle. The method is based on the observation (Bowman, 2005) that for a given electron energy, the proton momenta squared have a probability distribution with a slope proportional to the correlation coefficient *a*. From a 20 cm$^3$ neutron decay volume, a signal rate of 400 s$^{-1}$ is expected for full power of SNS. The Fierz term *b* is to be extracted from the electron energy spectrum. A new asymmetric version of the instrument is described in Alarcon *et al.* (2010). Later on, the project will evolve into the *abBA* instrument, where the electron and antineutrino asymmetries *A* and *B* will also be measured in a configuration similar to the previous PERKEO II instrument.

● At SNS the PANDA project intends to measure the proton asymmetry *C* in coincidence with the electrons, and dependent on the proton energy, see Alarcon *et al.*, 2008.

● The PERC project (Dubbers *et al.*, 2008a) is based on the belief that the highest precision in neutron decay can be reached with single-particle detection. PERC, short for Proton and Electron Radiation Channel, will not be another neutron decay spectrometer, but a facility that delivers well-defined beams of electrons and protons from neutron decay taking place inside a long section of a cold neutron guide at a rate of 10$^6$ decays per meter of guide. With PERC, one can measure the shapes and magnitudes of electron or proton energy spectra from polarized or unpolarized neutron decay; with or without electron spin analysis. The neutron decay coefficients *a*, *b*, *b′*, *A*, *B*, *C*, *N*, *G*, *Q* will be accessible from such single particle spectra. Our primary aim, however, are not these correlation coefficients but the weak amplitudes that the spectra depend on (within and beyond the SM), like $V_{ud}$, $g_A$, $g_S$, $g_T$, $\zeta$, $m_R$, $f_2$, $f_3$, $g_2$, etc.. The PERC construction period starts in 2011, and the facility will be open for dedicated user instruments at FRM-II several years later.

● At FRM II a project on neutron bound-beta decay is underway for the study of physics beyond the SM in the scalar/tensor and right-handed current sectors (Dollinger *et al.* 2006), see also end of Sec. D below. The neutron decay volume will be "in-pile", next to the reactor core inside a tangential through-going beam-tube. The hydrogen atoms from neutron decay leave the beam tube with a recoil energy of 326 eV. Selection of the various $n = 2$ hyperfine states of the atoms will be done by judicious Stark mixing and quenching via the unstable 2*P* state, plus by additional induced hyperfine RF transitions along the hydrogen flight path. Excited states with $n > 2$ will be depleted by laser ionization. Finally, the fast hydrogen atoms in a selected hyperfine state are ionized, velocity selected with a magnetic field, and accelerated onto a CsI scintillator. A first version of the apparatus aims at the detection of bound neutron decay without the analysis of the hyperfine states.

## 4. Comparison with muon decay data

Muon decay $\mu \to e + \bar{\nu}_e + \nu_\mu$ tests the purely leptonic part of the weak interaction. There are three observable vectors $\mathbf{p}_e$, $\mathbf{\sigma}_e$, $\mathbf{\sigma}_\mu$ in muon decay, besides the muon lifetime. From these, four correlations can be constructed (three scalar and one triple product) that have been measured over the years with precision similar to that obtained for the neutron correlation coefficients. In muon decay, also the shape of the electron spectrum depends sensitively on the coupling constants involved. This is in contrast to neutron decay, where the shapes of the unpolarized or polarized electron spectra from neutron decay depend very little on the relative sizes of the coupling constants involved; only the shapes of the proton spectra depend measurably on the coupling constants.

The Michel parameters with SM values $\rho = \tfrac{3}{4}$, $\eta = 0$ (from the shape of unpolarized muon decay spectra), plus the four correlation coefficients $\xi = 1$, $\delta = \tfrac{3}{4}$, and two transverse electron spin components (from polarized muon decay), have absolute errors $(4, 34, 35, 6, 80, 80) \times 10^{-4}$, respectively. The absolute errors for recent neutron measurements of *a*, *A*, *B*, *C*, *D*, *N*, *R*, are of comparable magnitude, namely, $(40, 4, 30, 9, 6, 120, 80) \times 10^{-4}$ respectively. The relative error of the muon lifetime, on the other hand, is two orders of magnitude smaller than that of the neutron lifetime, from which we profit when using $G_F$. For muon results beyond the SM, see Sec. D.

## C. Weak interaction results and the Standard Model

In this section, we discuss implications for the Standard Model of the neutron decay data presented in the preceding section. Global analyses of neutron and nuclear $\beta$-decay data were done by Severijns *et al.* (2006), who expand and supersede the analyses of Gaponov and Mostovoy (2000), and Glück *et al.* (1995). For a discussion of neutron decay in the framework of effective field theory, see Ando *et al.* (2004), and Gudkov *et al.* (2006). Reviews on low-energy weak interactions in a broader context were written by Herczeg (2001), and Erler and Ramsey-Musolf (2005).

### 1. Results from neutrons and other particles

The neutron decay parameters relevant for the SM are $\lambda = g_A / g_V$ and the upper left element $V_{ud}$ of the CKM matrix. The PDG 2010 value for $g_A/g_V$ was derived from the $\beta$-asymmetry $A$ and the Mostovoi et al. (2001) value for $\lambda$ to $\lambda = -1.2694 \pm 0.0028$, with a scale factor $S = 2.0$. With the updated value of $A$, Eq. (6.35), this becomes $\lambda = -1.2734 \pm 0.0021$, with a scale factor $S = 2.6$. Inclusion of the results for $a$, $B$, and $N$ does not change this result significantly, giving

$$\lambda = -1.2734 \pm 0.0019 \text{ (neutron)} \quad (6.42)$$

with a scale factor of 2.3 (we leave out the proton asymmetry $C$, which in part was based on the same data set as one of the $B$ measurements). Figure 23FIG. 23 shows the $\lambda$ values derived from various measurements. For the values labelled $F + D$ and $\tau_n$ on the abscissa of Fig. 23 see Eq. (6.43) and text to Eq. (6.52) below.

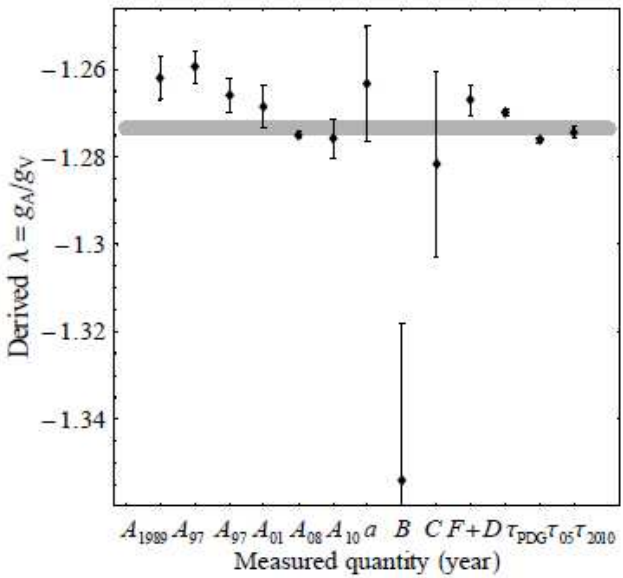

FIG. 23. The ratio of axial vector to vector coupling constants $\lambda = g_A / g_V$ (horizontal gray bar) as derived from various sources: from the neutron $\beta$-asymmetry $A$, neutrino asymmetry $B$, proton-asymmetry $C$, from the electron-neutrino correlation $a$, the hyperon correlations $F + D$, and indirectly with Eq. (6.10), from the neutron lifetime $\tau_n$ combined with nuclear $|V_{ud}|^2$. The width of the horizontal gray bar gives the upscaled error.

The nucleons are members of the flavor-$SU(3)$ baryon spin-½ octet, which they join with the strange hyperons (the $\Sigma, \Lambda, \Xi$-baryons with various states of electric charge). The $\beta$-decay parameters of this multiplet are linked to each other via the $SU(3)$ transformation properties. The neutron's $\lambda = g_A / g_V$ should be related to the hyperons' $F$ and $D$ $\beta$-decay parameters as $|\lambda| = F + D$. Indeed, a global evaluation of semileptonic hyperon decays by Cabibbo et al. (2004) gives

$$|\lambda| = F + D = 1.2670 \pm 0.0035 \text{ (hyperons)}. \quad (6.43)$$

Combination of this with the neutron value Eq. (6.42) causes a shift to

$$\lambda = -1.2719 \pm 0.0017 \text{ (baryons)}, \quad (6.44)$$

with $S = 2.2$.

Next, we derive the CKM matrix element $|V_{ud}|$ from neutron decay. The neutron lifetime $\tau_n$ from Eq. (6.31) and $\lambda$ from Eq. (6.42) inserted into Eq. (6.11) gives

$$|V_{ud}| = 0.9742 \pm 0.0012 \text{ (neutron)}, \quad (6.45)$$

which within error equals the value $|V_{ud}| = 0.9747 \pm 0.0015$ derived from the PDG 2010 data.

$|V_{ud}|$ can also be obtained from the rare decay $\pi^+ \to \pi^0 + e^+ + \nu_e$ with $1.0 \times 10^{-8}$ branching ratio, which gives (Počanić et al., 2004)

$$|V_{ud}| = 0.9728 \pm 0.0030 \text{ (pion)}. \quad (6.46)$$

Combination of this with the neutron value Eq. (6.45) causes a marginal shift to

$$|V_{ud}| = 0.9740 \pm 0.0010 \text{ (particles)}. \quad (6.47)$$

### 2. Results from nuclear $\beta$-decays

Altogether, the SM parameters $\lambda$ and $|V_{ud}|$ have the following main sources:

(1) Neutron decay correlations, in particular, the $\beta$-asymmetry $A$, are functions of $\lambda$ only;
(2) The neutron lifetime $\tau_n$ is a function of both $\lambda$ and $|V_{ud}|$.
(3) The nuclear $ft$ values for superallowed $0^+ \to 0^+$ $\beta$-transitions are a function of $|V_{ud}|$ only.

In the nuclear superallowed transitions, only the vector matrix element enters

$$ft = \frac{2\pi^2 (\hbar c)^7}{c(mc^2)^5} \frac{\ln 2}{G_F^2 |V_{ud}|^2}, \quad (6.48)$$

which conforms with Eq. (6.10) for half-life $t = \tau \ln 2$ and $\lambda = 0$. From any two of the three observables $A$, $\tau_n$, and $ft$ one can calculate the parameters $\lambda$ and $|V_{ud}|$ plus the remaining third observable, as is frequently done in all three combinations. Of course, it is preferable to use all measured observables and make a global fit.

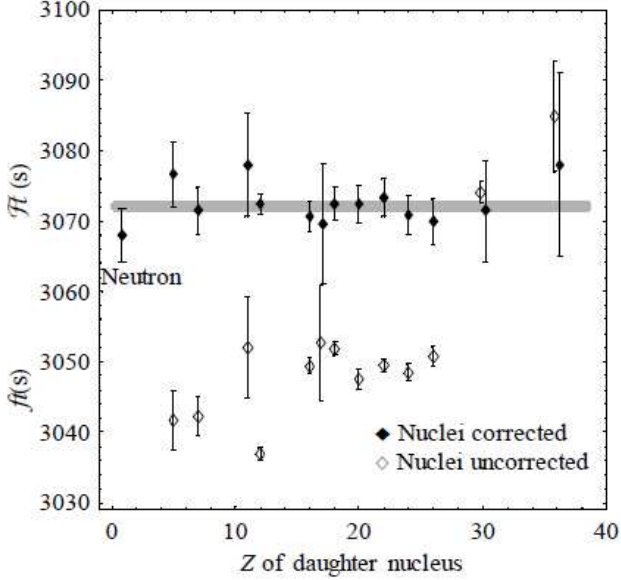

FIG. 24. : Measured $\mathcal{F}t$ values both from nuclear superallowed $0^+ \rightarrow 0^+$ $\beta$-transitions ($Z \geq 5$), and from the measured vector part of neutron decay ($Z = 1$). The superallowed data are shown before ($\Diamond$) and after ($\blacklozenge$) corrections for nuclear and radiative effects, as well as their mean value after correction (horizontal gray bar). For comparison, the neutron $\mathcal{F}t_{n\text{-Vector}}$ value is inserted as derived from Eq. (6.51) using only the latest two results for the neutron lifetime $\tau_n$ and the latest two results for $\lambda = g_A / g_V$.

For the nuclear decays, besides the rather well known radiative corrections, nuclear structure and isospin corrections must be applied, and much progress has been made in recent years. Figure 24 shows the nuclear data from Table IX of Hardy and Towner (2009), before ($ft$, $\Diamond$), and after corrections ($\mathcal{F}t$, $\blacklozenge$). From these data the mean value

$$\mathcal{F}t = (3071.81 \pm 0.83) \text{ s} \quad \text{(nuclear } \mathcal{F}t\text{)} \tag{6.49}$$

is derived with an error given by the width of the gray horizontal band in Fig. 24. From Eq. (6.48) then

$$|V_{ud}| = 0.97425 \pm 0.00022 \quad \text{(nuclear } \mathcal{F}t\text{)} \tag{6.50}$$

with an error five times smaller than the error from neutron plus pion. Notice, however, that in this evaluation the data point with the smallest error at $Z = 12$ is also the one with the largest correction.

For comparison, we also derive an $\mathcal{F}t$-value for the neutron, due solely to the vector part of neutron decay, which is obtained from Eqs. (6.11) and (6.48) as,

$$\mathcal{F}t_{n-\text{Vector}} = f^R \tfrac{1}{2}(1+3\lambda^2)\tau_n \ln 2 = (3068 \pm 3.8) \text{ s}$$
(neutron), (6.51)

with the radiatively corrected phase space factor $f^R = 1.71465 \pm 0.00015$. To demonstrate the potential sensitivity of neutron decay measurements we have used in Eq. (6.51) only the last two lifetime $\tau_n$ results (Serebrov et al., 2005, and Pichlmaier et al, 2010) and the last two $\beta$-asymmetry results (Abele, 2008, and Liu et al, 2010). This neutron $\mathcal{F}t$ value is inserted in Fig. 24 at $Z = 1$. To compare, the corresponding value derived from the PDG 2010 neutron data is $\mathcal{F}t_{n\text{-Vector}} = (3071 \pm 9)$ s.

For the neutron decay data the experimental error dominates, while for the nuclear decay data the error from theory dominates. Sometimes it is argued that the neutron derived value for $|V_{ud}|$ is more indirect because it needs input from the two measured quantities $\tau_n$ and $\lambda$. However, the nuclear derivation of $|V_{ud}|$ also requires input from various sources: from nuclear lifetimes, branching ratios, $Q$-values, plus nuclear theory.

Finally, using Eq. (6.11) one obtains from the neutron lifetime $\tau_n$ and the nuclear $|V_{ud}|$ an indirect value for $\lambda$ (see the last three entries in Fig. 23), which we can include in the world average to obtain

$$\lambda = g_A / g_V = -1.2735 \pm 0.0012$$
(all neutron data plus nuclear $|V_{ud}|$), (6.52)

with a scale factor $S = 3.1$. A few years ago, this value was 2.5 standard deviations lower and had a smaller error, $\lambda = g_A / g_V = -1.2699 \pm 0.0007$, with $S = 1.1$, see Severijns et al. (2006). This is because recent neutron decay data disagree significantly with the former world average. With $|V_{ud}| \approx 0.97$ fixed, the sensitivity of $\tau_n$ to $\lambda$ is $\partial_\lambda \tau_n / \tau_n = 1.3$ at $\tau \approx 880$ s lower than for $a$ and $A$, which was $-2.8$ and $-3.2$, respectively.

### 3. Unitarity of the CKM matrix

The measured value of $|V_{ud}|^2$ can be used to test the unitarity of the first row of the CKM matrix

$$|V_{ud}|^2 + |V_{us}|^2 + |V_{ub}|^2 = 1 - \Delta, \tag{6.53}$$

where unitarity in the SM requires $\Delta = 0$. The third term, $|V_{ub}|^2 \approx 10^{-5}$ as derived from $B$-decays, is negligible, so $\Delta$ mainly tests the old Cabibbo theory with $V_{ud} \approx \cos\theta_C$. Some years ago, a positive $\Delta = 0.0083(28)$ with $+3\sigma$ standard deviations from unitarity was observed for the neutron (Abele *et al.*, 2002), and similar deviations on the $+2\sigma$ and $+1\sigma$ level were observed from nuclear and pion decays, see the survey of Abele *et al.* (2004), which summarizes a workshop on this topic edited by Abele and Mund (2002).

Today it seems that the culprit for the deviation from unitarity was the $V_{us}$ value (with $V_{us} \approx \sin\theta_C$ in Cabibbo theory). Recent remeasurements of kaon, tau, and hyperon weak transitions shifted the measured $|V_{us}|f_+(0)$ by as much as six standard errors. When uncertainties from theory in the form factor $f_+(0)$ are included, this shifts the PDG 2004 value $|V_{us}| = 0.2200 \pm 0.0026$ to the PDG 2010 value $|V_{us}| = 0.2252 \pm 0.0009$.

With the value $|V_{ud}|$ from $0^+ \to 0^+$ decays, Eq. (6.50), this re-establishes unitarity on the level of one part per thousand, see the review Blucher and Marciano (2009), and the update by Towner and Hardy (2010), which give

1st row: $|V_{ud}|^2 + |V_{us}|^2 + |V_{ub}|^2 = 0.9999 \pm 0.0006$
(with nuclear $V_{ud}$), (6.54)

Using $|V_{ud}|$ alone from the neutron gives

1st row: $|V_{ud}|^2 + |V_{us}|^2 + |V_{ub}|^2 = 1.0000 \pm 0.0026$
(with neutron $V_{ud}$). (6.55)

High-energy results for the other rows and columns give (Ceccucci *et al.*, 2010)

2nd row: $|V_{cd}|^2 + |V_{cs}|^2 + |V_{cb}|^2 = 1.101 \pm 0.074$, (6.56)
1st col.: $|V_{ud}|^2 + |V_{cd}|^2 + |V_{td}|^2 = 1.002 \pm 0.005$, (6.57)
2nd col.: $|V_{us}|^2 + |V_{cs}|^2 + |V_{ts}|^2 = 1.098 \pm 0.074$. (6.58)

The sum of the three angles of the so-called unitarity triangle $\alpha + \beta + \gamma = (183 \pm 25)°$ is also consistent with the SM, as is the Jarlskog invariant $J = (3.05^{+0.19}_{-0.11}) \times 10^{-5}$, see Eq. (3.8).

From a model-independent analysis by Cirigliano *et al.* (2010a) follows that precision electroweak constraints alone would allow violations of unitarity as large as $\Delta = 0.01$, while the bound $\Delta = (-1 \pm 6) \times 10^{-4}$ from Eqs. (6.54) and (6.55) constrains contributions to the low-energy effective Lagrangian to $\Lambda > 11$ TeV at 90% C.L.

### 4. *Induced terms

For neutron decay there are no results yet on the small induced terms $f_2(0)$ for weak magnetism and $f_3(0)$ and $g_2(0)$ for second-class currents. As some of the formulae on these terms are hard to find in the literature, we shall discuss them in more detail.

One should test the SM prediction for the weak magnetism term $f_2(0)$ because it is at the heart of electroweak unification. Best suited for a measurement of weak magnetism is the typically 1% electron-energy dependence of the correlation coefficients. Past experiments on the $\beta$-asymmetry $A$ with PERKEO gave $f_2(0)$ as a free parameter only with an error of about seven times the expected effect.

To discuss the influence of such small corrections let us write an arbitrary correlation coefficient $X$ from Sec. A.4 within the SM as $X = s/\xi$, with normalization $\xi = 1 + 3\lambda^2$. For the $\beta$-asymmetry $A$, for instance, $s = -2(\lambda^2 + \lambda)$ from Eq. (6.15). When we have additional terms $\varepsilon \ll s$ in the numerator $s$, and $\varepsilon' \ll \xi$ in the denominator $\xi$, then the correlation coefficient is frequently written in one of the forms

$$X' \approx X(1 + \varepsilon/s - \varepsilon'/\xi) = X + (\xi\varepsilon - s\varepsilon')/\xi^2.$$
(6.59)

For the $\beta$-asymmetry $A$, for instance, the weak magnetism correction is

$$\varepsilon_{wm} = \tfrac{2}{3}M^{-1}[(2\lambda^2 + \lambda - \mu - 2\lambda\mu)W_0 - (11\lambda^2 + 7\lambda - \mu - 5\lambda\mu)W],$$
(6.60)

$$\varepsilon'_{wm} = 2M^{-1}[(-\lambda^2 + \lambda\mu)(W_0 + m^2/W) + 2(1 + 5\lambda^2 - 2\lambda\mu)W],$$
(6.61)

see Holstein *et al.* (1972), and, for more details, Holstein (1974). For the moment we write shortly $M$, $m$, $W$ for $Mc^2$, $m_e c^2$, $W_e$. In these equations, $\mu \equiv \kappa_p - \kappa_n + 1 = 4.706$, and the total electron endpoint energy is $W_0 = \tfrac{1}{2}(m_n - (m_p - m)^2/m_n)$. The solution Eq. (6.59) then agrees with the result in Appendix 3 of Wilkinson (1982).

To take into account $g_2(0)$ and $f_3(0)$ one must apply the following corrections to the $\beta$-asymmetry $A$

$$\varepsilon_{scc} = -(2/3M)[(1 + 2\lambda)W_0 + (\lambda - 1)W]g_2(0),$$ (6.62)
$$\varepsilon'_{scc} = (2/M)[\lambda(W_0 + m^2/W)g_2(0) + (m^2/W)f_3(0)],$$
(6.63)

also from Holstein (1974), in agreement with Gardner and Zhang (2001), who also present the analytic solution for the electron-neutrino correlation $a$. Analytic solutions for the proton asymmetry $C$ are given by Sjue (2005, Erratum 2010). Numerical weak magnetism corrections for various neutron decay correlation coefficients are tabulated in Glück (1998). If one wants to measure both weak magnetism $f_2$ and second-class amplitude $g_2$ without assuming the validity of CVC, then it is best to measure the precise spectral shapes for two different correlation coefficients.

For nuclei, the weak magnetism prediction is confirmed on the ±15% level, see Minamisono *et al.* (1998) and references therein. In particle physics, the best test of weak magnetism is from $\Sigma^- \to n e^- \bar{\nu}_e$ decay, also with error ±15%. For the proton, weak magnetism is tested at $q^2 = 0.1\,(\text{GeV/c})^2$ to about ±30% (Mueller *et al.*, 1997, see also Acha *et al.*, 2007). For a feature of the SM as basic as CVC this precision is not very impressive and should be improved whenever possible.

Radiative neutron decay $n \to p^+ + e^- + \bar{\nu}_e + \gamma$ was measured at NCNR-NIST by Nico *et al.* (2006), see also Cooper *et al.* (2010), and Beck *et al.* (2002) for an earlier search. The experiment used the NIST in-beam lifetime apparatus described above, complemented with a $\gamma$ detector. The bremsstrahlung probability was measured in triple $\gamma e^- p^+$-coincidence at a signal-to-noise ratio of about unity, in the $\gamma$-energy interval from 15 to 313 keV. The result was a branching ratio

$$\text{BR} = (3.09 \pm 0.32) \times 10^{-3},$$
(radiative neutron decay), (6.64)

in agreement with the theoretical value $\text{BR} = 2.85 \times 10^{-3}$.

In particle physics, the most well-studied such radiative decay is radiative kaon decay $K_L \to \pi^\pm e^\mp \nu_e \gamma$ with branching ratio $\text{BR} = (9.35 \pm 0.15) \times 10^{-3}$ for an expected $\text{BR} = (9.6 \pm 0.1) \times 10^{-3}$. The radiative decay of the pion $\pi^+ \to \mu^+ \nu_\mu \gamma$ occurs with $\text{BR} = (2.0 \pm 0.25) \times 10^{-4}$ for an expected $\text{BR} = 2.5 \times 10^{-4}$, and that of the muon $\mu^- \to e^- \bar{\nu}_e \nu_\mu \gamma$ with $\text{BR} = (0.014 \pm 0.004)$ for an expected $\text{BR} = 0.013$.

### D. Symmetry tests beyond the SM

Precision measurements of SM parameters are at the same time tests for physics beyond the SM. In the present section, we shall discuss limits on amplitudes beyond the SM derived from neutron decay, and shall compare them with limits derived from other nuclear or particle reactions. We shall not do this in a systematic way, as this was done recently by Severijns *et al.* (2006) for beta decay in general, but we limit the discussion to some basic issues.

We first turn to the question whether the basic symmetry of the electroweak interaction is really $V - A$. We then discuss whether the universe was created with a left-right asymmetry from the beginning, or whether parity violation arose as a spontaneous symmetry breaking in the course of one of the many phase and other transitions of the universe. We investigate only the occurence of one type of such exotic couplings at a time, and derive limits for the respective coupling parameters, which is reasonable as long as no such exotic signals are in sight.

We base our studies on the PDG 2010 data, recent shifts in then neutron lifetime $\tau_n$ and $\beta$-asymmetry $A$ are only a minor problem here, because the limits derived depend mainly on the antineutrino and proton asymmetries $B$ and $C$.

### 1. Is the electroweak interaction purely $V - A$?

The most general Lorentz invariant weak Hamiltonian can be constructed from five types of scalar products,

| Scalar×Scalar | (*S*), |
| Vector×Vector | (*V*), |
| Tensor×Tensor | (*T*), |
| Axial vector×Axial vector | (*A*), |
| Pseudoscalar×Pseudoscalar | (*P*), |

with 10 complex coupling constants. Neutron decay is not sensitive to pseudoscalar *P* amplitudes, due to its low energy release, so we limit discussion to possible scalar *S* and tensor *T* amplitudes. In fact, no one knows why nature seems to have chosen the $V - A$ variety, and, contrary to common belief, tests excluding *S* and *T* amplitudes are not very stringent.

The effect of *S-T* admixtures on neutron decay correlation coefficients due to supersymmetry were investigated by Profumo *et al.* (2007b), under the condition that simultaneously the tight bounds from the SM-forbidden lepton-flavor violating $\mu \to e\gamma$ decay (B.R. $< 1.2 \times 10^{-11}$) and the results from the muon $g - 2$ measurements (precision of *g*: $6 \times 10^{-10}$) are obeyed. These *S-T* admixtures generally are $\leq 10^{-4}$, but can reach order $\alpha / 2\pi \approx 10^{-3}$ in the limit of maximal left-right mixing.

From a global fit to all available nuclear and neutron data, Severijns *et al.* (2006) obtain the limits

$$|g_S / g_V| < 0.0026 \text{ and } |g_T / g_A| < 0.0066$$
(left-handed, 95% C.L., all data 2006) (6.65)

for a possible left-handed *S-T* sector. For a right-handed *S-T* sector the limits are less stringent,

$$|g_S / g_V| < 0.067 \text{ and } |g_T / g_A| < 0.081$$
(right-handed, 95% C.L., all data 2006). (6.66)

There should be a strong interest for better constraints on $g_S$ and $g_T$, in particular, in the right-handed $S$ and $T$ sectors, Eq. (6.66). However, as it sometimes happens, the lack of rapid progress over past decades led to a drop of interest in the problem, the caravan moved on although the problem is still with us.

a. *Left-handed S or T coupling*: One mode of access to scalar and tensor amplitudes is given by the so-called Fierz interference term $b$ in $\beta$-decay. The Fierz term enters unpolarized energy spectra as $d\Gamma \propto (1 + (m_e/W_e)b)\, dW_e$, with total electron energy $W_e = E_e + m_e$. It describes $g_S$-$g_V$ and $g_T$-$g_A$ interference, i.e., $b$ is linear in both $g_S$ and $g_T$. For a purely left-handed weak interaction,

$$b = 2\frac{S + 3\lambda^2 T}{1 + 3\lambda^2}, \text{ with } S = \frac{g_S}{g_V}, T = \frac{g_T}{g_A}. \quad (6.67)$$

As there are few neutron investigations on the Fierz term, we shall treat this in more detail. It is difficult to determine $b$ from the shape of unpolarized $\beta$-spectra, because experimental spectra are deformed by the differential nonlinearity of the apparatus, usually ten times larger than the typically one percent integral nonlinearity of the calibration curve of a $\beta$-detector.

The Fierz $b$ coefficient necessarily also enters all measured decay correlations $X_m$ in the form

$$X'_m = \frac{X_m}{1 + (m_e/W_e)b}. \quad (6.68)$$

This can be seen, for instance, in Eq. (6.34) for the measured neutron $\beta$-asymmetry $A_m$, where a $b$-term in the count rates would cancel in the numerator and add up in the denominator. Limits on $b$ can then be obtained from a fit to the measured $\beta$-asymmetry spectrum $A'_m(E_e)$, which, to our knowledge, has not been done before. This fit should be less sensitive to instrumental nonlinearities than the unpolarized spectrum, as only ratios of count rates are involved.

For demonstration, we have made a fit to some older $A_m(E_e)$ data of PERKEO, with $\lambda$ and the Fierz term $b$ as free parameters, which gave

$|b| < 0.19$ (neutron, 95% C.L., from $A_m(E_e)$), (6.69)

see the exclusion plot Fig. 25. Such evaluations should be done systematically for all available correlation spectra simultaneously, which may avoid the strong correlation between $b$ and $\lambda$ seen in Fig. 25.

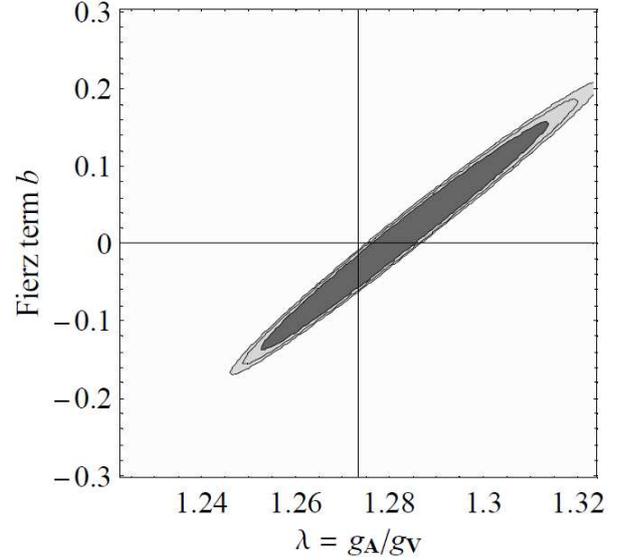

FIG. 25. Exclusion plot for the SM-forbidden Fierz interference terms $b$, derived from electron-energy dependence of the neutron $\beta$-asymmetry $A(E_e)$ with $1\sigma$ (dark gray), $2\sigma$ (gray), and $3\sigma$ contours (light gray) contours. The Standard Model prediction is $b = 0$. The vertical line indicates $\lambda$ from Eq. (6.42)

Often one does not have the full energy dependence $X_m(E_e)$ at hand, but only the published values of the correlation coefficients $X$. In this case the parameters $b$ and $\lambda$ can be derived from two or more such coefficients, see Glück *et al.* (1995). One can take for instance the measured values $a$ and $A$ and apply the appropriate energy averages $m_e \langle W_e^{-1} \rangle$ in Eq. (6.68), for numerical values see Appendix 6 of Wilkinson (1982). However, $a$ and $A$ are very collinear in $b$ and $\lambda$, too, and they constrain $b$ only to $-0.3 < b < 0.7$ at $2\sigma$. Nor does inclusion of the neutron lifetime $\tau_n$, with $|V_{ud}|^2$ as additional free parameter, give much better constraints.

Things are better if one includes the neutrino asymmetry $B$ and the proton asymmetry $C$, which are more "orthogonal" with respect to the Fierz terms than are $A$ and $a$. However, the neutrino asymmetry $B$ contains a second Fierz term $b'$ in the numerator,

$$B' = \frac{B + (m_e/W_e)b'}{1 + (m_e/W_e)b}, \quad (6.70)$$

with

$$b' = 2\lambda \frac{2\lambda T - S - T}{1 + 3\lambda^2} \quad (6.71)$$

from Jackson *et al.* (1957), with $S$ and $T$ defined as for Eq. (6.67). Therefore a third parameter is needed if one wants to extract the two basic quantities $g_S$ and $g_T$ in Eq. (6.67), for instance the newly measured proton asymmetry $C$, using Glück (1996). A fit to the coefficients $A$, $B$, $C$, with the $\chi^2$ minimum projected onto the $b-b'$ plane, gives

$$-0.3 < b, b' < 0.5$$
(95% C..L., neutron from $A$, $B$, $C$). (6.72)

For $B$ only the value $B = 0.9821 \pm 0.0040$ of Serebrov et al. (1998) was used, because $B$ and $C$ from Schumann et al. (2007, 2008) are partially based on the same data set. These limits translate into limits on the left-handed coupling constants $|g_T/g_A| < 0.20$ and $|g_S/g_V| < 0.25$, 95% C.L.

The neutron limits on the these quantities improve considerably if one includes the ratio of neutron and nuclear superallowed $ft$-values $\mathcal{F}t_n/\mathcal{F}t_{0\to0}$, with $\mathcal{F}t_n = f^R \tau_n \ln 2$, which gives

$$|b| < 0.03 \text{ and } |b'| < 0.02$$
(95% C.L., from $A$, $B$, $C$, $\mathcal{F}t_n/\mathcal{F}t_{0\to0}$), (6.73)

see Fig. 26. From this we derive the limits on the amplitudes $g_S$, $g_T$ for a left-handed $S$-$T$ sector

$$-0.23 < g_S/g_V < 0.08, \quad -0.02 < g_T/g_A < 0.05$$
(95% C.L., left-handed, Fierz interference). (6.74)

Similar limits for $g_S$ and $g_T$, based on a different neutron and nuclear data set, were derived by Faber et al. (2009), and by Konrad et al. (2010), see also the discussion in Glück et al. (1995).

In this last derivation one has to assume that the nuclear Fierz term is negligible compared to the neutron value, which is not necessarily fulfilled, the $1\sigma$ limit on $b$ from a systematic analysis of superallowed nuclear $ft$-values (Hardy and Towner, 2009) being

$$b = -0.0022 \pm 0.0026 \text{ (nuclear } \mathcal{F}t_{0\to0}), \quad (6.75)$$

while $b'$ is not accesssible from nuclear data.

Making use of Eqs. like (6.19) for further neutron decay correlations, one can isolate additional left-handed amplitudes linear in $g_S$ or $g_T$, and finds relations like

$$H + \frac{m_e}{W_e} a = -2 \frac{S}{1+3\lambda^2}, \quad K + A = 2 \frac{S - \lambda^2 T}{1+3\lambda^2}. \quad (6.76)$$

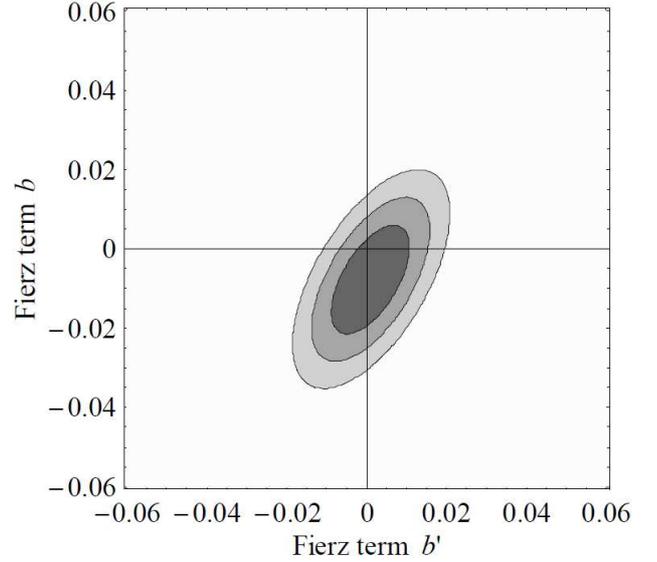

FIG. 26. Exclusion plot for the SM-forbidden Fierz interference terms $b$ and $b'$ derived from the PDG 2010 values of the parameters parameters $A$, $B$, $C$, and $\mathcal{F}t_n/\mathcal{F}t_{0\to0}$. The Standard Model prediction is $b = b' = 0$.

b. *Right-handed S or T coupling*: Scalar and tensor couplings could as well be right-handed, in which case the Fierz terms $b$ and $b'$ vanish. Right-handed $S$ and $T$ amplitudes enter the correlation coefficients only quadratically or bilinearly in $g_S$ and $g_T$. Therefore experimental limits on $S$ and $T$ couplings are less tight for right-handed than for left-handed couplings, see Eqs. (6.65) and (6.66). For the $\beta$-asymmetry $A$, for instance, in the notation of Eq. (6.59) one finds

$$\varepsilon_{ST} = -2\lambda(\lambda T^2 + ST), \quad \varepsilon'_{ST} = S^2 + 3\lambda^2 T^2$$
(right-handed $S$, $T$ coupling) (6.77)

For the electron helicity $G$, $\varepsilon_{ST} = \varepsilon'_{ST} = S^2 + 3\lambda^2 T^2$. In both cases, imaginary Coulomb corrections linear in $S$ and $T$ have been neglected.

Limits on right-handed $g_S$ and $g_T$ from neutron decay experiments are given in Schumann (2007) as

$$|g_S/g_V| < 0.15, \quad |g_T/g_A| < 0.10$$
(95% C.L., neutron right-handed). (6.78)

In the purely leptonic sector, bounds on $S$ and $T$ couplings from muon decay are of similar quality as the neutron bounds, with

$$|g_S/g_V| < 0.55, \quad |g_T/g_V| \equiv 0$$
(90% C.L., muon, left-handed), (6.79)

and

$|g_S/g_V| < 0.074$, $|g_T/g_V| < 0.021$
(90% C.L., muon, right-handed), (6.80)

see Fetscher and Gerber (2009).

c. *Bound $\beta$-decay*: Neutron decay into a hydrogen atom is one of the two rare allowed decay modes discussed in Sec. A.5.b above. In principle, one can also derive limits on scalar and tensor amplitudes from bound $\beta$-decay. The $W_i$ from Eq. (6.25) are linearly sensitive to scalar $S$ and tensor $T$ amplitudes in the combinations $g_S - g_T$, and $g_S + 3g_T$, see Faber *et al.* (2009).

**2. Was the universe left-handed from the beginning?**

Today, many models beyond the SM start with a left-right symmetric universe, and the left-handedness of the electroweak interaction arises as an "emergent property" due to a spontaneous symmetry breaking in the course of a phase transitions of the vacuum in the early universe. In the simplest case, the gauge group $SU(2)_L$ of the SM is replaced by $SU(2)_L \times SU(2)_R$, which then is spontaneously broken. This should lead to a mass splitting of the corresponding gauge boson, namely, the left-handed $W_1$ and the right-handed $W_2$, with masses $m_L$, $m_R$, and mass ratio squared $\delta = (m_L/m_R)^2 \ll 1$.

If, by additional symmetry breaking, the mass eigenstates $W_1$ and $W_2$ do not coincide with the electroweak eigenstates $W_L$ and $W_R$, then

$$W_L = W_1 \cos\zeta + W_2 \sin\zeta,$$
$$W_R = -W_1 \sin\zeta + W_2 \cos\zeta,$$  (6.81)

with left-right mixing angle $\zeta$ (where we omit an overall phase factor). When right-handed currents are admitted in the weak Hamiltonian, then neutron decay parameters depend not only on $|V_{ud}|$ and $\lambda$, but also on $\delta$ and $\zeta$, each parameter bilinear in $\delta$ and $\zeta$, but each in a different way. For references, see Severijns *et al.* (2006), Glück *et al.* (1995), and Dubbers *et al.* (1990) for earlier work.

a. *Limits on right-handed currents from neutron decay*: We use the framework of so-called manifest left-right symmetry, Beg *et al.* (1977) and Holstein and Treiman (1977). Right-handed $V+A$ currents there are obtained simply by replacing $-\gamma_5$ by $+\gamma_5$ in the formulae for the left-handed $V-A$ currents. We derive $\delta = (m_L/m_R)^2$, $\zeta$, and $\lambda$ from three measurements, for instance from $A$, $B$, $C$, which each depend differently on these parameters. For the $\beta$-asymmetry $A$, one finds

$$\varepsilon_{rhc} = 2\lambda(\delta-\zeta)(\delta+\zeta) + 2\lambda^2(\delta+\zeta)^2,$$
$$\varepsilon'_{rhc} = (\delta-\zeta)^2 + 3\lambda^2(\delta+\zeta),$$  (6.82)

and similarly for other correlation coefficients, see, for instance Appendix C.3. of Severijns *et al.* (2006).

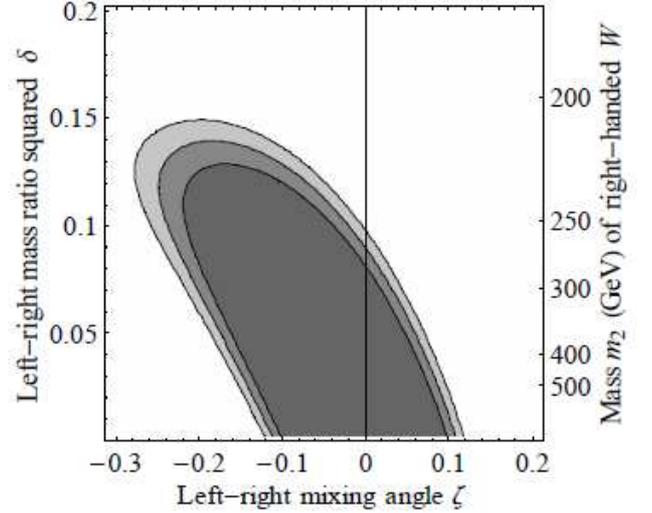

FIG. 27. Exclusion plot for the parameters of the manifest left-right symmetric model (beyond the SM), namely, the mass ratio squared $\delta = (m_R/m_L)^2$ of left and right-handed $W$ bosons, and their relative phase $\zeta$. The exclusion contours, defined as in Fig. 25, are based on the neutron decay correlation coefficients $A$, $B$, and $C$. The Standard Model prediction is $\delta = \zeta = 0$.

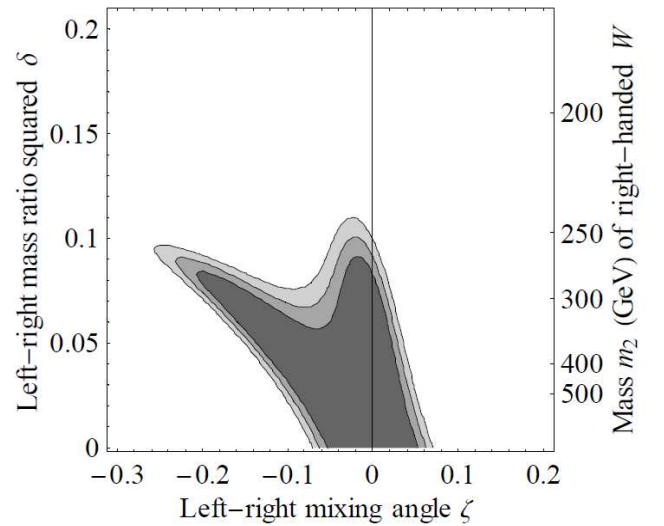

FIG. 28. Exclusion plot as in the preceding figure, but based on the coefficients $A$, $B$, $C$, and $\mathcal{F}t_n/\mathcal{F}t_{0\to0}$.

Figure 27 shows the result of such a fit, with $\lambda$, $\delta$, and $\zeta$ as free parameters, projected onto the $(\delta, \zeta)$ plane, using the PDG 2010 values of $A$, $B$, and $C$ (for $B$, as above, only the value of Serebrov *et al.* (1998) was used). If in addition we use the ratio of neutron to nuclear superallowed $\mathcal{F}t$ values, $\mathcal{F}t_n / \mathcal{F}t_{0 \to 0}$ from Eqs. (6.49) and (6.51), we obtain the exclusion plot of Fig. 28, both in good agreement with the SM expectation $\delta = \zeta = 0$. The limits on the mass $m_R$ of the right-handed $W_R$ and the mixing angle $\zeta$ are derived as

$$m_R > 250 \text{ GeV}, \quad -0.23 < \zeta < 0.06$$
(95% C.L., neutron). (6.83)

The neutron limits on the right-handed parameters $\delta$ and $\zeta$, impressive as they are, have not improved during past years. Again, the picture looks somewhat better if one uses only the latest results for $\tau_n$ and $\lambda$, but this we leave to future evaluations.

b. *Limits on right-handed currents from nuclear physics*: Severijns *et al.* (2006) reviewed searches for right-handed currents, which are a traditional topic also in nuclear physics. As $\mathcal{F}t' = \mathcal{F}t / (1 - 2\zeta)$ is linear in $\zeta$, tight limits are obtained for $\zeta$ from nuclear $0^+ \to 0^+$ transitions,

$$-0.0006 \leq \zeta \leq 0.0018 \quad (90\% \text{ C.L., nuclear } \mathcal{F}t).$$
(6.84)

From the longitudinal polarization of $\beta$-particles emitted from unpolarized and polarized nuclei, limits of order several times $10^{-4}$ are derived for the bilinear forms $\delta\zeta$ and $(\delta + \zeta)^2$, respectively, from which a lower limit on the mass of a right-handed $W$ is obtained of comparable size as that of the neutron, Eq. (6.83)

$$m_R > 320 \text{ GeV}$$
(90% C.L., nuclear $\beta$-helicity), (6.85)

for details see Severijns *et al.* (2006), and references therein.

c. *Limits on right-handed currents from high energy physics*: We can compare these results with those from direct high-energy searches for $W_R$. The limit for the mass of $W_R$ is

$$m_R > 715 \text{ GeV} \quad (90\% \text{ C.L., high energy}).$$
(6.86)

The limits derived from muon decay are

$$m_R > 549 \text{ GeV}, \quad |\zeta| < 0.022$$
(68% and 90% C.L., muon). (6.87)

We see that the limits on the mass of the right-handed $W_R$ derived from low-energy experiments are somewhat lower but of the same order as present limits from high-energy experiments. The limits from both sources complement each other, for instance the neutron tests the hadronic sector, while the muon tests the leptonic sector. The high-energy limits for direct production are valid for right-handed neutrinos $\nu_R$ with masses up to the mass of $W_R$, whereas low-energy limits require the $\nu_R$ mass to be below the relevant $\beta$-endpoint energy (the sum of the masses of the ordinary left-handed neutrinos being at most of order eV).

d. \**Right-handed currents from bound $\beta$-decay*: In neutron decay into a hydrogen atom, a unique experiment on right-handed currents may become feasible, see Byrne (2001) and references therein. The population of the hyperfine level $F = 1$, $M_F = -1$, is predicted to be $W_4 = 0$, Eq. (6.25), for any left-handed interaction, simply from angular momentum conservation. If a population $W_4 \neq 0$ is detected, then this would be a unique sign for the existence of right-handed currents, so to speak a "$g-2$ type" experiment for $V + A$ amplitudes. This, however, requires that optical transitions from $n > 2$ to $n = 2$ atomic states are suppressed.

Let us estimate the sensitivity of this method. In a left-right symmetric $V + A$ model as discussed above, with left-right mass ratio $\delta = (W_L / W_R)^2$ and phase $\zeta$, one expects a population $W_4 = \frac{1}{2}[(\lambda + 1)\delta + (\lambda - 1)\zeta]^2 / (1 + 3\lambda^2)$, where $\delta$ and $\zeta$ are linked to the antineutrino helicity $H_{\bar{\nu}}$ as $1 - H_{\bar{\nu}} = 2[(\delta - \zeta)^2 + 3\lambda^2(\delta + \zeta)^2] / (1 + 3\lambda^2)$. For either $\zeta^2 \ll \delta^2$ or $\zeta^2 \gg \delta^2$, one finds

$$W_4 = \frac{(1 + \lambda)^2}{4(1 + 3\lambda^2)}(1 - H_{\bar{\nu}}) = 3.1 \times 10^{-3}(1 - H_{\bar{\nu}}), \quad (6.88)$$

hence, the sensitivity to deviations of neutrino helicity from unity is weak, but, hopefully, background-free.

### 3. Time reversal invariance

In Sec. III, we extensively discussed *CP* violation. Under *CPT* invariance, *CP* violation goes hand in hand with *T* violation, and we need not discuss this topic again in any depth. Limits on time reversal invariance exist both from nuclear and particle physics, but none better than the neutron limit $D = (-4 \pm 6) \times 10^{-4}$ from Eq. (6.39).

In the SM, merely a $D < 10^{-12}$ would enter via the CKM phase $\varphi$, hence $D$ is sensitive to effects beyond the SM. Leptoquarks could induce a $D$ of size up to the present experimental limit. Left-right symmetry could produce a $D$ one order of magnitude, and supersymmetry a $D$ two orders of magnitude below the present bound, as cited in Lising *et al.* (2000). Final state effects begin only two orders of magnitude below the present neutron bound.

The $T$-violating $R$-coefficient for the neutron is $R = (8 \pm 16) \times 10^{-3}$, from Eq. (6.41), while for $Z^0$-decay it is

$$-0.022 < R < 0.039 \quad (90\% \text{ C.L., } Z^0), \quad (6.89)$$

from Abe *et al.* (1995), and for muon $\beta$-decay

$$R = (-1.7 \pm 2.5) \times 10^{-3} \quad (\text{muon}), \quad (6.90)$$

from Abe *et al.* (2004). The best $R$-coefficient from nuclear $\beta$-decay is that from the $\beta$-decay of polarized $^8$Li,

$$R = (-0.9 \pm 2.2) \times 10^{-3} \quad (^8\text{Li}), \quad (6.91)$$

from Huber *et al.* (2003).

For most theoretical models, EDM experiments are more sensitive to $CP$ and $T$ reversal violations than are nuclear and neutron correlation experiments. The neutron $R$-coefficient is discussed by Yamanaka *et al.* (2010) in effective field theory within the framework of the MSSM. They find interesting relations (their Table 2) between the expected size of $R$ and the sizes of other $T$ violating neutron decay correlation coefficients $L$, $S$, $U$, and $V$, as well as corrections to $B$, $H$, $K$, $N$, $Q$, and $W$.

Other tests beyond the Standard Model are possible in neutron decay, for instance on leptoquark exchange, exotic fermions or sterile neutrinos, and we refer to the theory reviews cited above. Thus, neutron $\beta$-decay provides a wealth of weak interaction parameters that are very useful to investigate extensions of the Standard Model up to high energy scales.

# VII. THE ROLE OF THE NEUTRON IN ELEMENT FORMATION

The next epoch in the history of the universe where neutrons play a dominant role is reached when the temperature has dropped to about 1 MeV. In Sec. A, we discuss the role of neutrons in the production of the first light elements during the first few minutes of the universe, a process called primordial nucleosynthesis. We adopt a simplified standard description of the process, for a more elegant derivation see the book by Kolb and Turner. In the subsequent Sec. B, the formation of the heavier elements in stellar processes that started half a billion years later (and is still going on) will be described.

## A. Primordial nucleosynthesis

### 1. The neutron density in the first three minutes

Up to a time $t \sim 1$ s, corresponding to a temperature $kT \sim 1$ MeV, the expanding universe was in thermal equilibrium. It was mainly filled with photons, three kinds of neutrinos/antineutrinos, electrons and positrons, and a tiny $\sim 10^{-10}$ fraction of baryons from $CP$ violating baryogenesis in the very early universe, frozen out as protons and neutrons during the quark-gluon phase transition at $t \sim 3 \times 10^{-5}$ s. At $t < 1$ sec, protons and neutrons were coupled to each other by the weak neutrino reactions

$$\begin{aligned} n + e^+ &\leftrightarrow p + \overline{\nu}_e, \\ n + \nu_e &\leftrightarrow p + e^- \end{aligned} \quad (7.1)$$

accompanied by neutron decay

$$n \rightarrow p + e^- + \overline{\nu}_e. \quad (7.2)$$

These processes involve all members $d$, $u$, $e$, $\nu_e$ of the first generation of particles. One can show that dark matter and dark energy did not influence primordial nucleosynthesis, and that the hot photon environment did not change the phase space available for free-neutron decay significantly.

The neutron mass exceeds the proton mass by $\Delta mc^2 = (m_p - m_n)c^2 = 1.293$ MeV. Therefore the reactions in Eq. (7.1) are exothermic when read from left to right, and endothermic in the opposite sense. Free neutron decay, Eq. (7.2), proceeded only to the right because three-particle encounters are unlikely at this stage. These reactions kept neutrons and protons in chemical equilibrium. Their number ratio ($\approx$ mass ratio) at temperature $T$ is approximately given by the Boltzmann law $n/p = \exp(-\Delta m/kT)$, where for simplicity we use the same symbol for the number of protons and neutrons as for the particles themselves.

The dashed line in Fig. 29 gives the neutron mass fraction $n/(n+p)$ from the Boltzmann law as a function of temperature and time, see Eq. (2.4). On the right of this figure, for $t \ll 1$ s, $n \approx p$ and the neutron mass fraction goes to one half. If, for $t > 1$ s, neutrons and protons were kept in equilibrium, the neutrons would disappear with falling temperature within a few seconds, and Boltzmann's law would predict that the universe would consist entirely of protons, and no element other than hydrogen would populate the universe today.

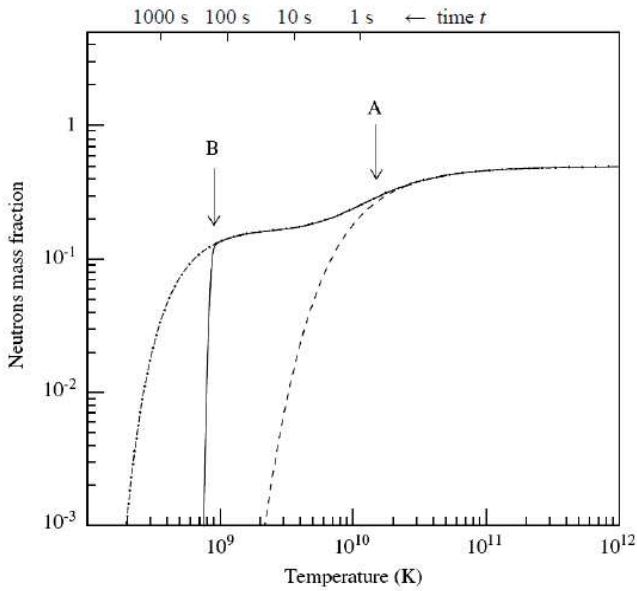

FIG. 29. Neutron mass fraction $n/(n+p)$ in the early universe, as a function of temperature $T$ and time $t$ (upper scale) after the big bang. For the dashed line it is assumed that neutrons and protons are permanently in thermal equilibrium, with $n/p$ following Boltzmann's law $n/p = \exp(-\Delta m/kT)$. For the dash-dotted line it is assumed that neutrons decouple from the protons near arrow A and decay freely. Nucleosynthesis sets in near arrow B as soon as deuteron break-up by photons ceases. The solid line is from a full network calculation with all relevant nuclear reactions. From Coc (2009).

Fortunately, the weak neutrino cross sections in Eq. (7.1) are small enough that, in the expanding and cooling universe, these reactions were stopped early enough at freeze-out time $t_f \approx 2$ s when still a sizable number of neutrons was present, with

$$n/p\big|_{t=t_f} = \exp(-\Delta m/kT_f) \quad (7.3)$$

where $T_f = T(t_f)$ is the freeze-out temperature. As we shall see below, this ratio is $n/p \approx 1/6$, with $kT_f \approx 0.7$ MeV. At this time and temperature (arrow A in Fig. 29), the expanding system fell out of thermal equilibrium because nucleons and neutrinos did not find each other anymore, and the universe became transparent to neutrinos.

The strong-interaction cross sections are much larger than the weak neutrino cross sections in Eq. (7.1), so the nucleons in the expanding universe still had sufficient time to find each other and fuse. However, nuclear fusion of protons and neutrons only started about three minutes after freeze-out, because the first nuclei formed in the hot universe were unstable in the universe's strong photon field. (Nucleosynthesis was delayed mainly by the high entropy of the thermal $\gamma$ radiation, due to radiation dominance, and less by the low binding energy of the deuteron.) This suppression of nucleosynthesis was effective until the temperature had dropped to below 0.1 MeV (arrow B in Fig. 29). By that time, the neutron to proton ratio Eq. (7.3) had dropped to a new value $n'/p'$, with $n+p = n'+p'$, due to free-neutron $\beta^-$-decay and to continuing neutrino interactions, which latter have a phase space available that is much larger than for neutron decay. The time interval between the arrows A and B in Fig. 29 is $t_d \approx 150$ s (Weinberg's famous "first three minutes"). The process of $^4$He production stopped shortly after when almost all neutrons had ended up bound in $^4$He.

We can, in a rough approximation, determine the early neutron-to-proton fraction Eq. (7.3), and with it the freeze-out temperature $T_f$, directly from the observed primordial $^4$He mass fraction $Y_p$. We just have to take into account that essentially all neutrons finally wound up in $^4$He nuclei (which is due to the large binding energy of $^4$He). As about half of the $^4$He mass is from neutrons, and all neutrons are bound in $^4$He, the $^4$He mass fraction is twice the neutron mass fraction, $Y_p = 2n'/(n'+p')$ at $t = 150$ s. Using Eq. (7.3), and the approximation $n' \approx n\exp(-t_d/\tau_n)$, this gives

$$Y_p = \frac{\text{He-4 mass}}{\text{total mass}} \approx \frac{2n'}{n'+p'} = \frac{2}{1+p'/n'}$$
$$\approx \frac{2\exp(-t_d/\tau_n)}{1+\exp(\Delta m/kT_f)}. \quad (7.4)$$

The $^4$He mass fraction $Y_p$ can be observed in regions of the universe with very low metallicity, i.e., very low density of elements beyond $A = 4$, which points to a very old age of the population observed. The measured values then are extrapolated to zero metallicity. (Astrophysicists call metals what nuclear physicists call heavy ions – everything beyond helium.) The observed value is $Y_p \approx 25\%$. At the beginning of nucleosynthesis the neutron to proton ratio becomes $n'/p' = Y_p/(2-Y_p) \approx 1/7$, from Eq. (7.4), from which one can calculate back $n/p \approx 1/6$ at the time of decoupling of the weak interaction, which gives $kT_f \approx 0.7$ MeV from Eq. (7.3) that we have mentioned above.

## 2. Nuclear reactions in the expanding universe

Altogether, the four main isotopes formed in primordial nucleosynthesis via various fusion reactions are $^2$H, $^3$He, $^4$He, and $^7$Li. From the known nuclear reaction data for these fusion processes one can calculate the mass ratios $^2$H/H, $^3$He/H, $Y_p$, and $^7$Li/H. The path to heavier nuclides is blocked by the absence of stable $A=5$ and $A=8$ nuclei and by the entropy-driven time delay in nucleosynthesis mentioned above (at later time temperature is lower and Coulomb barriers are more difficult to overcome).

For numerical and observational results on primordial abundances, we use the recently published analyses by Iocco *et al.* (2009) and Cyburt *et al.* (2008). A simplified but still rather long analytical derivation of the nucleosynthesis process, precise to several percent, was given by Bernstein *et al.* (1989). In the following, we shall present an even simpler description that will help us understanding the main parameter dependences of the $^4$He abundance $Y_p$, with main emphasis on its dependence on the neutron lifetime.

Let $\Gamma$ be the rate, at which the reactions Eq. (7.1) occur. As it will turn out, the reaction rate $\Gamma(t)$ decreases with time much faster than the expansion rate, given by the Hubble function $\mathcal{H}(t)$. As a rule of thumb, which works surprisingly well, the system falls out of equilibrium at crossover $\Gamma(t_f) = \mathcal{H}(t_f)$. (At crossover, the time $\Gamma^{-1}$ between successive reactions equals the momentary age $t \sim \mathcal{H}^{-1}$ of the universe, and less than one reaction is expected within $t$.) At the time of crossover, the freeze-out temperature is $T_f = T(t_f)$, so we derive $T_f$ from the condition

$$\Gamma(T_f) = \mathcal{H}(T_f). \tag{7.5}$$

To calculate the reaction rate $\Gamma$ in the expanding universe on the left-hand side, we must carefully integrate the appropriate rate equations over time and over neutrino energy $E_\nu$. Instead, we make the simplified ansatz of a momentary total neutrino reaction rate proportional to the local neutron density $\Gamma \approx \sigma_\nu n_\nu c$. The neutrino number density is $n_\nu = \rho_r/E_\nu \propto T^3$ from Eq. (2.5), and we have set $E_\nu \propto T$, dropping all thermal averages and writing $E_\nu$ for $\langle E_\nu \rangle$. The neutrino-nucleon cross section $\sigma_\nu$ is too small to be measured precisely in the laboratory. Instead we must calculate $\sigma_\nu \propto 1/\tau_n$ from the measured neutron lifetime $\tau_n$, which is based on the fact that the processes in Eqs. (7.1) and (7.2) all have the same Feynman diagram. Furthermore, for $E_\nu \gg m_e c^2$, $\sigma_\nu$ is known to grow with the square of neutrino energy $E_\nu$, so $\sigma_\nu \propto T^2/\tau_n$, and altogether $\Gamma \propto T^5/\tau_n$.

The expansion rate $\mathcal{H}$ on the right-hand side of Eq. (7.5) is, from Eqs. (2.1) and (2.5), $\mathcal{H} \propto N^{1/2}T^2$. The freeze-out condition (7.5) then gives $T_f^5/\tau_n \propto \Gamma = \mathcal{H} \propto N^{1/2}T_f^2$. At freeze-out time the number of relativistic degrees of freedom (dof) is $N = \frac{1}{4}(22 + 7N_\nu)$. This contains 2 dof for the 2 helicity states of the photon, $4 \times \frac{7}{8}$ dof for the 2 helicity states of $e^+$ and $e^-$, and $2N_\nu \times \frac{7}{8}$ dof for the 1 helicity state of $N_\nu$ types of neutrinos and antineutrinos. The relativistic weight factors employed are 1 for bosons and $\frac{7}{8}$ for fermions. This gives

$$T_f \approx const \times (22 + 7N_\nu)^{1/6}\tau_n^{1/3}, \tag{7.6}$$

where *const* is some known combination of natural constants and masses. This equation is not meant to provide us with precise values of $T_f$ and $Y_p$. Instead, we use it to investigate the sensitivity of $Y_p$ to the parameters $\tau_n$ and $N_\nu$.

## 3. Neutron lifetime, particle families, and baryon density of the universe

To find the sensitivity of the $^4$He abundance $Y_p$ to the value of the neutron lifetime, we form $(\partial Y_p/Y_p)/(\partial \tau_n/\tau_n)$ from Eq. (7.4), with $T_f$ from Eq. (7.6). At time $t_d = 150$ s, with $\tau_n = 882$ s and $Y_p \approx 25\%$, this sensitivity becomes $\Delta Y_p/Y_p = +(0.50 + 0.17)\Delta\tau_n/\tau_n$, where the first number +0.50 is mainly due to the dependence of the neutrino cross section on the neutron lifetime, and the second number +0.17 is due to free-neutron decay up to time $t_d$. This result is just several percent below the elaborate results of Iocco *et al.* (2009) and Bernstein *et al.* (1989), who both find

$$\Delta Y_p/Y_p = +0.72\,\Delta\tau_n/\tau_n. \tag{7.7}$$

The plus sign in Eq. (7.7) is as expected, because a longer neutron lifetime gives a smaller neutrino cross section and an earlier freeze-out time with more neutrons and hence more helium. Furthermore, in the first three minutes free-neutron decay is slower for longer $\tau_n$ and leads to more helium.

From Eq. (7.7) we can derive how well we need to know the neutron lifetime, when we require that the error due to the uncertainty in $\tau_n$ is small compared to the observational error of $Y_p$. As discussed in Sec. VI.B.1, measurements of the neutron lifetime over past years differ by almost one percent. Hence, in the extreme case that the PDG 2010 average would need to be shifted by $\Delta \tau_n / \tau_n = -0.008$, this would shift the calculated value $Y_p = 0.2480 \pm 0.0003$, as adopted by Iocco et al. (2009), by $\Delta Y_p = -0.0015$.

From Eq. (7.6), we also find the sensitivity to changes in the number $N_\nu$ of light species. Taking the derivative of $Y_p$ in Eq. (7.6) with respect to $N_\nu$ we obtain $\Delta Y_p = +0.010 \, \Delta N_\nu$ for $N_\nu = 3$ neutrino species and $Y_P \approx 1/4$. Bernstein et al. find the same value +0.010, to which they add +0.004 to account for a small $N_\nu$-dependence of the deuterium bottleneck, so $\Delta Y_p = +0.014 \, \Delta N_\nu$, or

$$\Delta Y_p / Y_p = +0.17 \, \Delta N_\nu / N_\nu. \qquad (7.8)$$

Again, the sign is positive because energy density, Eq. (2.5), increases with the number of light species $N_\nu$, which speeds up expansion, Eq. (2.1), and leads to earlier freeze-out at higher neutron number.

For the number of light neutrino species we use the value $N_\nu = 3$, with $\Delta N_\nu / N_\nu = 0.0027$, when we take $N_\nu$ and $\Delta N_\nu$ from the $Z^0$-resonance, which gives $N_\nu = 2.9840 \pm 0.0082$. From Eq. (7.8) we then find $\Delta Y_p = \pm 0.0001$.

As already mentioned, the ratio $\eta = n_b / n_\gamma$ of baryon (i.e., nucleon) to photon density, Eq. (3.1), is another important parameter because a high photon density delays nucleosynthesis. Bernstein et al. find $\Delta Y_p = +0.009 \, \Delta \eta / \eta$ at $Y_P \approx 1/4$, for $\Delta \eta \ll \eta$, in agreement with Iocco et al., who find

$$\Delta Y_p / Y_p = +0.039 \, \Delta \eta / \eta \qquad (7.9)$$

at $\eta = 6.22 \times 10^{-10}$. (The relation $\eta = 273 \, \Omega_b h^2$ can be used to compare results from different sources). The sign in Eq. (7.9) is positive because a larger $\eta = n_b / n_\gamma$ means a lower number of photons, hence less delay in nucleosynthesis and more helium formation. The baryon-to-photon ratio obtained from the cosmic microwave background satellite data (WMAP 5-year result, Dunkley et al., 2009) is $\eta = (6.225 \pm 0.170) \times 10^{-10}$, or $(\Delta \eta / \eta)_{\text{obs}} = 0.027$. From Eq. (7.9) we then find $\Delta Y_p = \pm 0.00025$ due to the error in the baryon density of the universe.

To sum up, big bang nucleosynthesis is a parameter-free theory because the three parameters $\tau_n$, $N_\nu$, and $\eta$ are all known from measurement. The error in the neutron lifetime dominates the error in the calculated $Y_p$, and may require a shift of $\Delta Y_p \approx -0.0015$. Iocco et al. adopted the observed value $Y_{p\,\text{obs}} = 0.250 \pm 0.003$, and the reanalysis by Cyburt et al. (2008) gives a very similar value, $Y_{p\,\text{obs}} = 0.252 \pm 0.003$ (while earlier authors prefer wider error margins). Hence, the theoretical uncertainty from the neutron lifetime may be about half the observational error of $Y_p$. This is not dramatic, but the field keeps progressing, and if we want the errors due to nuclear inputs to remain negligible, then the problems with the neutron lifetime should be fixed.

### 4. Light-element abundances

There is also good agreement between the predicted abundance of deuterium $(^2\text{H/H})_{\text{calc}} = (2.53 \pm 0.12) \times 10^{-5}$, calculated with $N_\nu = 3$ and the $\eta$ from WMAP, and the observed abundance $(^2\text{H/H})_{\text{obs}} = (2.87 \pm 0.22) \times 10^{-5}$, see, for instance, Iocco et al. (2009). For $^3$He the calculated abundance is $(^3\text{He/H})_{\text{calc}} = (1.02 \pm 0.04) \times 10^{-5}$, while only an upper bound can be given for the observed abundance $(^3\text{He/H})_{\text{obs}} < 1 \times 10^{-5}$. The agreement is not as good for $^7$Li, with $(^7\text{Li/H})_{\text{calc}} = (4.7 \pm 0.5) \times 10^{-10}$ vs. $(^7\text{Li/H})_{\text{obs}} = (1.9 \pm 1.3) \times 10^{-10}$, which is possibly due to $^7$Li fusion and consumptions in stars. The update by Cyburt et al. (2008) comes, within errors, to the same results. The overall concordance of the data is astonishingly good, and constitutes a strong pillar of the standard big bang model.

One can also use the number of neutrino species $N_\nu$ and/or the baryon content $\eta$ as free parameters in primordial nucleosynthesis calculations. When both $N_\nu$ and $\eta$ are free parameters, from Fig. 12 of Iocco et al. one finds

$$N_\nu = 3.18 \pm 0.22 \text{ and } \eta = (5.75 \pm 0.55) \times 10^{-10}$$
(from nucleosynthesis), $\qquad (7.10)$

compatible with $N_\nu = 3$ and the WMAP values

$N_\nu = 3.18 \pm 0.44$ and $\eta = (6.22 \pm 0.17) \times 10^{-10}$
(from microwave background). (7.11)

The two independent values for $\eta$ can be combined to $\eta = (6.08 \pm 0.14) \times 10^{-10}$ as used in Eq. (3.1). These studies are interesting because with Eqs. (7.10) and (7.11) we compare $N_\nu$ at time $t = 1\,\text{s}$ with $N_\nu$ measured today at $t = 1.4 \times 10^{10}\,\text{y}$, and we compare $\eta$ and $N_\nu$ at time $t = 1\,\text{s}$ with $\eta$ at a time $t = 3.7 \times 10^5\,\text{y}$ when the universe became transparent to photons, and find that the numbers have not changed considerably in between.

From these data on primordial nucleosynthesis many stringent and often unique constraints on new physics can be derived, for instance on extra dimensions, on the time dependence of fundamental constants, extra light species, and others, as is also discussed by Iocco *et al.* (2009), and references therein. To sum up, the early universe is a rich laboratory for the study of the fundamental interactions with a continuing need for precise nuclear and neutron data.

## B. Stellar nucleosynthesis

More than half a billion years later, the first heavier elements were created in protostars: The first conglomerates of primordial hydrogen and helium atoms as formed under the catalytic action of dark matter. As we shall see, in the ensuing stellar processes of element formation, neutron reactions are dominant, too. To understand these processes of stellar nucleosynthesis more and better neutron-nuclear data are needed.

### 1. Stellar nucleosynthesis and the *s*-process

Under gravitational contraction, the star heats up until nuclear fusion sets in. This nuclear burning, which prevents further contraction of the star, starts with the breeding of $^4$He via the *p-p* chain in the sun, or via the CNO-cycle in heavier stars.

In past years, nuclear processes within the sun have received much attention, in particular, in the context of the so-called solar-neutrino problem: Only one third to one half of the neutrinos generated by weak-interaction processes in the sun arrive as such on earth, the others are converted into other neutrino flavors by so-called neutrino-oscillation processes.

The *p-p* chain starts with the weak-interaction fusion reaction $p + p \to {}^2\text{H} + e^+ + \nu_e$. To come close to each other, the Pauli principle requires the incoming protons to have antiparallel spins, while the outgoing deuteron $^2$H is known to have spin $j = 1$. Therefore *p-p* fusion is a $|\Delta j| = 1$ Gamow-Teller transition, whose cross section is proportional to $g_A^2$, where $g_A$ is the axial vector weak-coupling constant as defined in Sec.VI.A.3.

The most frequently studied solar neutrino process is the $\beta^+$-decay ${}^8\text{B} \to {}^8\text{Be} + e^+ + \nu_e$, whose continuous neutrino spectrum reaches up to 14 MeV. The $^8$B neutrino flux $\Phi_8$ is the dominant component in the original Homestake neutrino experiment (detection threshold 0.8 MeV) and is the only component seen in the Kamiokande experiments (threshold 4.5 MeV). The $^8$B neutrino flux decreases with $g_A$ as $\Phi_8 \propto g_A^{-5.2}$, see for instance Eq. (6c) in Adelberger *et al.* (1998), so

$$\Delta\Phi_8 / \Phi_8 = -5.2\, \Delta g_A / g_A \qquad (7.12)$$

(in a coupled multi-component reaction system, it is not unusual that the partial flux of one reaction component decreases with increasing coupling).

The cross sections of the weak reactions in the sun are too small to be measured directly, so again $g_A = (-1.2734 \pm 0.0019) g_V$ must be taken from free-neutron decay data, Eq. (6.42). Neutrino detection in solar neutrino experiments also needs input from experimental neutron decay data, in particular, when it relies on the inverse neutron decay reaction $p + \bar{\nu}_e \to n + e^+$. Neutrino fluxes are both calculated and measured with typically 10% precision, therefore the present quality of neutron decay data is quite sufficient, in spite of the strong sensitivity of $\Phi_8$ to the error of $g_A$.

After the *p-p* chain and the CNO cycle, other exothermic fusion reactions take place, for instance ${}^{12}\text{C} + {}^{12}\text{C} \to {}^{20}\text{Ne} + {}^4\text{He}$, followed by successive $\alpha$-captures along the $N = Z$ diagonal of the nuclear chart, up to the doubly magic isotope $^{56}$Ni, which decays back to stable $^{56}$Fe within several days via two successive $\beta^+$-transitions.

When $^{56}$Fe is reached, nuclear burning stops because $^{56}$Fe has the highest mean binding energy per nucleon of all isotopes $(E_B/A = 8.6\,\text{MeV})$, and hence for $^{56}$Fe both fission and fusion processes are endothermic. Iron, with an atomic ratio Fe/H $\approx 10^{-4}$, therefore is one of the most abundant heavier species in the universe, see the insert of Fig. 30. Beyond iron, almost all elements up to the heaviest actinides originate in stellar processes by successive neutron captures, followed by $\beta^-$-decays back towards the stable valley of isotopes. The capture of neutrons is the preferred reaction because neutrons see no Coulomb barrier. So here, too, neutrons play a decisive role, and a great amount of input data is required from experimental neutron and nuclear physics. We give only a short overview of this rich and interesting field because many excellent reviews exist on the subject, for instance Langanke and Martínez-Pinedo (2003), Kratz *et al.* (2007), and references therein.

Several different nucleosynthesis processes must exist in order to explain the observed nuclear abundances beyond $^{56}$Fe. The first such process is called the slow or *s*-process, other processes will be discussed in the following section. In the interior of stars like the sun, late in their lives as red giants, neutron fluxes of $10^{8\pm3}$ cm$^{-2}$s$^{-1}$ are created in $(\alpha,n)$ reactions like $^{13}$C + $^{4}$He $\rightarrow$ $^{18}$O + $n$, at temperatures of tens to hundreds of keV, and over time scales of thousands of years. As the neutron capture rate is small compared to the typical $\beta^-$-decay rate, the *s*-process usually leads to no more than one additional neutron away from the stable valley, as shown by the dark zigzag line along the valley of stability in Fig. 30, which shows an excerpt of the nuclear chart.

To understand the *s*-process, we must know the neutron capture cross sections and their energy dependence from $kT = 0.3$ keV up to several 100 keV. Such neutron cross sections are studied at dedicated pulsed neutron sources, mostly based on the $(p,n)$ reactions at small proton accelerators, or by $(\gamma,n)$ reactions with up to 100 MeV bremsstrahlung $\gamma$'s produced at large linear electron accelerators. Today the most prolific pulsed sources of fast neutrons are spallation sources with annexed time-of-flight facilities. To obtain the dependence of the capture cross section on neutron energy, prompt capture $\gamma$-rates are measured in dependence of the neutron's time-of-flight to the probe, details are given in the review by Käppeler *et al.* (2010). Besides the neutron cross sections one must also know the properties and decay modes of low-lying excited nuclear levels. For instance, the isotope $^{176}$Lu plays an important role as a nuclear thermometer for the *s*-process, and spectroscopic and lifetime measurements at LOHENGRIN at ILL were used to constrain temperatures and neutron fluxes during the *s*-process (Doll *et al.*, 1999). Such measurements then can be used to constrain stellar models.

## 2. Explosive nucleosynthesis and the *r*-process

The stellar *s*-process cannot reach all known stable isotopes, see Fig. 30, and stops when bismuth is reached, so other processes of nucleosynthesis must exist. The rapid or *r*-process takes place in regions of extremely high neutron densities, many orders of magnitude higher than for the *s*-process. In these high fluxes, multiple successive neutron captures take place, until extremely neutron rich isotopes with very short $\beta^-$ lifetimes are reached, up to 30 neutrons away from the valley of stability. The path of this *r*-process is not far away from the so-called neutron drip-line, beyond which nuclei are no longer bound systems.

The *r*-process must be due to an explosive event and takes no longer than about one second. When it stops, the entire *r*-process path is populated with neutron-rich nuclides that decay back to the stable valley under $\beta^-$ emission, see the dashed diagonal arrows in Fig. 30. The *r*-process path reaches up to the heaviest transuranium elements, which undergo spontaneous fission into two lighter nuclei located far from stability in the intermediate nuclear mass region.

The precise location of the *r*-process path on the nuclear chart depends on the prevailing neutron density and temperature. The vertical double line in Fig. 30 shows nuclear shell closure at magic neutron number $N = 50$. At shell closure, neutron cross sections are very small, and hence neutron absorption rates become smaller than the corresponding $\beta^-$-decay rates. When this happens, no more neutrons are added, and the *r*-process proceeds along the isotone (thick vertical line in Fig. 30) towards the valley of stability, up to the kink where $\beta^-$ lifetimes are again long enough for neutron captures to resume (thick diagonal line). As isotopes crowd at shell closure ("waiting point"), a peak in the element abundance appears near $A = 80$, marked "r" in the inset to Fig. 30.

A similar abundance peak, marked "s" in the inset, appears at somewhat higher $Z$ where the magic $N = 50$ isotone reaches the valley of stability, where neutron shell closure affects the *s*-process. More such double peaks are seen near $A = 130$ and $A = 200$, due to the respective $N = 82$ and $N = 126$ neutron shell closures. From these observed abundance peaks one can reconstruct geometrically the location of the waiting point regions on the nuclear chart. They are found to lie around $^{80}$Zn, $^{130}$Cd, and $^{195}$Tm for magic neutron numbers $N = 50$, $82$, and $126$, respectively.

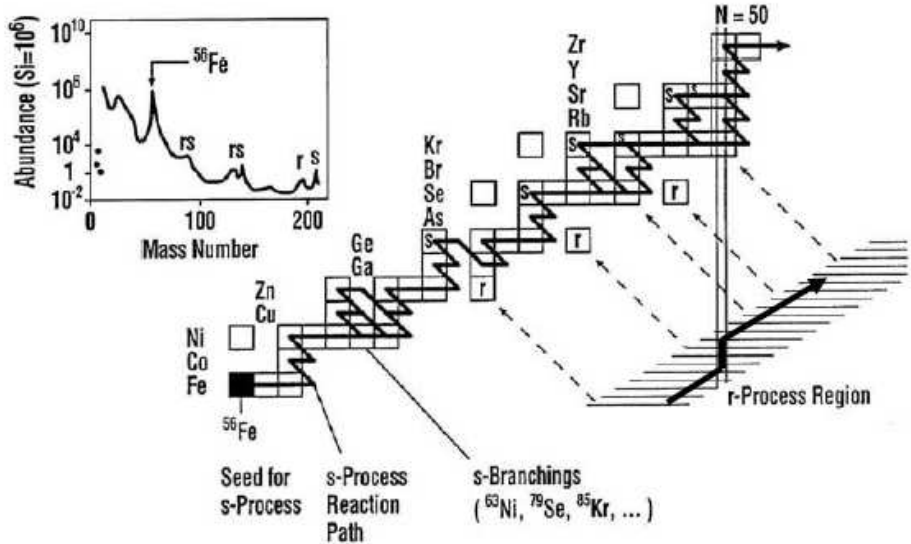

FIG. 30. Pathways of nucleosynthesis in the nuclear chart $Z$ vs. $N$. The slow or *s*-process proceeds near the valley of stability via neutron captures (horizontal lines) and subsequent $\beta^-$-decays (diagonal lines). The rapid or *r*-process takes place far-off stability after multiple neutron captures (hatched area). *s*-only stable isotopes are marked s, *r*-only isotopes are marked r. The isolated open squares are proton-rich stable isotopes reached only by the so-called *rp* or *p*-processes. The inset shows relative element abundances in the solar system, for the small peaks marked r and s see text. Adapted from Käppeler *et al.* (2010).

To give an example, the isotope $^{80}$Zn has 10 more neutrons than the last stable zinc isotope. Baruah *et al.* (2008) measured the masses of short-lived isotopes up to $^{81}$Zn with a precision of $\Delta m/m \approx 10^{-8}$, after deceleration of the neutron rich isotopes and capture in a Penning trap. From the measured masses of neighboring isotopes, the neutron separation energies were derived. From this one calculates the boundaries $T \geq 10^9$ K and $n \geq 10^{20}$ g/cm$^3$ for the neutron temperature and density at the location of the *r*-process.

Some nuclides can only be reached by the *r*-process, others only by the *s*-process, and many by both, see Fig. 30. Some nuclides on the proton rich side of the stable valley are reached neither by the *r*-process nor by the *s*-process. Their existence must be due either to a rapid proton capture, called an *rp*-process, or to $(\gamma,n)$ or $(\gamma,\alpha)$ photodisintegration in a very hot photon bath, called the *p*-process.

While the *s*-process is well understood, the *r*-, *rp*-, and *p*-processes are much less so. A possible site of these processes is in supernova explosions, where core collapse may cause extreme neutron fluxes in the expanding envelope. An alternative site is in binary systems that consist of a normal star feeding matter onto a companion neutron star. To clarify the nature of the *r*-process, responsible for about half the heavy nuclei in the solar system, one needs better nuclear data on isotopes far-off stability, which come from radioactive ion beam and from neutron facilities.

At ILL, the fission product spectrometer LOHENGRIN gives both the fission yields (Bail *et al.*, 2008) as well as spectroscopic information on a multitude of neutron-rich fission products. Bernas *et al.* (1991) did an early study of the start of the *r*-process via the $^{68}$Fe and $^{68,69}$Co isotopes, which are 9 and 10 neutrons away from stability. More recently, Rząca-Urban *et al.* (2007) studied the level scheme of the iodine isotope $^{138}$I, 11 neutrons away from the stable isotope $^{127}$I. For a review, see also Bernas (2001). With all these data on neutron rich isotopes one can hope that, in the course of time, one will be able to elucidate the nature of element formation and recycling processes going on in the universe.

To conclude, element-production in the universe, both primordial and stellar, is full of interesting and often unexplored neutron physics, and requires a great amount of precision data from nuclear accelerators and from fast and slow-neutron sources.

## VIII. CONCLUSIONS

Neutrons play an important role in the history of the universe for two reasons. Firstly, there is a high neutron abundance in the universe, and neutrons play a dominant role in the formation of the chemical elements, both shortly after the big bang, and during the ongoing stellar evolution. The second, more subtle reason is that today's experiments with cold and ultracold neutrons, together with high-energy collider experiments, shed light on processes going on at extremely high energy scales or temperatures in the early universe. These processes often take place at times much earlier than the time when neutrons made their first appearance, which was several microseconds after the start of the universe in a (modified) big bang model.

Limits on the electric dipole moment of the neutron, together with those of the electron and the proton, severely constrain *CP* and time reversal nonconserving theories beyond the Standard Model, in particular, supersymmetric models as the Minimal Supersymmetric Standard Model (MSSM), as is shown in Figs. 8 and 9. At the same time, these limits also severely constrain the ways how the process of baryogenesis could have happened, which is responsible for the evident dominance of matter over antimatter in the universe. How tight these constraints are today is shown in Figs. 9, 10, and 11.

In the next generation of EDM experiments, the experimental sensitivity will have passed the region where most models on electroweak baryogenesis predict a nonzero EDM. If these upcoming experiments find a neutron EDM, this will be a strong hint that the creation of matter-antimatter asymmetry has occurred around the electroweak scale. If they do not see an EDM, other ways of baryogenesis, as leptogenesis or other models at intermediate energy scales or close to the inflationary scale, are more probable.

The last few years have seen the emergence of a new field of neutron studies, the measurement of quantum states of ultracold neutrons in the earth's gravitational field, as shown in Figs. 14 and 15. These and other neutron investigations make it possible to test Newton's gravitational law down to atomic and subatomic distances. Deviations from Newton's law are required in new models with "compactified" extra spatial dimensions that elegantly solve the so-called hierarchy problem, the riddle of why the gravitational force is so much weaker than all other known forces.

These neutron measurements constrain new forces down to the sub-picometer range, see Fig. 16. Other neutron measurements constrain spin-dependent exotic forces down to the micrometer range, see Fig. 17. In this way, they start closing the last "window" where one could still find the axion, a particle needed to constrain strong *CP* violation, and a favorite candidate for dark matter. What will come next? The first experiments on neutron gravitational levels have just exploited the first prototype UCN instruments, and much progress can be expected from the next generation of instruments now under construction. In particular, transitions between gravitational quantum levels were measured in a recent experiment. In addition, both EDM and gravitational experiments will gain considerably once the new UCN sources under construction at several places will come into operation.

In the $\beta$-decay of free neutrons, many weak-interaction parameters are accessible to experiment, ten of which at present are under study, many of which are measured to a precision that competes well with the precision reached in muon decay. Neutron decay data are needed in several fields of science because today they are the only source to predict precisely semileptonic weak interaction parameters needed in particle physics, astrophysics, and cosmology. Neutron data further permit precise tests of several basic symmetries of our world. Figures 25 and 26 show exclusion plots derived from neutron decay on the so-called Fierz interference term in $\beta$-decay, which would signal interactions with underlying scalar and tensor symmetries. Processes with these symmetries are not foreseen in the Standard Model, which has pure $V-A$ symmetry, dominant in the universe for unknown reasons. Figures 27 and 28 show that from neutron decay data one can exclude a right-handed *W*-boson up to masses of rather high energy. The existence of such right-handed interactions would help to understand how the evident but still unexplained left-right asymmetry of the universe came into being. Still higher effective scales $\Lambda > 10$ TeV can be excluded from the present model-independent limits on violation of unitarity in CKM quark mixing.

The precision of main neutron decay parameters has strongly increased in recent years. Some errors seem to have been underestimated in the past, though with the recent advent of new results, in particular, on the neutron lifetime $\tau_n$ and the $\beta$-asymmetry *A*, data now seem to consolidate. Four new neutron decay observables, namely, *C*, *N*, *R*, and the branching ratio for the radiative decay $n \rightarrow p \, e \, \nu_e \, \gamma$, have become accessible recently, and more will follow. Constraints on models beyond the $V-A$ Standard Model (left-right symmetric, scalar, tensor, and other couplings) today all come from low-energy experiments, though are still less stringent than many people think. These models beyond the SM will be investigated at a much higher level of precision than is possible today, and multi-TeV energy scales become accessible with beams of milli-eV to nano-eV neutrons.

Ultracold neutrons, used in recent neutron-mirror neutron oscillation experiments, could further signal the presence of a mirror world, invoked to solve the still unexplained left-right asymmetry of the universe. Neutron processes were also used to test Einstein's mass-energy relation to an unprecedented accuracy, 50 times better than before. Furthermore, neutron-nucleon and neutron-nuclear weak interaction studies will provide another angle of attack onto the strong interaction.

Finally, in the last chapter, the role of the neutron in the formation of the elements was investigated, both during the "first three minutes", and, later on, during stellar processes. Here neutron and nuclear physics experiments provide whole networks of data needed to understand these processes of nucleosynthesis, with a continued strong need for better data. Our excursion led us from the first instances of the big bang all the way down to presently ongoing stellar processes, and showed how neutron physics, together with exciting developments in other fields, can contribute to the understanding of the emergence of the world as it is today.

## ACKNOWLEDGEMENTS

The authors thank T. Soldner, H. Abele, S. Baeßler, M. Bartelmann, C. Clarke, K. Leung, B. Märkisch, H. Mest, U. Schmidt, K. Schreckenbach, S. Sjue, and O. Zimmer for helpful comments and discussions. Part of this work is sponsored by the DFG priority program SPP 1491.